\documentclass[journal]{IEEEtran}

\usepackage{cite}

\ifCLASSINFOpdf
\else
   \usepackage[dvips]{graphicx}
\fi
\usepackage[cmex10]{amsmath}
\usepackage{amssymb}
\usepackage{balance}
\usepackage{enumitem}

\newtheorem{definition}{Definition}
\newtheorem{assumption}{Assumption}
\newtheorem{example}{Example}
\newtheorem{theorem}{Theorem}
\newtheorem{lemma}{Lemma}
\newtheorem{proposition}{Proposition}
\newtheorem{corollary}{Corollary}
\newtheorem{remark}{Remark}

\newcommand{\ato}{\overset{\mathrm{a.s.}}{\to}}
\newcommand{\aeq}{\overset{\mathrm{a.s.}}{=}}
\newcommand{\ag}{\overset{\mathrm{a.s.}}{>}}
\newcommand{\al}{\overset{\mathrm{a.s.}}{<}}

\begin{document}
%
\title{Bayes-Optimal Convolutional AMP}
%
%
%

\author{Keigo~Takeuchi,~\IEEEmembership{Member,~IEEE}
\thanks{
The author was in part supported by the Grant-in-Aid 
for Scientific Research~(B) (JSPS KAKENHI Grant Numbers 18H01441 and 
21H01326), Japan. 
The material in this paper was presented in part at 2019 IEEE 
International Symposium on Information Theory and submitted in part 
to 2021 IEEE International Symposium on Information Theory. 
}
\thanks{K.~Takeuchi is with the Department of Electrical and Electronic Information Engineering, Toyohashi University of Technology, Toyohashi 441-8580, Japan (e-mail: takeuchi@ee.tut.ac.jp).}
}

\markboth{IEEE transactions on information theory,~Vol.~, No.~,}%
{Takeuchi: Bayes-Optimal Convolutional AMP}
%

\IEEEpubid{0000--0000/00\$00.00~\copyright~2020 IEEE}


\maketitle

\begin{abstract}
This paper proposes Bayes-optimal convolutional approximate 
message-passing (CAMP) for signal recovery in compressed sensing. 
CAMP uses the same low-complexity matched filter (MF) for interference 
suppression as approximate message-passing (AMP). To improve the 
convergence property of AMP for ill-conditioned sensing matrices, 
the so-called Onsager correction term in AMP is replaced by a convolution 
of all preceding messages. The tap coefficients in the convolution 
are determined so as to realize asymptotic Gaussianity of estimation 
errors via state evolution (SE) under the assumption of orthogonally 
invariant sensing matrices. An SE equation is derived to optimize the 
sequence of denoisers in CAMP. The optimized CAMP is proved 
to be Bayes-optimal for all orthogonally invariant sensing matrices 
if the SE equation converges to a fixed-point and 
if the fixed-point is unique. 
For sensing matrices with low-to-moderate condition 
numbers, CAMP can achieve the same performance as high-complexity 
orthogonal/vector AMP that requires the linear minimum mean-square error 
(LMMSE) filter instead of the MF.
\end{abstract}


%
\IEEEpeerreviewmaketitle

\section{Introduction}
\subsection{Compressed Sensing}
\IEEEPARstart{C}{ompressed} sensing (CS)~\cite{Donoho06,Candes061} is 
a powerful technique for recovering sparse signals from compressed 
measurements. Under the assumption of linear measurements, CS is formulated 
as estimation of a sparse signal vector $\boldsymbol{x}\in\mathbb{R}^{N}$ 
from a compressed measurement vector $\boldsymbol{y}\in\mathbb{R}^{M}$ 
($M\leq N)$ and a sensing matrix $\boldsymbol{A}\in\mathbb{R}^{M\times N}$, 
given by 
\begin{equation} \label{model} 
\boldsymbol{y} = \boldsymbol{A}\boldsymbol{x} + \boldsymbol{w}, 
\end{equation}   
where $\boldsymbol{w}\in\mathbb{R}^{M}$ is an unknown additive noise vector. 

For simplicity in information-theoretic discussion~\cite{Wu10}, suppose that 
the signal vector $\boldsymbol{x}$ has independent and identically 
distributed (i.i.d.) elements. Sparsity of signals is measured with the 
R\'enyi information dimension~\cite{Renyi59} of each signal element. 
When each signal takes non-zero real values with probability $\rho\in[0, 1]$, 
the information dimension is equal to~$\rho$. In the noiseless case 
$\boldsymbol{w}=\boldsymbol{0}$, Wu and Verd\'u~\cite{Wu10} proved that, 
if and only if the compression rate $\delta=M/N$ is equal to or larger than 
the information dimension, there are some sensing matrix $\boldsymbol{A}$ and 
method for signal recovery such that the signal vector $\boldsymbol{x}$ can be  
recovered with negligibly small error probability in the large 
system limit, where $M$ and $N$ tend to infinity with the compression rate 
$\delta$ kept constant. Thus, an important issue in CS is a construction of 
practical sensing matrices and a low-complexity algorithm for signal recovery 
achieving the information-theoretic compression limit. 

Important examples of sensing matrices are zero-mean i.i.d.\ 
sensing matrices~\cite{Candes05} and random sensing matrices with 
orthogonal rows~\cite{Candes062}. The information-theoretic compression 
limit of zero-mean i.i.d.\ sensing matrices was analyzed with the non-rigorous 
replica method~\cite{Tanaka02,Guo051}---a tool developed in statistical 
mechanics~\cite{Mezard87,Nishimori01}. The compression limit is characterized 
via a potential function called free energy. The results themselves were 
rigorously justified in \cite{Reeves16,Reeves19,Barbier16,Barbier19} while 
the justification of the replica method is still open. It is a simple 
exercise to prove that the compression limit for zero-mean i.i.d.\ sensing 
matrices is equal to the R\'enyi information dimension in the noiseless case, 
by using a relationship between the information dimension and 
mutual information~\cite[Theorem~6]{Wu11}.  

\IEEEpubidadjcol

Random sensing matrices with orthogonal rows can be constructed 
efficiently in terms of both time and space complexity while zero-mean 
i.i.d.\ sensing matrices require ${\cal O}(MN)$ time and memory 
for matrix-vector multiplication.  
When the fast Fourier transform or fast Walsh-Hadamard transform is used, 
the matrix-vector multiplication needs ${\cal O}(N\log N)$ time and 
${\cal O}(N)$ memory. Thus, random sensing matrices with orthogonal rows 
are preferable from a practical point of view.  
 
The class of orthogonally invariant matrices includes zero-mean i.i.d.\ 
Gaussian matrices and Haar orthogonal matrices~\cite{Hiai00,Tulino04}, of 
which the latter is regarded as an idealized model of random matrices with 
orthogonal rows. The class allows us to analyze the information-theoretic 
compression limit in signal recovery. 
The replica method~\cite{Takeda06,Tulino13} was used 
to analyze the compression limit for orthogonally invariant sensing matrices. 
The replica results themselves were justified in \cite{Barbier18}. 
In particular, Haar orthogonal matrices achieve the Welch lower 
bound~\cite{Welch74} and were proved to be optimal for 
Gaussian~\cite{Rupf94} and general~\cite{Kitagawa10} signals. 
In the noiseless case, of course, Haar orthogonal sensing matrices achieve 
the compression rate that is equal to the R\'enyi information dimension. 

In practical systems, the measurement vector is subject not only to 
additive noise but also to multiplicative noise. A typical example is 
fading in wireless communication systems~\cite{Tse05,Goldsmith05}. 
The effective sensing matrix containing fading influence may be 
ill-conditioned even if a Haar orthogonal sensing matrix is used. 
Such effective sensing matrices can be modeled as orthogonally invariant 
matrices. Thus, an ultimate algorithm for signal recovery is required to be low 
complexity and Bayes-optimal for all orthogonally invariant sensing matrices.

\subsection{Message-Passing}
A promising solution to signal recovery is message-passing (MP). 
Approximate message-passing (AMP)~\cite{Donoho09} is a low-complexity and 
powerful algorithm for signal recovery from zero-mean i.i.d.\ sub-Gaussian 
measurements. Bayes-optimal AMP is regarded as an exact large-system 
approximation of loopy belief propagation (BP)~\cite{Kabashima03}. The main 
feature of AMP is the so-called Onsager correction to realize asymptotic 
Gaussianity of the estimation errors before denoising. The Onsager 
correction originates from that in the Thouless-Anderson-Palmer (TAP) 
equation~\cite{Thouless77} for a solvable  spin glass model with i.i.d.\ 
interaction between all spins~\cite{Sherrington75}. The Onsager correction 
cancels intractable dependencies of the current estimation error on past 
estimation errors due to i.i.d.\ dense sensing matrices. 

The convergence property of AMP was analyzed rigorously via state 
evolution (SE)~\cite{Bayati11,Bayati15}, inspired by Bolthausen's conditioning 
technique~\cite{Bolthausen14}. SE is a dense counterpart of density 
evolution~\cite{Richardson08} in sparse systems. SE tracks a few 
state variables to describe rigorous dynamics of MP in the large system 
limit. SE analysis in \cite{Bayati11,Bayati15} implies that AMP is 
Bayes-optimal for zero-mean i.i.d.\ sub-Gaussian sensing matrices when the 
compression rate $\delta$ is larger than a certain value called BP 
threshold~\cite{Takeuchi15}. Spatial 
coupling~\cite{Kudekar11,Krzakala12,Donoho13,Takeuchi15} is needed to 
realize the optimality of AMP for any compression rate. However, this paper 
does not consider spatial coupling since spatial coupling is a universal 
technique~\cite{Takeuchi15} to improve the performance of MP. 

A disadvantage of AMP is that AMP fails to converge when the sensing matrix 
is non-zero mean~\cite{Caltagirone14} or ill-conditioned~\cite{Rangan191}.  
To solve this issue, orthogonal AMP (OAMP)~\cite{Ma17} and vector 
AMP~\cite{Rangan17,Rangan192} were proposed. The two MP algorithms are 
equivalent to each other. Bayes-optimal OAMP/VAMP can be regarded as an exact 
large-system approximation of expectation propagation 
(EP)~\cite{Minka01,Cespedes14,Takeuchi171,Takeuchi201}. 
Rigorous SE analysis~\cite{Rangan17,Rangan192,Takeuchi171,Takeuchi201}   
proved that OAMP/VAMP is Bayes-optimal for orthogonally invariant sensing 
matrices when the compression rate is larger than BP threshold. While 
non-zero mean matrices are outside the class of orthogonally invariant 
matrices, numerical simulations in \cite{Rangan192} indicated that OAMP/VAMP 
can treat the non-zero mean case. 

A prototype of OAMP/VAMP was originally proposed by Opper and 
Winther~\cite[Appendix~D]{Opper05}. Historically, they~\cite{Opper01} 
generalized the Onsager correction in the TAP equation~\cite{Thouless77} from   
zero-mean i.i.d.\ spin interaction to orthogonally invariant interaction.  
Their method was formulated as the expectation-consistency (EC) 
approximation~\cite{Opper05}. The EC approximation itself does not produce MP 
algorithms but a potential function of which a local minimum should be solved 
with some MP algorithm. OAMP/VAMP can be derived from an EP-type 
iteration--called a single loop algorithm~\cite{Opper05}---to solve a local 
minimum of the EC potential. See \cite[Appendix~A]{Tatsuno21} 
for the derivation of OAMP/VAMP via the EC approximation. 
 
The main weakness of OAMP/VAMP is a per-iteration requirement of the linear 
minimum mean-square error (LMMSE) filter, of which the time complexity is 
${\cal O}(M^{3}+M^{2}N)$ per iteration. The singular-value decomposition (SVD) 
of the sensing matrix allows us to circumvent the use of the LMMSE 
filter~\cite{Rangan192}. However, the complexity of the SVD itself is high 
in general. The performance of OAMP/VAMP degrades 
significantly when the LMMSE filter is replaced by the low-complexity 
matched filter (MF)~\cite{Ma17} used in AMP. Thus, OAMP/VAMP can be applied 
only to limited problems in which the SVD of the sensing matrix is 
computed efficiently.  

In summary, it is still open to construct a low-complexity and Bayes-optimal 
MP algorithm for all orthogonally invariant sensing matrices. The purpose of 
this paper is to tackle the design issue of such ultimate MP algorithms. 

\subsection{Methodology}
The main idea of this paper is to extend the class of MP algorithms.  
Conventional MP algorithms use update rules that depend only on messages 
in the latest iteration. Long-memory MP algorithms considered in this paper 
are allowed to depend on messages in all preceding iterations. 

This class of long-memory MP algorithms was motivated by SE  
analysis of AMP for orthogonally invariant sensing matrices~\cite{Takeuchi19}. 
When the asymptotic singular-value distribution of the sensing matrix is equal 
to that of zero-mean i.i.d.\ Gaussian matrices, the error model of AMP was 
proved to be an instance of a general error model~\cite{Takeuchi19}, 
in which each error depends on errors in all preceding iterations.   
This result implies that the Onsager correction in AMP uses messages in all 
preceding iterations to realize the asymptotic Gaussianity of the current 
estimation error while the representation itself of the correction term looks 
as if only messages in the latest iteration are utilized. Inspired by this 
observation, we consider long-memory MP algorithms as a starting point.  

The idea of long-memory MP was originally proposed in Opper, \c{C}akmak, 
and Winther's paper~\cite{Opper16} to solve the TAP 
equations for spin glass models with orthogonally invariant interaction.    
Their methodology was based on non-rigorous dynamical functional theory. 
After the initial submission of this paper, their results were rigorously 
justified via SE in \cite{Fan20}. 

The proposed design of long-memory MP consists of three steps: A first step 
is an establishment of rigorous SE for analyzing the dynamics of 
long-memory MP algorithms for orthogonally invariant sensing 
matrices. This step has been already established in \cite{Takeuchi19} 
by generalizing conventional SE analysis~\cite{Rangan192,Takeuchi201} to 
the long-memory case. The SE analysis provides a sufficient condition for 
a long-memory MP algorithm to have Gaussian-distributed estimation errors 
in the large system limit.  
The main advantage in the SE analysis is to provide a systematic design 
of long-memory MP that satisfies the asymptotic Gaussianity in estimation 
errors while the class of long-memory MP is slightly smaller than in 
\cite{Opper16,Fan20}.    

A second step is to modify the Onsager correction in AMP so as to satisfy the 
sufficient condition for the asymptotic Gaussianity. A solvable class of 
long-memory MP was proposed in \cite{Takeuchi202}, where the Onsager 
correction was defined as a convolution of messages in all preceding 
iterations. The tap coefficients in the convolution were determined so as to 
satisfy the sufficient condition. Thus, long-memory MP proposed 
in \cite{Takeuchi202} was called convolutional AMP (CAMP) and 
is the main object of this paper. 

This paper generalizes CAMP in \cite{Takeuchi202}, motivated by an 
implementation of OAMP/VAMP based on conjugate gradient 
(CG)~\cite{Takeuchi172}. OAMP/VAMP applies the LMMSE filter to a message 
$\boldsymbol{z}\in\mathbb{R}^{M}$ 
after interference subtraction. The LMMSE filter is decomposed into a 
noise-whitening filter and MF. In principle, CG approximates the output of 
the noise-whitening filter with a vector in the Krylov subspace spanned by 
$\{\boldsymbol{z}, \boldsymbol{A}\boldsymbol{A}^{\mathrm{T}}\boldsymbol{z}, 
(\boldsymbol{A}\boldsymbol{A}^{\mathrm{T}})^{2}\boldsymbol{z}, 
\ldots\}$ i.e.\ a finite weighted sum of 
$\{(\boldsymbol{A}\boldsymbol{A}^{\mathrm{T}})^{j}\boldsymbol{z}\}$. 
On the other hand, messages in the original CAMP~\cite{Takeuchi202} are in 
the $0$th Krylov subspace $\{\alpha\boldsymbol{z}: \alpha\in\mathbb{R}\}$ 
since only the MF is used. To fill this gap, we generalize a convolution 
of all preceding messages in the original CAMP~\cite{Takeuchi202} to that 
of affine transforms of the preceding messages. 

The last step is to optimize the sequence of denoisers in CAMP. 
The last step is a new contribution of this paper, submitted 
to \cite{Takeuchi21}. The optimization requires information on 
the distribution of the estimation errors before denoising in each 
iteration. Since the estimation errors  are asymptotically 
Gaussian-distributed, we need to track the dynamics of the variance of 
the estimation errors. To analyze this dynamics, we utilize the SE analysis 
established in the first step.

\subsection{Contributions}
The contributions of this paper are sixfold: A first contribution 
(Theorem~\ref{theorem_SE} in Section~\ref{sec2}) is to propose a general error 
model for long-memory MP and prove the asymptotic Gaussianity of estimation 
errors in the general error model via rigorous SE under the assumption of 
orthogonally invariant sensing matrices. The general error model contains 
both error models of AMP and OAMP/VAMP.  

A second contribution (Section~\ref{sec3A}) is the addition of a convolution 
proportional to $\boldsymbol{A}\boldsymbol{A}^{\mathrm{T}}$ to the Onsager 
correction in \cite{Takeuchi202}, according to the above-mentioned 
argument on the Krylov subspace. This addition improves the convergence 
property of CAMP. 

A third contribution (Theorem~\ref{theorem_CAMP} in Section~\ref{sec3C}) is 
to design tap coefficients in the convolution so as to guarantee the 
asymptotic Gaussianity of estimation errors for all orthogonally invariant 
sensing matrices. Part of the tap coefficients are used to realize the 
asymptotic Gaussianity. The remaining coefficients can be utilized to improve 
the convergence property of CAMP.   

A fourth contribution (Theorem~\ref{theorem_solution} in Section~\ref{sec3C}) 
is to present the designed tap coefficients in closed-form. 
This closed-form representation circumvents numerical instability in solving 
the tap coefficients numerically. The third and fourth contributions are 
based on the same proof strategy as in \cite{Takeuchi202}. 
 
A fifth contribution (Theorems~\ref{theorem_CAMP_SE} and 
\ref{theorem_fixed_point} in Section~\ref{sec3D}) 
is to optimize the sequence of denoisers in CAMP. An SE equation is derived to 
describe the dynamics of the variance of the estimation errors before 
denoising in CAMP. The SE equation is a two-dimensional 
nonlinear difference equation. By analyzing the fixed-point of the SE equation, 
we prove that optimized CAMP is Bayes-optimal for all orthogonally invariant 
sensing matrices if the SE equation converges to a fixed-point and 
if the fixed-point is unique. 

The last contribution (Section~\ref{sec4}) is numerical evaluation of CAMP. 
The remaining parameters in the Bayes-optimal CAMP are optimized numerically 
to improve the convergence property. Numerical simulations show that the CAMP 
can converge for sensing matrices with larger condition numbers than the 
original CAMP~\cite{Takeuchi202} when the design parameters are optimized.  
The CAMP can achieve the same performance as OAMP/VAMP for sensing matrices 
with low-to-moderate condition numbers while it is inferior to OAMP/VAMP for 
high condition numbers. 

\subsection{Organization}
The remainder of this paper is organized as follows: After summarizing the 
notation used in this paper, we present a unified SE framework for analyzing 
long-memory MP under the assumption of orthogonally invariant sensing 
matrices in Section~\ref{sec2}. This section corresponds to the first step 
for proposing Bayes-optimal CAMP. 

In Section~\ref{sec3}, we propose CAMP with design parameters. 
This section corresponds to the remaining two steps for establishing 
Bayes-optimal CAMP. The proposed 
CAMP is more general than in \cite{Takeuchi202}. We utilize the SE framework 
established in Section~\ref{sec2} to determine the tap coefficients in 
CAMP that guarantee the asymptotic Gaussianity of estimation errors. To 
design the remaining design parameters, we derive an SE equation to optimize 
the performance of signal recovery. 

Section~\ref{sec4} presents numerical results. The remaining design parameters 
in CAMP are optimized via numerical simulations. 
The optimized CAMP is compared to conventional AMP and OAMP/VAMP via the 
SE equation and numerical simulations. Section~\ref{sec5} concludes this 
paper. The details for the proofs of the main theorems are presented in  
appendices. 

\subsection{Notation}
For a matrix $\boldsymbol{M}$, the transpose of $\boldsymbol{M}$ is 
denoted by $\boldsymbol{M}^{\mathrm{T}}$. The notation 
$\mathrm{Tr}(\boldsymbol{A})$ represents the trace of a square matrix 
$\boldsymbol{A}$. For a symmetric matrix $\boldsymbol{A}$, the minimum 
eigenvalue of $\boldsymbol{A}$ is written as 
$\lambda_{\mathrm{min}}(\boldsymbol{A})$. 
The notation $\mathcal{O}_{M\times N}$ denotes 
the space of all possible $M\times N$ matrices with orthonormal columns for 
$M\geq N$ and orthonormal rows for $M<N$. In particular, 
$\mathcal{O}_{N\times N}$ reduces to the space $\mathcal{O}_{N}$ of 
all possible $N\times N$ orthogonal matrices. 

For a vector $\boldsymbol{v}$, 
the notation $\mathrm{diag}(\boldsymbol{v})$ denotes the diagonal matrix 
of which the $n$th diagonal element is equal to $v_{n}=[\boldsymbol{v}]_{n}$. 
The norm $\|\boldsymbol{v}\|=\sqrt{\boldsymbol{v}^{\mathrm{T}}\boldsymbol{v}}$ 
represents the Euclidean norm. 
For a matrix $\boldsymbol{M}_{i}$ with an index~$i$, the $t$th column 
of $\boldsymbol{M}_{i}$ is denoted by $\boldsymbol{m}_{i,t}$. Furthermore, 
we write the $n$th element of $\boldsymbol{m}_{i,t}$ as $m_{i,t,n}$. 

The Kronecker delta is denoted by $\delta_{\tau,t}$ while the Dirac delta 
function is represented as $\delta(\cdot)$. We write the Gaussian 
distribution with mean $\boldsymbol{\mu}$ and covariance $\boldsymbol{\Sigma}$ 
as $\mathcal{N}(\boldsymbol{\mu},\boldsymbol{\Sigma})$. The notations 
$\ato$ and $\aeq$ denote almost sure convergence and equivalence, 
respectively.   

We use the notational convention $\sum_{t=t_{1}}^{t_{2}}\cdots=0$ and 
$\prod_{t=t_{1}}^{t_{2}}\cdots=1$ for $t_{1}>t_{2}$. 
For any multivariate function $\phi:\mathbb{R}^{t}\to\mathbb{R}$, the 
notation $\partial_{t'}\phi$ for $t'=0,\ldots,t-1$ denotes the partial 
derivative of $\phi$ with respect to the $t'$th variable $x_{t'}$, 
\begin{equation}
\partial_{t'}\phi = \frac{\partial\phi}{\partial x_{t'}}
(x_{0},\ldots,x_{t-1}). 
\end{equation}
For any vector $\boldsymbol{v}\in\mathbb{R}^{N}$, 
the notation $\langle \boldsymbol{v}\rangle=N^{-1}\sum_{n=1}^{N}v_{n}$ represents 
the arithmetic mean of the elements. 
For any scalar function $f:\in\mathbb{R}\to\mathbb{R}$, the notation 
$f(\boldsymbol{v})$ means the element-wise application of $f$ to a vector 
$\boldsymbol{v}$, i.e.\ $[f(\boldsymbol{v})]_{n}=f(v_{n})$. 

For a sequence $\{p_{t}\}_{t=0}^{\infty}$, we define the Z-transform of 
$\{p_{t}\}$ as 
\begin{equation}
P(z) = \sum_{t=0}^{\infty}p_{t}z^{-t}. 
\end{equation}
For two sequences $\{p_{t}, q_{t}\}_{t=0}^{\infty}$, 
we define the convolution operator $*$ as 
\begin{equation}
p_{t+i}*q_{t+j} = \sum_{\tau=0}^{t}p_{\tau+i}q_{t-\tau+j}, 
\end{equation}
with $p_{t}=0$ and $q_{t}=0$ for $t<0$. For finite-length sequences 
$\{p_{t}\}_{t=0}^{T}$ of length~$T+1$, we transform them into infinite-length 
sequences by adding $p_{t}=0$ and $q_{t}=0$ for all $t>T$. 

For two arrays $\{a_{t',t}, b_{t',t}: t', t=0,\ldots, \infty\}$, 
we write the two-dimensional convolution as 
\begin{equation}
a_{t'+i,t+j}*b_{t'+k,t+l} 
= \sum_{\tau'=0}^{t'}\sum_{\tau=0}^{t}a_{\tau'+i,\tau+j}b_{t'-\tau'+k,t-\tau+l}, 
\end{equation}
where $a_{t',t}=0$ and $b_{t',t}=0$ are defined for $t'<0$ or $t<0$. 

Whether a convolution is one-dimensional can be distinguished as follows: 
A convolution is one-dimensional, such as $a_{t+i}*b_{t+j}$, when both operands 
contain only one identical subscript. On the other hand, a convolution is 
two-dimensional, such as $(a_{t'}a_{t+i})*b_{t'+j,t}$, when both operands 
include two identical subscripts.

\section{Unified Framework} \label{sec2}
\subsection{Definitions and Assumptions}
We define the statistical properties of the random variables in the 
measurement model~(\ref{model}). 
The performance of MP is commonly measured in terms of the mean-square error 
(MSE). Nonetheless, we follow \cite{Bayati11} to consider 
a general performance measure in terms of separable and pseudo-Lipschitz 
functions while we assume the separability and Lipschitz-continuity 
for denoisers. 

\begin{definition}
A vector-valued function $\boldsymbol{f}=(f_{1},\ldots,f_{N})^{\mathrm{T}}
:\mathbb{R}^{N\times t}\to\mathbb{R}^{N}$ is said to be 
separable if $[\boldsymbol{f}(\boldsymbol{x}_{1},\ldots,\boldsymbol{x}_{t})]_{n}
=f_{n}(x_{1,n},\ldots,x_{t,n})$ holds for all 
$\boldsymbol{x}_{i}\in\mathbb{R}^{N}$. 
\end{definition}  

\begin{definition}
A function $f:\mathbb{R}^{t}\to\mathbb{R}$ is said to be pseudo-Lipschitz 
of order~$k$~\cite{Bayati11} if there are some Lipschitz constant $L>0$ and 
some order $k\in\mathbb{N}$ such that 
for all $\boldsymbol{x}\in\mathbb{R}^{t}$ and $\boldsymbol{y}\in\mathbb{R}^{t}$ 
\begin{equation}
|f(\boldsymbol{x}) - f(\boldsymbol{y})| 
\leq L(1 + \|\boldsymbol{x}\|^{k-1} + \|\boldsymbol{y}\|^{k-1})
\|\boldsymbol{x} - \boldsymbol{y}\|. 
\end{equation}
\end{definition}

By definition, any pseudo-Lipschitz function of order~$k=1$ is 
Lipschitz-continuous. A vector-valued function 
$\boldsymbol{f}=(f_{1},\ldots,f_{N})^{\mathrm{T}}$ is pseudo-Lipschitz if all 
element functions $\{f_{n}\}$ are pseudo-Lipschitz. 

\begin{definition}
A separable pseudo-Lipschitz function 
$\boldsymbol{f}:\mathbb{R}^{N\times t}\to\mathbb{R}^{N}$ is said to be proper 
if the Lipschitz constant $L_{n}>0$ of the $n$th function $f_{n}$ satisfies 
\begin{equation}
\limsup_{N\to\infty}\frac{1}{N}\sum_{n=1}^{N}L_{n}^{j}<\infty
\end{equation}
for any $j\in\mathbb{N}$. 
\end{definition}

A proper pseudo-Lipschitz function allows us apply a proof strategy for 
pseudo-Lipschitz functions with $n$-independent Lipschitz constant $L_{n}=L$ 
to the $n$-dependent case straightforwardly.    
The space of all possible separable and proper pseudo-Lipschitz functions of 
order~$k$ is denoted by $\mathcal{PL}(k)$. We have the inclusion relation 
$\mathcal{PL}(k)\subset\mathcal{PL}(k')$ for all $k<k'$ since 
$\|\boldsymbol{x}\|^{k}\leq\|\boldsymbol{x}\|^{k'}$ holds for 
$\|\boldsymbol{x}\|\gg1$.




We assume statistical properties of the signal vector associated 
with separable and proper pseudo-Lipschitz functions of order~$k\geq 2$. 
Note that the integer~$k$ in the following assumptions is an identical 
parameter that is equal to the order of separable and proper 
pseudo-Lipschitz functions used in SE to measure the performance of MP. 
If the MSE is considered, the integer~$k$ is set to $2$.  

\begin{assumption} \label{assumption_x}
The signal vector $\boldsymbol{x}$ satisfies the following strong law of 
large numbers:   
\begin{equation}
\langle \boldsymbol{f}(\boldsymbol{x})\rangle
- \mathbb{E}\left[
 \langle \boldsymbol{f}(\boldsymbol{x})\rangle
\right]\ato0 
\end{equation}
as $N\to\infty$ for any separable and proper pseudo-Lipschitz function 
$\boldsymbol{f}:\mathbb{R}^{N}\to\mathbb{R}^{N}$ of order~$k\geq2$. Furthermore, 
$\boldsymbol{x}$ has zero-mean and bounded $(2k-2+\epsilon)$th moments 
for some $\epsilon>0$.
\end{assumption}

Assumption~\ref{assumption_x} follows from the classical strong law of large 
numbers when $\boldsymbol{x}$ has i.i.d.\ elements. 

\begin{definition}
An orthogonal matrix $\boldsymbol{V}\in\mathcal{O}_{N}$ is said to be 
Haar-distributed~\cite{Hiai00} if $\boldsymbol{V}$ is orthogonally invariant, 
i.e.\ $\boldsymbol{V}\sim\boldsymbol{\Phi}\boldsymbol{V}\boldsymbol{\Psi}$ 
for all orthogonal matrices $\boldsymbol{\Phi}, 
\boldsymbol{\Psi}\in\mathcal{O}_{N}$ independent of $\boldsymbol{V}$. 
\end{definition}
\begin{assumption} \label{assumption_A} 
The sensing matrix $\boldsymbol{A}$ is right-orthogonally invariant, i.e.\ 
$\boldsymbol{A}\sim\boldsymbol{A}\boldsymbol{\Psi}$ for any orthogonal 
matrix $\boldsymbol{\Psi}\in\mathcal{O}_{N}$ independent of $\boldsymbol{A}$. 
More precisely, the orthogonal matrix $\boldsymbol{V}\in\mathcal{O}_{N}$ 
in the SVD $\boldsymbol{A}=\boldsymbol{U}\boldsymbol{\Sigma}
\boldsymbol{V}^{\mathrm{T}}$ is Haar-distributed and independent of 
$\boldsymbol{U}\boldsymbol{\Sigma}$. Furthermore, the empirical eigenvalue 
distribution of $\boldsymbol{A}^{\mathrm{T}}\boldsymbol{A}$ converges almost 
surely to a compactly supported deterministic distribution with unit first 
moment in the large system limit. 
\end{assumption}

The assumption of unit first moment implies the almost sure convergence 
$N^{-1}\mathrm{Tr}(\boldsymbol{A}^{\mathrm{T}}\boldsymbol{A})\ato1$ 
in the large system limit. Assumption~\ref{assumption_A} holds when 
$\boldsymbol{A}$ has zero-mean i.i.d.\ Gaussian elements with variance 
$M^{-1}$. As shown in SE, the asymptotic Gaussianity of 
estimation errors in MP depends heavily on the Haar assumption of 
$\boldsymbol{V}$. Intuitively, the orthogonal transform $\boldsymbol{V}
\boldsymbol{a}$ of a vector $\boldsymbol{a}\in\mathbb{R}^{N}$ is distributed as 
$N^{-1/2}\|\boldsymbol{a}\|\boldsymbol{z}$ in which 
$\boldsymbol{z}\sim\mathcal{N}(\boldsymbol{0},\boldsymbol{I}_{N})$ is a 
standard Gaussian vector and independent of $\|\boldsymbol{a}\|$. When  
the amplitude $N^{-1/2}\|\boldsymbol{a}\|$ tends to a constant as $N\to\infty$, 
the vector $\boldsymbol{V}\boldsymbol{a}$ looks like a Gaussian vector. 
This is a rough intuition on the asymptotic Gaussianity of estimation 
errors. 

\begin{assumption} \label{assumption_w}
The noise vector $\boldsymbol{w}$ is orthogonally 
invariant, i.e.\ $\boldsymbol{w}\sim\boldsymbol{\Phi}\boldsymbol{w}$ for 
any orthogonal matrix $\boldsymbol{\Phi}\in\mathcal{O}_{M}$ independent of 
$\boldsymbol{w}$. Furthermore, $\boldsymbol{w}$ has zero-mean,  
$\lim_{M\to\infty}M^{-1}\|\boldsymbol{w}\|^{2}\aeq\sigma^{2}>0$, and 
bounded $(2k-2+\epsilon)$th moments for some $\epsilon>0$.
\end{assumption}

Assumption~\ref{assumption_w} holds when $\boldsymbol{w}\sim
\mathcal{N}(\boldsymbol{0},\sigma^{2}\boldsymbol{I}_{M})$ is an additive white 
Gaussian noise (AWGN) vector. It holds for 
$\boldsymbol{U}^{\mathrm{T}}\boldsymbol{w}$ when the sensing matrix 
$\boldsymbol{A}$ is 
left-orthogonally invariant, i.e.\ $\boldsymbol{A}\sim\boldsymbol{\Phi}
\boldsymbol{A}$ for any orthogonal matrix $\boldsymbol{\Phi}\in\mathcal{O}_{M}$ 
independent of $\boldsymbol{A}$.

\subsection{General Error Model} 
We propose a unified framework of SE for analyzing MP algorithms 
that have asymptotically Gaussian-distributed estimation errors for 
orthogonally invariant sensing matrices. Instead of starting with concrete 
MP algorithms, we consider a general class of error models. The proposed class 
does not necessarily contain the error models of all possible long-memory 
MP algorithms. However, it is a natural class of error models that allows 
us to prove the asymptotic Gaussianity of estimation errors for orthogonally 
invariant sensing matrices via a generalization of conventional 
SE~\cite{Takeuchi201}.  

Let $\boldsymbol{h}_{t}\in\mathbb{R}^{N}$ and 
$\boldsymbol{q}_{t+1}\in\mathbb{R}^{N}$ denote error vectors in iteration~$t$ 
before and after denoising, respectively. We assume that the error vectors 
are recursively given by 
\begin{equation} \label{b}
\boldsymbol{b}_{t} = 
\boldsymbol{V}^{\mathrm{T}}\tilde{\boldsymbol{q}}_{t}, \quad 
\tilde{\boldsymbol{q}}_{t} 
= \boldsymbol{q}_{t} - \sum_{t'=0}^{t-1}
\langle\partial_{t'}\boldsymbol{\psi}_{t-1}\rangle\boldsymbol{h}_{t'}, 
\end{equation}
\begin{equation} \label{m}
\boldsymbol{m}_{t} 
= \boldsymbol{\phi}_{t}(\boldsymbol{B}_{t+1}, \tilde{\boldsymbol{w}}; 
\boldsymbol{\lambda}),  
\end{equation} 
\begin{equation} \label{h}
\boldsymbol{h}_{t} 
= \boldsymbol{V}\tilde{\boldsymbol{m}}_{t}, 
\quad \tilde{\boldsymbol{m}}_{t} 
= \boldsymbol{m}_{t} - \sum_{t'=0}^{t}
\langle\partial_{t'}\boldsymbol{\phi}_{t}\rangle\boldsymbol{b}_{t'},
\end{equation}
\begin{equation} \label{q}
\boldsymbol{q}_{t+1} 
= \boldsymbol{\psi}_{t}
(\boldsymbol{H}_{t+1}, \boldsymbol{x}), 
\end{equation}
with $\boldsymbol{q}_{0}=-\boldsymbol{x}$. In (\ref{b}), 
the orthogonal matrix $\boldsymbol{V}\in\mathcal{O}_{N}$ consists of the 
right-singular vectors in the SVD 
$\boldsymbol{A}=\boldsymbol{U}\boldsymbol{\Sigma}
\boldsymbol{V}^{\mathrm{T}}$, with $\boldsymbol{U}\in\mathcal{O}_{M}$. 
In (\ref{m}) and (\ref{q}), we have defined $\boldsymbol{B}_{t+1}
=(\boldsymbol{b}_{0},\ldots,\boldsymbol{b}_{t})$ and 
$\boldsymbol{H}_{t+1}=(\boldsymbol{h}_{0},\ldots,\boldsymbol{h}_{t})$. 
Furthermore, $\boldsymbol{\lambda}\in\mathbb{R}^{N}$ is the vector of 
eigenvalues of $\boldsymbol{A}^{\mathrm{T}}\boldsymbol{A}$. 
The vector $\tilde{\boldsymbol{w}}\in\mathbb{R}^{N}$ is given by  
\begin{equation} \label{w_tilde}
\tilde{\boldsymbol{w}}
=\begin{bmatrix}
\boldsymbol{U}^{\mathrm{T}}\boldsymbol{w} \\ 
\boldsymbol{0}
\end{bmatrix}, 
\end{equation} 
where $\boldsymbol{w}$ is the additive noise vector in (\ref{model}). 

The vector-valued functions 
$\boldsymbol{\phi}_{t}:\mathbb{R}^{N\times(t+3)}\to\mathbb{R}^{N}$ and 
$\boldsymbol{\psi}_{t}:\mathbb{R}^{N\times(t+2)}\to\mathbb{R}^{N}$ are assumed 
to be separable, nonlinear, and proper Lipschitz-continuous. 
\begin{assumption} \label{assumption_thresholding}
The functions $\boldsymbol{\phi}_{t}$ and $\boldsymbol{\psi}_{t}$ are 
separable. The nonlinearities $\boldsymbol{\phi}_{t}\neq\sum_{t'=0}^{t}
\boldsymbol{D}_{t'}\boldsymbol{b}_{t'}$ and 
$\boldsymbol{\psi}_{t}\neq\sum_{t'=0}^{t}\tilde{\boldsymbol{D}_{t'}}
\boldsymbol{h}_{t'}$ 
hold for all diagonal matrices $\{\boldsymbol{D}_{t'}, 
\tilde{\boldsymbol{D}}_{t'}\}$. The function $\boldsymbol{\phi}_{t}$ is 
Lipschitz-continuous with respect to the first $t+2$ variables and proper 
while $\boldsymbol{\psi}_{t}$ is proper Lipschitz-continuous with respect to 
all variables.  
\end{assumption}

It might be possible to relax Assumption~\ref{assumption_thresholding} to 
the non-separable case~\cite{Berthier19,Ma19,Fletcher19}. For simplicity, 
however, this paper postulates separable denoisers. 
The nonlinearity is a technical condition for circumventing the zero norm  
$N^{-1}\|\tilde{\boldsymbol{q}}_{t}\|^{2}=0$ or 
$N^{-1}\|\tilde{\boldsymbol{m}}_{t}\|^{2}=0$, which implies error-free 
estimation $N^{-1}\|\boldsymbol{b}_{t}\|^{2}=0$ or 
$N^{-1}\|\boldsymbol{h}_{t}\|^{2}=0$.  

By definition, the $n$th function $\phi_{t,n}$ has a $\lambda_{n}$-dependent 
Lipschitz constant $L_{n}=L_{n}(\lambda_{n})$. Thus, the proper assumption for 
$\boldsymbol{\phi}_{t}$ may be regarded as a 
condition on the asymptotic eigenvalue distribution of 
$\boldsymbol{A}^{\mathrm{T}}
\boldsymbol{A}$, as well as a condition on the denoiser 
$\boldsymbol{\phi}_{t}$. For example, $\boldsymbol{\phi}_{t}$ is proper 
when the asymptotic eigenvalue distribution has a compact support and when 
the Lipschitz constant $L_{n}(\lambda_{n})$ itself is a pseudo-Lipschitz 
function of $\lambda_{n}$.  

The main feature of the general error model is in the definitions of 
$\tilde{\boldsymbol{q}}_{t}$ and $\tilde{\boldsymbol{m}}_{t}$. The 
second terms on the right-hand sides (RHSs) of (\ref{b}) and (\ref{h}) 
are correction terms to realize the asymptotic Gaussianity of 
$\{\boldsymbol{b}_{t}\}$ and $\{\boldsymbol{h}_{t}\}$. The correction terms 
are a modification of conventional correction that allows us to prove  
the asymptotic Gaussianity via a natural generalization~\cite{Campese15} 
of Stein's lemma used in conventional SE~\cite{Takeuchi201}. See 
Lemma~\ref{lemma_Stein} in Appendix~\ref{proof_theorem_SE} for the details.  

The following examples imply that the general error model~(\ref{b})--(\ref{q}) 
contains those of OAMP/VAMP and AMP. 

\begin{example}
Consider OAMP/VAMP~\cite{Ma17,Rangan192} with 
a sequence of scalar denoisers $f_{t}:\mathbb{R}\to\mathbb{R}$: 
\begin{equation} \label{module_A_mean}
\boldsymbol{x}_{\mathrm{A}\to \mathrm{B},t} = \boldsymbol{x}_{\mathrm{B}\to \mathrm{A},t} 
+ \gamma_{t}\boldsymbol{A}^{\mathrm{T}}\boldsymbol{W}_{t}^{-1}
(\boldsymbol{y} - \boldsymbol{A}\boldsymbol{x}_{\mathrm{B}\to \mathrm{A},t}), 
\end{equation}
\begin{equation}
v_{\mathrm{A}\to \mathrm{B},t} = \gamma_{t} - v_{\mathrm{B}\to \mathrm{A},t}, 
\end{equation}
\begin{equation}
\boldsymbol{W}_{t} 
= \sigma^{2}\boldsymbol{I}_{M} 
+ v_{\mathrm{B}\to \mathrm{A},t}\boldsymbol{A}\boldsymbol{A}^{\mathrm{T}},
\end{equation}
\begin{equation} \label{gamma_t}
\gamma_{t}^{-1} 
= \frac{1}{N}\mathrm{Tr}\left(
 \boldsymbol{W}_{t}^{-1}\boldsymbol{A}\boldsymbol{A}^{\mathrm{T}} 
\right), 
\end{equation}
\begin{equation} \label{module_B_mean} 
\boldsymbol{x}_{\mathrm{B}\to \mathrm{A},t+1} 
= v_{\mathrm{B}\to \mathrm{A},t+1}\left(
 \frac{f_{t}(\boldsymbol{x}_{\mathrm{A}\to \mathrm{B},t})}
 {\xi_{t} v_{\mathrm{A}\to \mathrm{B},t}} 
 - \frac{\boldsymbol{x}_{\mathrm{A}\to \mathrm{B},t}}{v_{\mathrm{A}\to \mathrm{B},t}}  
\right), 
\end{equation}
\begin{equation} \label{module_B_var}
\frac{1}{v_{\mathrm{B}\to \mathrm{A},t+1}} 
= \frac{1}{\xi_{t}v_{\mathrm{A}\to \mathrm{B},t}} 
- \frac{1}{v_{\mathrm{A}\to \mathrm{B},t}}, 
\end{equation}
with $\xi_{t} =\langle f_{t}'(\boldsymbol{x}_{\mathrm{A}\to \mathrm{B}}^{t})\rangle$.  

It is an exercise to prove that the error model of the OAMP/VAMP is an 
instance of the general error model with
\begin{equation}
[\boldsymbol{\phi}_{t}(\boldsymbol{b}_{t}, \tilde{\boldsymbol{w}}; 
\boldsymbol{\lambda})]_{n} 
= b_{t,n} 
- \frac{\gamma_{t}\lambda_{n}b_{t,n}
-\gamma_{t}\sqrt{\lambda_{n}}\tilde{w}_{n}}
{\sigma^{2}+v_{\mathrm{B}\to\mathrm{A},t}\lambda_{n}}, 
\end{equation}
\begin{equation}
\boldsymbol{\psi}_{t}(\boldsymbol{h}_{t}, \boldsymbol{x}) 
= \frac{f_{t}(\boldsymbol{x}+\boldsymbol{h}_{t}) - \boldsymbol{x}}
{1 - \xi_{t}}, 
\end{equation}
by using the fact that $\xi_{t}$ converges almost surely to a constant in the 
large system limit~\cite{Rangan192,Takeuchi201}.  
The two separable functions $\boldsymbol{\psi}_{t}$ and $\boldsymbol{\phi}_{t}$ 
for the OAMP/VAMP depend only on the vectors $\boldsymbol{b}_{t}$ and 
$\boldsymbol{h}_{t}$ in the latest iteration. 
\end{example}

\begin{example}
Consider AMP~\cite{Donoho09} with 
a sequence of scalar denoisers $f_{t}:\mathbb{R}\to\mathbb{R}$: 
\begin{equation} \label{xt_AMP}
\boldsymbol{x}_{t+1} 
= f_{t}(\boldsymbol{x}_{t} + \boldsymbol{A}^{\mathrm{T}}\boldsymbol{z}_{t}), 
\end{equation}
\begin{equation} \label{zt_AMP}
\boldsymbol{z}_{t} 
= \boldsymbol{y} - \boldsymbol{A}\boldsymbol{x}_{t} 
+ \frac{\xi_{t-1}}{\delta}\boldsymbol{z}_{t-1}. 
\end{equation}

Suppose that the empirical eigenvalue distribution of $\boldsymbol{A}^{\mathrm{T}}
\boldsymbol{A}$ is equal to that for zero-mean i.i.d.\ Gaussian matrix 
$\boldsymbol{A}$ in the large system limit. Then, 
the error model of the AMP was proved in \cite{Takeuchi19} to be an 
instance of the general error model with 
\begin{IEEEeqnarray}{rl}
\boldsymbol{\phi}_{t}
&= (\boldsymbol{I}_{N}-\boldsymbol{\Lambda})\boldsymbol{b}_{t} 
- \frac{\xi_{t-1}}{\delta}\boldsymbol{b}_{t-1} 
+ \mathrm{diag}(\{\sqrt{\lambda_{n}}\})\tilde{\boldsymbol{w}} 
\nonumber \\
+& \xi_{t-1}\left\{
 \left(
  1+\frac{1}{\delta}
 \right)\boldsymbol{I}_{N} - \boldsymbol{\Lambda}
\right\}\boldsymbol{\phi}_{t-1}
- \frac{\xi_{t-1}\xi_{t-2}}{\delta}\boldsymbol{\phi}_{t-2}, 
\end{IEEEeqnarray} 
\begin{equation}
\boldsymbol{\psi}_{t}(\boldsymbol{h}_{t}, \boldsymbol{x}) 
= f_{t}(\boldsymbol{x}+\boldsymbol{h}_{t}) - \boldsymbol{x}, 
\end{equation} 
with $\boldsymbol{\Lambda}=\mathrm{diag}(\boldsymbol{\lambda})$ and 
$\xi_{t} =\langle f_{t}'(\boldsymbol{x}+\boldsymbol{h}_{t})\rangle$.  
Note that $\boldsymbol{\phi}_{t}$ is a function of $\boldsymbol{B}_{t+1}$ while 
$\boldsymbol{\psi}_{t}$ is a function of $\boldsymbol{h}_{t}$. 
\end{example}

\subsection{State Evolution}
A rigorous SE result for the general error model~(\ref{b})--(\ref{q}) is 
presented in the large system limit. 

\begin{theorem} \label{theorem_SE}
Suppose that Assumptions~\ref{assumption_x}--\ref{assumption_thresholding} 
hold. Then, the following properties hold for all $t=0,\ldots$ and 
$t'=0,\ldots, t$ in the large system limit: 
\begin{enumerate}
\item \label{property1} 
The inner products 
$N^{-1}\tilde{\boldsymbol{m}}_{t}^{\mathrm{T}}\tilde{\boldsymbol{m}}_{t'}$ and 
$N^{-1}\tilde{\boldsymbol{q}}_{t}^{\mathrm{T}}\tilde{\boldsymbol{q}}_{t'}$ converge 
almost surely to some constants $\pi_{t,t'}\in\mathbb{R}$ and 
$\kappa_{t,t'}\in\mathbb{R}$, respectively. 

\item Suppose that 
$\tilde{\boldsymbol{\psi}}_{t}(\boldsymbol{H}_{t+1},\boldsymbol{x}):
\mathbb{R}^{N\times(t+2)}\to\mathbb{R}^{N}$ is a separable and proper  
pseudo-Lipschitz function of order~$k$, 
that $\tilde{\boldsymbol{\phi}}_{t}(\boldsymbol{B}_{t+1}, 
\tilde{\boldsymbol{w}}; \boldsymbol{\lambda}):
\mathbb{R}^{N\times(t+3)}\to\mathbb{R}^{N}$ is separable, pseudo-Lipschitz 
of order~$k$ with respect to the first $t+2$ variables, and proper, and that 
$\boldsymbol{Z}_{t+1}=(\boldsymbol{z}_{0},\ldots,\boldsymbol{z}_{t})
\in\mathbb{R}^{N\times(t+1)}$ denotes a zero-mean Gaussian random matrix with 
covariance $\mathbb{E}[\boldsymbol{z}_{\tau}\boldsymbol{z}_{\tau'}^{\mathrm{T}}]
=\pi_{\tau,\tau'}\boldsymbol{I}_{N}$ for all $\tau, \tau'=0,\ldots,t$, while 
a zero-mean Gaussian random matrix 
$\tilde{\boldsymbol{Z}}_{t+1}=(\tilde{\boldsymbol{z}}_{0},\ldots,
\tilde{\boldsymbol{z}}_{t})\in\mathbb{R}^{N\times(t+1)}$ has covariance 
$\mathbb{E}[\tilde{\boldsymbol{z}}_{\tau}\tilde{\boldsymbol{z}}_{\tau'}^{\mathrm{T}}]
=\kappa_{\tau,\tau'}\boldsymbol{I}_{N}$. Then, 
\begin{equation} \label{psi_tilde}
\langle\tilde{\boldsymbol{\psi}}_{t}(\boldsymbol{H}_{t+1}, \boldsymbol{x})\rangle
- \mathbb{E}\left[
 \langle\tilde{\boldsymbol{\psi}}_{t}(\boldsymbol{Z}_{t+1}, 
 \boldsymbol{x})\rangle 
\right]\ato 0, 
\end{equation}
\begin{equation} \label{phi_tilde}
\langle\tilde{\boldsymbol{\phi}}_{t}(\boldsymbol{B}_{t+1}, 
\tilde{\boldsymbol{w}};\boldsymbol{\lambda})\rangle
- \mathbb{E}\left[
 \langle\tilde{\boldsymbol{\phi}}_{t}(\tilde{\boldsymbol{Z}}_{t+1}, 
 \tilde{\boldsymbol{w}};\boldsymbol{\lambda})\rangle 
\right]\ato 0.  
\end{equation}
In evaluating the expectation in (\ref{phi_tilde}), 
$\boldsymbol{U}^{\mathrm{T}}\boldsymbol{w}$ in (\ref{w_tilde}) 
follows the zero-mean Gaussian distribution with covariance 
$\sigma^{2}\boldsymbol{I}_{M}$.  
In particular, for $k=1$   
\begin{equation} \label{deriv_psi_tilde}
\langle\partial_{t'}\tilde{\boldsymbol{\psi}}_{t}(\boldsymbol{H}_{t+1}, 
\boldsymbol{x})\rangle
- \mathbb{E}\left[
 \langle\partial_{t'}\tilde{\boldsymbol{\psi}}_{t}(\boldsymbol{Z}_{t+1}, 
 \boldsymbol{x})\rangle 
\right]\ato 0, 
\end{equation}
\begin{equation} \label{deriv_phi_tilde}
\langle\partial_{t'}\tilde{\boldsymbol{\phi}}_{t}(\boldsymbol{B}_{t+1}, 
\tilde{\boldsymbol{w}};\boldsymbol{\lambda})\rangle
- \mathbb{E}\left[
 \langle\partial_{t'}\tilde{\boldsymbol{\phi}}_{t}(\tilde{\boldsymbol{Z}}_{t+1}, 
 \tilde{\boldsymbol{w}};\boldsymbol{\lambda})\rangle 
\right]\ato 0. 
\end{equation}

\item Suppose that 
$\tilde{\boldsymbol{\psi}}_{t}(\boldsymbol{H}_{t+1},\boldsymbol{x}):
\mathbb{R}^{N\times(t+2)}\to\mathbb{R}^{N}$ is separable and 
proper Lipschitz-continuous, and that  
$\tilde{\boldsymbol{\phi}}_{t}(\boldsymbol{B}_{t+1}, 
\tilde{\boldsymbol{w}};\boldsymbol{\lambda}):
\mathbb{R}^{N\times(t+3)}\to\mathbb{R}^{N}$ is separable, 
Lipschitz-continuous with respect to the first $t+2$ variables, and proper. 
Then,   
\begin{equation}  \label{h_orthogonality} 
\frac{1}{N}\boldsymbol{h}_{t'}^{\mathrm{T}}\left(
 \tilde{\boldsymbol{\psi}}_{t}
 - \sum_{\tau=0}^{t}\left\langle
  \partial_{\tau}\tilde{\boldsymbol{\psi}}_{t}
 \right\rangle\boldsymbol{h}_{\tau}
\right)
\ato0, 
\end{equation}
\begin{equation}  \label{b_orthogonality} 
\frac{1}{N}\boldsymbol{b}_{t'}^{\mathrm{T}}\left(
 \tilde{\boldsymbol{\phi}}_{t}
 - \sum_{\tau=0}^{t}\left\langle
  \partial_{\tau}\tilde{\boldsymbol{\phi}}_{t}
 \right\rangle\boldsymbol{b}_{\tau}
\right)
\ato0. 
\end{equation}
\end{enumerate}
\end{theorem}
\begin{IEEEproof}
See Appendix~\ref{proof_theorem_SE}. 
\end{IEEEproof}

Properties~(\ref{psi_tilde}) and (\ref{phi_tilde}) are used to evaluate 
the performance of MP by specifying the functions 
$\tilde{\boldsymbol{\psi}}_{t}$ and $\tilde{\boldsymbol{\phi}}_{t}$ according 
to a performance measure. An important observation is the asymptotic 
Gaussianity of 
$\boldsymbol{H}_{t+1}$ and $\boldsymbol{B}_{t+1}$. In evaluating the performance 
of MP, we can replace them with tractable Gaussian random matrices 
$\boldsymbol{Z}_{t+1}$ and $\tilde{\boldsymbol{Z}}_{t+1}$.  

The asymptotic Gaussianity originates from the definitions of 
$\tilde{\boldsymbol{q}}_{t}$ and $\tilde{\boldsymbol{m}}_{t}$ in 
(\ref{b}) and (\ref{h}). Properties~(\ref{h_orthogonality}) and 
(\ref{b_orthogonality}) imply the asymptotic orthogonality 
$N^{-1}\boldsymbol{h}_{t'}^{\mathrm{T}}\tilde{\boldsymbol{q}}_{t+1}\ato0$ 
and $N^{-1}\boldsymbol{b}_{t'}^{\mathrm{T}}\tilde{\boldsymbol{m}}_{t}\ato0$.  
This orthogonality is used to prove that the distributions of 
$\boldsymbol{H}_{t+1}$ and $\boldsymbol{B}_{t+1}$ are asymptotically Gaussian.  

Properties~(\ref{h_orthogonality}) and (\ref{b_orthogonality}) can be 
regarded as computation formulas to evaluate 
$N^{-1}\boldsymbol{h}_{t'}^{\mathrm{T}}\tilde{\boldsymbol{\psi}}_{t}$ and 
$N^{-1}\boldsymbol{b}_{t'}^{\mathrm{T}}\tilde{\boldsymbol{\phi}}_{t}$. They 
can be computed via linear combinations of 
$\{N^{-1}\boldsymbol{h}_{t'}^{\mathrm{T}}\boldsymbol{h}_{\tau}\}_{\tau=0}^{t}$  
and $\{N^{-1}\boldsymbol{b}_{t'}^{\mathrm{T}}\boldsymbol{b}_{\tau}\}_{\tau=0}^{t}$.   
In particular, (\ref{b}), (\ref{h}), and Property~\ref{property1}) in 
Theorem~\ref{theorem_SE} imply 
$N^{-1}\boldsymbol{h}_{t'}^{\mathrm{T}}\boldsymbol{h}_{\tau}\ato\pi_{t',\tau}$ and 
$N^{-1}\boldsymbol{b}_{t'}^{\mathrm{T}}\boldsymbol{b}_{\tau}\ato\kappa_{t',\tau}$. 
Furthermore, the coefficients in the linear combinations can be computed 
with (\ref{deriv_psi_tilde}) and (\ref{deriv_phi_tilde}). From these 
observations, the SE equations of the general error model are given as 
dynamical systems with respect to $\{\pi_{t,t'}, \kappa_{t,t'}\}$ in general.  

We do not derive SE equations with respect to $\{\pi_{t,t'}, 
\kappa_{t,t'}\}$ in a general form. Instead, we derive SE equations 
after specifying MP. The usefulness of Theorem~\ref{theorem_SE} is clarified 
in deriving SE equations. 

\section{Signal Recovery} \label{sec3}
\subsection{Convolutional Approximate Message-Passing} \label{sec3A}
Let $\boldsymbol{x}_{t}\in\mathbb{R}^{N}$ denote an estimator of 
the signal vector~$\boldsymbol{x}$ in iteration~$t$. CAMP computes the 
estimator $\boldsymbol{x}_{t}$ recursively as 
\begin{equation} \label{xt} 
\boldsymbol{x}_{t+1} 
= f_{t}(\boldsymbol{x}_{t} + \boldsymbol{A}^{\mathrm{T}}\boldsymbol{z}_{t}), 
\end{equation}
\begin{equation} \label{zt} 
\boldsymbol{z}_{t} 
= \boldsymbol{y} - \boldsymbol{A}\boldsymbol{x}_{t} 
+ \sum_{\tau=0}^{t-1}\xi_{\tau}^{(t-1)}(
\theta_{t-\tau}\boldsymbol{A}\boldsymbol{A}^{\mathrm{T}}
- g_{t-\tau}\boldsymbol{I}_{M})\boldsymbol{z}_{\tau}, 
\end{equation}
with the initial condition $\boldsymbol{x}_{0}=\boldsymbol{0}$, 
where $\xi_{\tau}^{(t-1)}=\prod_{t'=\tau}^{(t-1)}\xi_{t'}$ is the product of 
$\{\xi_{t'}\}$ given by 
\begin{equation}
\xi_{t} = 
\left\langle
 f_{t}'(\boldsymbol{x}_{t} + \boldsymbol{A}^{\mathrm{T}}\boldsymbol{z}_{t}) 
\right\rangle. 
\end{equation}

In (\ref{xt}) and (\ref{zt}), $\boldsymbol{A}$ and $\boldsymbol{y}$ are the 
sensing matrix and the measurement vector in (\ref{model}), respectively. 
The functions $\{f_{t}: \mathbb{R}\to\mathbb{R}\}$ are a sequence of 
Lipschitz-continuous denoisers. 
The tap coefficients $\{g_{\tau}\in\mathbb{R}\}$ and 
$\{\theta_{\tau}\in\mathbb{R}\}$ in the convolution are design parameters. 
The parameters $\{\theta_{\tau}\}$ are optimized to improve the performance of 
the CAMP while $\{g_{\tau}\}$ are determined so as to realize the asymptotic 
Gaussianity of the estimation errors via Theorem~\ref{theorem_SE}. 

To impose the initial condition $\boldsymbol{x}_{0}=\boldsymbol{0}$, 
it is convenient to introduce the notational convention 
$f_{-1}(\cdot)=0$, which is used throughout this paper.

The CAMP is a generalization of AMP~\cite{Donoho09} and reduces to AMP when 
$g_{1}=-\delta^{-1}$, $g_{\tau}=0$ for $\tau>1$, and $\theta_{\tau}=0$ hold. 
Also, as a generalization of CAMP in \cite{Takeuchi202}, the affine transform 
$(\theta_{t-\tau}\boldsymbol{A}\boldsymbol{A}^{\mathrm{T}}
-g_{t-\tau}\boldsymbol{I}_{M})
\boldsymbol{z}_{\tau}$ has been applied before the convolution. 
Nonetheless, the proposed MP is called CAMP simply. 
In particular, the MP algorithm reduces to the original 
CAMP~\cite{Takeuchi202} when $\theta_{\tau}=0$ is assumed.

\begin{remark} \label{remark1}
The design parameters $\{\theta_{\tau}\}$ are not required and can be set to 
zero for sensing matrices with identical non-zero singular values since 
$\boldsymbol{A}\boldsymbol{A}^{\mathrm{T}}$ reduces 
to the identity matrix with the exception of a constant factor. 
Thus, non-zero parameters $\{\theta_{\tau}\}$ should be introduced only for the 
case of non-identical singular values. 
\end{remark}

\subsection{Error Model} 
To design the parameters $g_{\tau}$ and $\theta_{\tau}$ via 
Theorem~\ref{theorem_SE}, we derive an error model of the CAMP. 
Let $\boldsymbol{h}_{t}=\boldsymbol{x}_{t} 
+ \boldsymbol{A}^{\mathrm{T}}\boldsymbol{z}_{t}-\boldsymbol{x}$ and 
$\boldsymbol{q}_{t+1}=\boldsymbol{x}_{t+1}-\boldsymbol{x}$ denote the error 
vectors before and after denoising $f_{t}$, respectively. Then, we have 
\begin{equation}
\boldsymbol{q}_{t+1} = f_{t}(\boldsymbol{x} + \boldsymbol{h}_{t}) 
- \boldsymbol{x} \equiv\psi_{t}(\boldsymbol{h}_{t}, \boldsymbol{x}),  
\end{equation}
\begin{equation} \label{q_tilde_CAMP}
\tilde{\boldsymbol{q}}_{t+1}=\boldsymbol{q}_{t+1}-\xi_{t}\boldsymbol{h}_{t}. 
\end{equation}
Using the notational convention $f_{-1}(\cdot)=0$, we obtain the initial 
condition $\boldsymbol{q}_{0}=-\boldsymbol{x}$ imposed in the general 
error model.  

We define $\boldsymbol{m}_{t}=\boldsymbol{V}^{\mathrm{T}}\boldsymbol{h}_{t}$  
and $\boldsymbol{b}_{t}=\boldsymbol{V}^{\mathrm{T}}\tilde{\boldsymbol{q}}_{t}$ 
to formulate the error model of the CAMP in a form corresponding to 
the general error model~(\ref{b})--(\ref{q}). Substituting the definition  
$\boldsymbol{h}_{t}=\boldsymbol{x}_{t} 
+ \boldsymbol{A}^{\mathrm{T}}\boldsymbol{z}_{t}-\boldsymbol{x}$ into 
$\boldsymbol{m}_{t}=\boldsymbol{V}^{\mathrm{T}}\boldsymbol{h}_{t}$ yields 
\begin{equation} 
\boldsymbol{m}_{t} 
= \boldsymbol{V}^{\mathrm{T}}\boldsymbol{q}_{t} 
+ \boldsymbol{\Sigma}^{\mathrm{T}}\boldsymbol{U}^{\mathrm{T}}\boldsymbol{z}_{t}, 
\end{equation}
where we have used the definition 
$\boldsymbol{q}_{t}=\boldsymbol{x}_{t}-\boldsymbol{x}$ and 
the SVD $\boldsymbol{A}=\boldsymbol{U}\boldsymbol{\Sigma}
\boldsymbol{V}^{\mathrm{T}}$. We utilize the definitions (\ref{q_tilde_CAMP}), 
$\boldsymbol{b}_{t}=\boldsymbol{V}^{\mathrm{T}}\tilde{\boldsymbol{q}}_{t}$, and 
$\boldsymbol{m}_{t}=\boldsymbol{V}^{\mathrm{T}}\boldsymbol{h}_{t}$ to obtain  
\begin{equation} \label{Vq}
\boldsymbol{V}^{\mathrm{T}}\boldsymbol{q}_{t}  
= \boldsymbol{b}_{t} + \xi_{t-1}\boldsymbol{m}_{t-1}. 
\end{equation}
Combining these two equations yields 
\begin{equation} \label{suz}
\boldsymbol{\Sigma}^{\mathrm{T}}\boldsymbol{U}^{\mathrm{T}}\boldsymbol{z}_{t}
= \boldsymbol{m}_{t} 
- \boldsymbol{b}_{t} - \xi_{t-1}\boldsymbol{m}_{t-1}. 
\end{equation}

To obtain a closed-form equation with respect to $\boldsymbol{m}_{t}$, 
we left-multiply (\ref{zt}) by $\boldsymbol{\Sigma}^{\mathrm{T}}
\boldsymbol{U}^{\mathrm{T}}$ and use (\ref{model}) to have   
\begin{IEEEeqnarray}{rl}
\boldsymbol{\Sigma}^{\mathrm{T}}\boldsymbol{U}^{\mathrm{T}}\boldsymbol{z}_{t} 
&= - \boldsymbol{\Lambda}\boldsymbol{V}^{\mathrm{T}}\boldsymbol{q}_{t} 
+ \boldsymbol{\Sigma}^{\mathrm{T}}\boldsymbol{U}^{\mathrm{T}}\boldsymbol{w} 
\nonumber \\
+ \sum_{\tau=0}^{t-1}&\xi_{\tau}^{(t-1)}(\theta_{t-\tau}\boldsymbol{\Lambda}
- g_{t-\tau}\boldsymbol{I}_{M})
\boldsymbol{\Sigma}^{\mathrm{T}}\boldsymbol{U}^{\mathrm{T}}\boldsymbol{z}_{\tau}, 
\end{IEEEeqnarray}
with $\boldsymbol{\Lambda}=
\boldsymbol{\Sigma}^{\mathrm{T}}\boldsymbol{\Sigma}$. Substituting 
(\ref{Vq}) and 
(\ref{suz}) into this expression, we arrive at 
\begin{IEEEeqnarray}{rl}
\boldsymbol{m}_{t} 
= (\boldsymbol{I}_{N} - \boldsymbol{\Lambda})&
(\boldsymbol{b}_{t} + \xi_{t-1}\boldsymbol{m}_{t-1}) 
+ \boldsymbol{\Sigma}^{\mathrm{T}}\boldsymbol{U}^{\mathrm{T}}\boldsymbol{w} 
\nonumber \\
+ \sum_{\tau=0}^{t-1}&\xi_{\tau}^{(t-1)}(\theta_{t-\tau}\boldsymbol{\Lambda} 
- g_{t-\tau}\boldsymbol{I}_{M}) \nonumber \\
&\cdot(\boldsymbol{m}_{\tau} 
- \boldsymbol{b}_{\tau} - \xi_{\tau-1}\boldsymbol{m}_{\tau-1}), 
 \label{m_CAMP} 
\end{IEEEeqnarray}
where any vector with a negative index is set to zero. 
This expression implies that $\boldsymbol{\phi}_{t}$ for the CAMP depends on 
all messages $\boldsymbol{B}_{t+1}$. 

We note that Assumption~\ref{assumption_thresholding} holds under 
Assumption~\ref{assumption_A} since the denoiser $f_{t}$ has been 
assumed to be Lipschitz-continuous. 

\subsection{Asymptotic Gaussianity} \label{sec3C}
We compare the obtained error model with the general error 
model~(\ref{b})--(\ref{q}). The only difference is in (\ref{h}): 
The correction $\tilde{\boldsymbol{m}}_{t}$ of $\boldsymbol{m}_{t}$ is 
used to define $\boldsymbol{h}_{t}$ in the general error model while  
no correction is performed in the error model of the CAMP. 
Thus, the general error model contains the error model of the CAMP when 
$\langle\partial_{t'}\boldsymbol{m}_{t}\rangle=0$ holds for all 
$t'=0,\ldots,t$. In the CAMP, the parameters $\{g_{\tau}\}$ are determined 
so as to guarantee $\langle\partial_{t'}\boldsymbol{m}_{t}\rangle=0$ 
in the large system limit.  

Let $\mu_{j}$ denote the $j$th moment of the asymptotic eigenvalue 
distribution of $\boldsymbol{A}^{\mathrm{T}}\boldsymbol{A}$, given by 
\begin{equation} \label{moment}
\mu_{j} = \lim_{M=\delta N\to\infty}
\frac{1}{N}\mathrm{Tr}(\boldsymbol{\Lambda}^{j}). 
\end{equation}
Assumption~\ref{assumption_A} implies $\mu_{1}=1$. 
We define a coupled dynamical system $\{g_{\tau}^{(j)}\}$ determined via 
the tap coefficients $\{g_{\tau}\}$ and $\{\theta_{\tau}\}$ as  
\begin{equation} \label{g_0}
g_{0}^{(j)} = \mu_{j+1} - \mu_{j},  
\end{equation}
\begin{IEEEeqnarray}{rl}
g_{1}^{(j)}
=& g_{0}^{(j)} - g_{0}^{(j+1)}
- g_{1}(g_{0}^{(j)} + \mu_{j})  
\nonumber \\
&+ \theta_{1}(g_{0}^{(j+1)} + \mu_{j+1}),   
\label{g_1}
\end{IEEEeqnarray}
\begin{IEEEeqnarray}{rl}
g_{\tau}^{(j)}
=& g_{\tau-1}^{(j)} - g_{\tau-1}^{(j+1)} 
- g_{\tau}\mu_{j} + \theta_{\tau}\mu_{j+1}
\nonumber \\
&+ \sum_{\tau'=0}^{\tau-1}(\theta_{\tau-\tau'}g_{\tau'}^{(j+1)} 
- g_{\tau-\tau'}g_{\tau'}^{(j)})
\nonumber \\
&- \sum_{\tau'=1}^{\tau-1}(\theta_{\tau-\tau'}g_{\tau'-1}^{(j+1)} 
- g_{\tau-\tau'}g_{\tau'-1}^{(j)}) \label{g_tau} 
\end{IEEEeqnarray}
for $\tau>1$.  

\begin{theorem} \label{theorem_CAMP} 
Suppose that Assumptions~\ref{assumption_x}--\ref{assumption_w} 
hold, that the denoiser $f_{t}$ is Lipschitz-continuous, and that 
the tap coefficients $\{g_{\tau}\}$ and $\{\theta_{\tau}\}$ in the CAMP satisfy  
\begin{equation} \label{g_1_tap}
g_{1} = \theta_{1}(g_{0}^{(1)} + 1) - g_{0}^{(1)} , 
\end{equation}
\begin{equation} 
g_{\tau}
= \theta_{\tau} - g_{\tau-1}^{(1)} 
+ \sum_{\tau'=0}^{\tau-1}\theta_{\tau-\tau'}g_{\tau'}^{(1)}
- \sum_{\tau'=1}^{\tau-1}\theta_{\tau-\tau'}g_{\tau'-1}^{(1)}, 
\label{g_tau_tap}
\end{equation}
where $\{g_{\tau}^{(1)}\}$ is governed by the dynamical 
system~(\ref{g_0})--(\ref{g_tau}). 
Then, $\langle\partial_{t'}\boldsymbol{m}_{t}\rangle\to0$ holds in the large 
system limit, i.e.\ the error model of the CAMP is included into the general 
error model. 
\end{theorem}
\begin{IEEEproof}
Let 
\begin{equation}
g_{t',t}^{(j)} 
= -\lim_{M=\delta N\to\infty}\left\langle 
 \boldsymbol{\Lambda}^{j}\partial_{t'}\boldsymbol{m}_{t}
\right\rangle. 
\end{equation}
It is sufficient to prove $g_{t',t}^{(j)}\aeq \xi_{t'}^{(t-1)}g_{t-t'}^{(j)}+o(1)$ 
and $g_{\tau}^{(0)}=0$ under the notational convention $\xi_{t'}^{(t)}=1$ 
for $t'>t$. The latter property $g_{\tau}^{(0)}=0$ follows from 
(\ref{g_0}) for $\tau=0$, (\ref{g_1}) and 
(\ref{g_1_tap}) for $\tau=1$, and from (\ref{g_tau}) and (\ref{g_tau_tap}). 
See Appendix~\ref{proof_theorem_CAMP} for the proof of the former property. 
\end{IEEEproof}

Throughout this paper, we assume that the tap coefficients $\{g_{\tau}\}$ and 
$\{\theta_{\tau}\}$ satisfy (\ref{g_1_tap}) and (\ref{g_tau_tap}). Thus, 
Theorem~\ref{theorem_SE} implies that  
the asymptotic Gaussianity is guaranteed for the CAMP. In principle, it is 
possible to compute the tap coefficients by solving the coupled dynamical 
system~(\ref{g_0})--(\ref{g_tau_tap}) numerically for a given moment sequence 
$\{\mu_{j}\}$. However, numerical evaluation indicated that the 
dynamical system is unstable against numerical errors when the moment 
sequence $\{\mu_{j}\}$ is a diverging sequence. Thus, we need 
a closed-form solution to the tap coefficients.  

To present the closed-form solution, we define the $\eta$-transform of the 
asymptotic eigenvalue distribution of 
$\boldsymbol{A}^{\mathrm{T}}\boldsymbol{A}$~\cite{Tulino04} as 
\begin{equation}
\eta(x) = \lim_{M=\delta N\to\infty}\frac{1}{N}\mathrm{Tr}\left\{
 \left(
  \boldsymbol{I}_{N} + x\boldsymbol{A}^{\mathrm{T}}\boldsymbol{A}
 \right)^{-1} 
\right\}. 
\end{equation}
By definition, we have the power-series expansion 
\begin{equation} \label{eta_transform}
\eta(x) = \lim_{M=\delta N\to\infty}\frac{1}{N}\sum_{n=1}^{N}\frac{1}{1+x\lambda_{n}}
= \sum_{j=0}^{\infty}\mu_{j}(-x)^{j}   
\end{equation}
for $|x|<1/\max\{\lambda_{n}\}$. 
Let $G(z)$ denote the generating function of the tap coefficients 
$\{g_{\tau}\}$ given by  
\begin{equation} \label{generating_g}
G(z) = \sum_{\tau=0}^{\infty}g_{\tau}z^{-\tau}, 
\quad g_{0}=1.
\end{equation}
Similarly, we write the generating function of 
$\{\theta_{\tau}\}$ with $\theta_{0}=1$ as $\Theta(z)$. 

\begin{theorem} \label{theorem_solution}
Suppose that the tap coefficients $\{g_{\tau}\}$ and $\{\theta_{\tau}\}$ 
satisfy (\ref{g_1_tap}) and (\ref{g_tau_tap}). Then, 
the generating functions $G(z)$ and $\Theta(z)$ of   
$\{g_{\tau}\}$ and $\{\theta_{\tau}\}$ satisfy 
\begin{equation} \label{closed_form} 
\eta\left(
 \frac{1 - (1-z^{-1})\Theta(z)}{(1-z^{-1})G(z)}
\right) 
= (1-z^{-1})\Theta(z), 
\end{equation}
where $\eta$ denotes the $\eta$-transform of the asymptotic eigenvalue 
distribution of $\boldsymbol{A}^{\mathrm{T}}\boldsymbol{A}$. 
\end{theorem}
\begin{IEEEproof}
See Appendix~\ref{proof_theorem_solution}. 
\end{IEEEproof}

Suppose that the $\eta$-transform is given. Since the $\eta$-transform has 
the inverse function, from Theorem~\ref{theorem_solution} we have 
$(1-z^{-1})G(z)=[1-(1-z^{-1})\Theta(z)]/\eta^{-1}((1-z^{-1})\Theta(z))$ for a 
fixed generating function $\Theta(z)$. Each tap coefficient $g_{\tau}$ can be 
computed by evaluating the coefficient of the $\tau$th-order term in 
$G(z)$. 

\begin{corollary} \label{corollary1}
Suppose that the sensing matrix $\boldsymbol{A}$ has independent Gaussian 
elements with mean $\sqrt{\gamma/M}$ and variance $(1-\gamma)/M$ for any 
$\gamma\in[0, 1)$. Then, the tap coefficient $g_{t}$ is given by 
\begin{equation} \label{g_AMP}
g_{t} = \left(
 1 - \frac{1}{\delta}
\right)\theta_{t} 
+ \frac{1}{\delta}\sum_{\tau=0}^{t}(\theta_{\tau}-\theta_{\tau-1})\theta_{t-\tau}
\end{equation}
for fixed tap coefficients $\{\theta_{t}\}$. 
\end{corollary}
\begin{IEEEproof}
We shall evaluate the generating function $G(z)$. 
The R-transform $R(x)$~\cite[Section 2.4.2]{Tulino04} of the asymptotic 
eigenvalue distribution of $\boldsymbol{A}^{\mathrm{T}}\boldsymbol{A}$ is given by 
\begin{equation}
R(x) = \frac{\delta}{\delta - x}. 
\end{equation}
Using Theorem~\ref{theorem_solution} and the relationship between the 
R-transform and the $\eta$-transform~\cite[Eq.~(2.74)]{Tulino04} 
\begin{equation} \label{relationship} 
\eta(x) = \frac{1}{1 + xR(-x\eta(x))}, 
\end{equation}
we obtain 
\begin{equation}
G(z)  
= \left[
 1 - \frac{1}{\delta} + \frac{(1-z^{-1})}{\delta}\Theta(z)
\right]\Theta(z), 
\end{equation}
which implies the time-domain expression (\ref{g_AMP}). 
\end{IEEEproof}

In particular, we consider the original CAMP $\theta_{\tau}=0$ for $\tau>0$. 
In this case, we have $g_{1}=-\delta^{-1}$ and $g_{\tau}=0$. As remarked in 
\cite{Takeuchi202}, 
the original CAMP reduces to the AMP for the i.i.d.\ Gaussian sensing matrix.  

\begin{corollary} \label{corollary2}
Suppose that the sensing matrix $\boldsymbol{A}$ has $M$ identical non-zero 
singular values for $M\leq N$, i.e.\ 
$\boldsymbol{A}\boldsymbol{A}^{\mathrm{T}}=\delta^{-1}\boldsymbol{I}_{M}$. 
Then, the tap coefficient $g_{t}$ in the original CAMP $\theta_{t}=0$ for $t>0$ 
is given by $g_{\tau}=1-\delta^{-1}$ for all $\tau\geq1$. 
\end{corollary}
\begin{IEEEproof}
We evaluate the generating function $G(z)$. 
By definition, the $\eta$-transform is given by 
\begin{equation}
\eta(x) = \frac{1}{N}\left(
 \frac{M}{1+x\delta^{-1}} + N-M
\right) = 1 - \delta + \frac{\delta^{2}}{\delta+x}. 
\end{equation}
Using Theorem~\ref{theorem_solution} and $\Theta(z)=1$ yields 
\begin{equation}
G(z)
= \frac{1-\delta^{-1}z^{-1}}{1-z^{-1}}
= 1 + \sum_{j=1}^{\infty}\left(
 1 - \frac{1}{\delta} 
\right)z^{-j},
\end{equation}
which implies $g_{\tau}=1-\delta^{-1}$ for all $\tau\geq1$. 
\end{IEEEproof} 

\begin{corollary} \label{corollary3}
Suppose that the sensing matrix $\boldsymbol{A}$ has non-zero singular 
values $\sigma_{0}\geq\cdots\geq\sigma_{M-1}>0$ satisfying condition number 
$\kappa=\sigma_{0}/\sigma_{M-1}>1$, $\sigma_{m}/\sigma_{m-1}=\kappa^{-1/(M-1)}$, 
and $\sigma_{0}^{2}=N(1-\kappa^{-2/(M-1)})/(1-\kappa^{-2M/(M-1)})$. 
Assume $\theta_{t}=0$ for all $t>t_{1}$ for some $t_{1}\in\mathbb{N}$. 
Let $\alpha_{0}^{(j)}=1$ and 
\begin{equation}
\alpha_{t}^{(j)} 
= \left\{
\begin{array}{cl}
\frac{C^{t/j}}{(t/j)!}\bar{\theta}_{j}^{t/j} & 
\hbox{if $t$ is divisible by $j$,} \\ 
0 & \hbox{otherwise} 
\end{array}
\right.
\end{equation}
for $t\in\mathbb{N}$ and $j\in\{1,\ldots,t_{1}\}$, 
with $\bar{\theta}_{t}=\theta_{t-1}-\theta_{t}$ and  
$C=2\delta^{-1}\ln\kappa$. 
Define $p_{0}=\bar{q}_{0}=1$ and 
\begin{equation}
p_{t}
= - \frac{\beta_{t}^{(t_{1})}}{\kappa^{2}-1},  
\end{equation}
\begin{equation}
\bar{q}_{t} 
= \frac{1}{\bar{\theta}_{1}}\left(
 \frac{\beta_{t+1}^{(t_{1})}}{C} 
 - \sum_{\tau=1}^{t_{1}}\bar{\theta}_{\tau+1}\bar{q}_{t-\tau} 
\right)
\end{equation}
for $t>0$, 
with $\beta_{t}^{(t_{1})}=\alpha_{t}^{(1)}*\alpha_{t}^{(2)}*\cdots*\alpha_{t}^{(t_{1})}$. 
Then, the tap coefficient $g_{t}$ is recursively given by
\begin{equation} \label{g_CAMP}
g_{t} = p_{t} - \sum_{\tau=1}^{t}q_{\tau}g_{t-\tau}, 
\end{equation}
with 
\begin{equation}
q_{t} = \bar{q}_{t} - \bar{q}_{t-1}. 
\end{equation}
\end{corollary}
\begin{IEEEproof}
We first evaluate the inverse of the $\eta$-transform. By definition, 
$\sigma_{m}^{2}=\kappa^{-2m/(M-1)}\sigma_{0}^{2}$ holds. Thus, we have 
\begin{IEEEeqnarray}{rl}
\mu_{j} 
=& \frac{1}{N}\sum_{m=0}^{M-1}\sigma_{m}^{2j}
= \sigma_{0}^{2j}\frac{1 - \kappa^{-2jM/(M-1)}}{N(1 - \kappa^{-2j/(M-1)})}
\nonumber \\
\to& \left(
 \frac{C}{1-\kappa^{-2}}
\right)^{j}
\frac{1-\kappa^{-2j}}{Cj}
\end{IEEEeqnarray}
in the large system limit, where we have used the convergence 
$N(1-\kappa^{-a/(M-1)})\to\delta^{-1}a\ln\kappa$ for any $a\in\mathbb{R}$. 
We note the series-expansion $\ln(1+x)=\sum_{j=1}^{\infty}(-1)^{j-1}j^{-1}x^{j}$ 
for $|x|<1$ to obtain 
\begin{equation}
\eta(x) = 1 + \sum_{j=1}^{\infty}(-x)^{j}\mu_{j} 
= 1 - \frac{1}{C}\ln\left(
 \frac{\kappa^{2}-1 + \kappa^{2}Cx}
 {\kappa^{2}-1 + Cx}
\right), 
\end{equation}
which implies the inverse function 
\begin{equation}
\eta^{-1}(x)
= \frac{(\kappa^{2}-1)\{e^{C(1 - x)} - 1\}}
{C\{\kappa^{2} - e^{C(1 - x)}\}}. 
\end{equation}

We next evaluate the generating function $G(z)$. 
Using Theorem~\ref{theorem_solution} yields $G(z)=P(z)/Q(z)$, with 
\begin{equation}
P(z) 
= \frac{\kappa^{2} - e^{C\bar{\Theta}(z)}}{\kappa^{2}-1}, 
\end{equation}
\begin{equation}
Q(z)
= (1-z^{-1})\bar{Q}(z), \quad 
\bar{Q}(z) 
= \frac{e^{C\bar{\Theta}(z)} - 1}{C\bar{\Theta}(z)}, 
\end{equation}
\begin{equation}
\bar{\Theta}(z) = \sum_{t=1}^{\infty}\bar{\theta}_{t}z^{-t}. 
\end{equation}

Finally, we derive a time-domain expression of $G(z)$. 
It is an exercise to 
confirm that the series-expansions of $P(z)$ 
and $\bar{Q}(z)$ have the coefficients $p_{t}$ and $\bar{q}_{t}$ 
for the $t$th-order terms, respectively. Then, the Z-transform of 
(\ref{g_CAMP}) is equal to $P(z)/Q(z)$. 
\end{IEEEproof}

The sequences $\{p_{\tau}\}$ and $\{q_{\tau}\}$ in Corollary~\ref{corollary3} 
define the generating functions $P(z)$ and $Q(z)$ with $p_{0}=q_{0}=1$, 
respectively, which satisfy 
$G(z)=P(z)/Q(z)$. Thus, we derive an SE equation in time domain in terms of 
$\{p_{\tau}, q_{\tau}\}$, rather than $\{g_{\tau}\}$.

\subsection{SE Equation} \label{sec3D}
We design the tap coefficients $\{\theta_{\tau}\}$ so as to 
minimize the MSE $N^{-1}\|\boldsymbol{x}_{t}-\boldsymbol{x}\|^{2}$ for 
the CAMP estimator $\boldsymbol{x}_{t}$ in the large system limit. 
For that purpose, we derive an SE equation that describes  
the dynamics of the MSE. For simplicity, we assume i.i.d.\ signals. 

The CAMP has no closed-form SE equation with respect to the MSEs  
$N^{-1}\|\boldsymbol{x}_{t}-\boldsymbol{x}\|^{2}$ in general. Instead, 
it has a closed-form SE equation with respect to the correlations 
\begin{equation} \label{correlation}
d_{t'+1,t+1}= \mathbb{E}\left[
 \{f_{t'}(x_{1}+z_{t'}) - x_{1}\}\{f_{t}(x_{1}+z_{t}) - x_{1}\}
\right], 
\end{equation}
where $\{z_{t}\}$ denote zero-mean Gaussian random variables with 
covariance $a_{t',t}=\mathbb{E}[z_{t'}z_{t}]$. In particular, 
$d_{t+1,t+1}$ corresponds to the MSE of the CAMP estimator in iteration~$t$.  

As an asymptotic alternative to $\xi_{t}$, we use the following 
quantity:  
\begin{equation} \label{xi_bar}
\bar{\xi}_{t} = \mathbb{E}\left[
 f_{t}'(x_{1} + z_{t})
\right], 
\end{equation}
which is a function of $a_{t,t}$. 
The notation $\bar{\xi}_{t'}^{(t)}$ is defined in the same manner as in 
$\xi_{t'}^{(t)}$. 

\begin{theorem} \label{theorem_CAMP_SE} 
Assume that Assumptions~\ref{assumption_x}--\ref{assumption_w} hold, 
that the denoiser $f_{t}$ is Lipschitz-continuous, and 
that the signal vector $\boldsymbol{x}$ has i.i.d.\ elements. 
Suppose that the generating functions $G$ and $\Theta$ 
for the tap coefficients $\{g_{\tau}\}$ 
and $\{\theta_{\tau}\}$---given in (\ref{generating_g})---satisfy the 
condition~(\ref{closed_form}) in Theorem~\ref{theorem_solution}. 
\begin{itemize}
\item Define generating functions $A(y,z)$, $D(y,z)$, and $\Sigma(y,z)$ as 
\begin{equation}
A(y,z) 
= \sum_{t',t=0}^{\infty}\frac{a_{t',t}}{\bar{\xi}_{0}^{(t'-1)}\bar{\xi}_{0}^{(t-1)}}
y^{-t'}z^{-t},
\end{equation}
\begin{equation}
D(y,z) 
= \sum_{t',t=0}^{\infty}\frac{d_{t',t}}{\bar{\xi}_{0}^{(t'-1)}\bar{\xi}_{0}^{(t-1)}}
y^{-t'}z^{-t},
\end{equation}
\begin{equation}
\Sigma(y,z) 
= \sum_{t',t=0}^{\infty}\frac{\sigma^{2}}{\bar{\xi}_{0}^{(t'-1)}\bar{\xi}_{0}^{(t-1)}}
y^{-t'}z^{-t}. 
\end{equation}
Then, the correlation $N^{-1}(\boldsymbol{x}_{t'}-\boldsymbol{x})^{\mathrm{T}}
(\boldsymbol{x}_{t}-\boldsymbol{x})$ converges almost 
surely to $d_{t',t}$ in the large system limit, which satisfies the following 
SE equation in terms of the generating functions:
\begin{IEEEeqnarray}{rl} 
F_{G,\Theta}(y,z)A(y,z) 
=& \left\{
 G(z)\Delta_{\Theta} - \Theta(z)\Delta_{G} 
\right\}D(y,z) 
\nonumber \\
&+ \left(
 \Delta_{\Theta_{1}}  - \Delta_{\Theta}   
\right)\Sigma(y,z), 
\label{SE_equation_generating_tmp}
\end{IEEEeqnarray}
with 
\begin{IEEEeqnarray}{rl}
F_{G,\Theta}(y,z) 
=& (y^{-1}+z^{-1}-1)[G(z)\Delta_{\Theta} - \Theta(z)\Delta_{G}]
\nonumber \\
&+ \Delta_{G_{1}} - \Delta_{G}, \label{Operator} 
\end{IEEEeqnarray}
where the notations $S_{1}(z)=z^{-1}S(z)$ and $\Delta_{S}
=[S(y) - S(z)]/(y^{-1}-z^{-1})$ have been used for 
any generating function $S(z)$. 

\item Suppose that $G(z)$ is represented as $G(z)=P(z)/Q(z)$ for the 
generating functions $P(z)$ and $Q(z)$ of some sequences $\{p_{\tau}\}$ and 
$\{q_{\tau}\}$ with $p_{0}=1$ and $q_{0}=1$. Let $r_{t}=q_{t}*\theta_{t}$.  
Then, the SE equation~(\ref{SE_equation_generating_tmp}) reduces to 
\begin{IEEEeqnarray}{l}
\sum_{\tau'=0}^{t'}\sum_{\tau=0}^{t}\bar{\xi}_{t'-\tau'}^{(t'-1)}\bar{\xi}_{t-\tau}^{(t-1)}
\Big\{ \mathfrak{D}_{\tau',\tau}a_{t'-\tau',t-\tau}  
\nonumber \\
- (p_{\tau}*r_{\tau'+\tau+1} - r_{\tau}*p_{\tau'+\tau+1})d_{t'-\tau',t-\tau}
\nonumber \\
- \sigma^{2}\left[
 (q_{\tau'}q_{\tau})*(\theta_{\tau'+\tau} - \theta_{\tau'+\tau+1})
\right]\Big\} 
=0, \label{SE_equation}  
\end{IEEEeqnarray}
where all variables with negative indices are set to zero, with 
\begin{IEEEeqnarray}{rl}
\mathfrak{D}_{\tau',\tau}
&= (p_{\tau'+\tau} - p_{\tau'+\tau+1})*q_{\tau} + (p_{\tau} - p_{\tau-1})*q_{\tau'+\tau+1} 
\nonumber \\
+& (p_{\tau-1}-p_{\tau})*r_{\tau'+\tau+1} + (r_{\tau} - r_{\tau-1})*p_{\tau'+\tau+1}
 \nonumber \\
+& p_{\tau}*(r_{\tau'+\tau} - \delta_{\tau',0}r_{\tau}) 
- r_{\tau}*(p_{\tau'+\tau} - \delta_{\tau',0}p_{\tau}). 
\end{IEEEeqnarray}
In solving the SE equation~(\ref{SE_equation}), we impose 
the initial condition $d_{0,0}=1$ and boundary conditions 
$d_{0,\tau+1}=d_{\tau+1,0}=-\mathbb{E}[x_{1}\{f_{\tau}(x_{1}+z_{\tau})-x_{1}\}]$ 
for any $\tau$.  
\end{itemize}
\end{theorem}
\begin{IEEEproof}
See Appendix~\ref{proof_theorem_CAMP_SE}. 
\end{IEEEproof}

The SE equation~(\ref{SE_equation}) in time domain is useful for numerical 
evaluation of $\{a_{t',t}\}$ while the generating-function 
representation~(\ref{SE_equation_generating_tmp}) is used in fixed-point 
analysis. To apply Corollary~\ref{corollary3}, we have represented the 
generating function $G(z)$ as $G(z)=P(z)/Q(z)$. If $G(z)$ is given directly, 
the functions $P(z)=G(z)$ and $Q(z)=1$ can be used. In this case, we have 
$p_{\tau}=g_{\tau}$, $q_{\tau}=\delta_{\tau,0}$, and $r_{\tau}=\theta_{\tau}$.   

Note that $d_{t'+1,t+1}$ given in (\ref{correlation}) is a function of 
$\{a_{t',t}, a_{t',t'}, a_{t,t}\}$, 
so that the SE equation~(\ref{SE_equation}) in time domain is a nonlinear 
difference equation with respect to $\{a_{t',t}\}$ for given tap 
coefficients $\{g_{\tau}\}$ and $\{\theta_{\tau}\}$.  
Theorem~\ref{theorem_CAMP_SE} allows us to compute the MSEs 
$a_{t,t}$ and $d_{t+1,t+1}$ before and after denoising.  

The SE equation~(\ref{SE_equation}) in time domain can be solved recursively 
by extracting the term $\mathfrak{D}_{0,0}a_{t',t}$ for $\tau'=\tau=0$ 
in the sum and moving the other terms to the RHS. More precisely, 
we can solve the SE equation~(\ref{SE_equation}) as follows:   

\begin{enumerate}
\item Let $t=0$ and solve $a_{0,0}$ with the SE equation~(\ref{SE_equation}) 
and the initial condition $d_{0,0}=1$. 

\item \label{step2}
Suppose that $\{a_{\tau',\tau}, d_{\tau',\tau}\}$ have been 
obtained for all $\tau',\tau=0,\ldots,t$. Use the boundary condition 
$d_{0,t+1}$ in Theorem~\ref{theorem_CAMP_SE} and compute $d_{\tau,t+1}$ with
the definition~(\ref{correlation}) for all $\tau=1,\ldots,t+1$ 
while the symmetry $d_{t+1,\tau}=d_{\tau,t+1}$ is used in the lower triangular 
elements. 

\item Compute $a_{\tau,t+1}$ with the SE equation~(\ref{SE_equation}) in the   
order $\tau=0,\ldots,t+1$ while the symmetry $a_{t+1,\tau}=a_{\tau,t+1}$ is used 
in the upper triangular elements. 

\item If some termination conditions are satisfied, output $\{a_{\tau',\tau}, 
d_{\tau',\tau}\}$. Otherwise, update 
$t:=t+1$ and go back to Step~\ref{step2}). 
\end{enumerate}

We can define the Bayes-optimal denoiser $f_{t}$ via the 
MSE $d_{t+1,t+1}$ in the large system limit. A denoiser $f_{t}$ is 
said to be Bayes-optimal if $f_{t}=\mathbb{E}[x_{1}|x_{1}+z_{t}]$ is    
the posterior mean of $x_{1}$ given an AWGN observation $x_{1} + z_{t}$ with 
$z_{t}\sim\mathcal{N}(0,a_{t,t})$. We write the Bayes-optimal denoiser 
as $f_{t}(\cdot)=f_{\mathrm{opt}}(\cdot; a_{t,t})$. 

The boundary condition $d_{0,\tau+1}$ in Theorem~\ref{theorem_CAMP_SE} 
has a simple representation for 
the Bayes-optimal denoiser $f_{\mathrm{opt}}$. Since the posterior mean estimator 
$f_{\mathrm{opt}}(x_{1}+z_{\tau}; a_{\tau,\tau})$ is uncorrelated with the 
estimation error $f_{\mathrm{opt}}(x_{1}+z_{\tau}; a_{\tau,\tau})-x_{1}$, we 
obtain 
\begin{IEEEeqnarray}{rl}
d_{0,\tau+1} 
=& \mathbb{E}[\{f_{\mathrm{opt}}(x_{1}+z_{\tau}; a_{\tau,\tau}) - x_{1} 
- f_{\mathrm{opt}}(x_{1}+z_{\tau}; a_{\tau,\tau})\} \nonumber \\
&\cdot\{f_{\mathrm{opt}}(x_{1}+z_{\tau}; a_{\tau,\tau})-x_{1}\}] 
\nonumber \\
=& \mathbb{E}[\{f_{\mathrm{opt}}(x_{1}+z_{\tau}; a_{\tau,\tau})-x_{1}\}^{2}]
= d_{\tau+1,\tau+1}. 
\end{IEEEeqnarray}

\begin{theorem} \label{theorem_fixed_point} 
Consider the Bayes-optimal denoiser under the same assumptions as in 
Theorem~\ref{theorem_CAMP_SE}. Suppose 
that the SE equation~(\ref{SE_equation}) in time domain converges to a 
fixed-point $\{a_{\mathrm{s}}, d_{\mathrm{s}}\}$, 
i.e.\ $\lim_{t',t\to\infty}a_{t',t}=a_{\mathrm{s}}$ and 
$\lim_{t',t\to\infty}d_{t',t}=d_{\mathrm{s}}$. If $\Theta(\xi_{\mathrm{s}}^{-1})=1$ 
and $1+(\xi_{\mathrm{s}}-1)d\Theta(\xi_{\mathrm{s}}^{-1})/(dz^{-1})\neq0$ 
hold for $\xi_{\mathrm{s}}=d_{\mathrm{s}}/a_{\mathrm{s}}$, then 
the fixed-point $\{a_{\mathrm{s}}, d_{\mathrm{s}}\}$ of the SE 
equation~(\ref{SE_equation}) satisfies 
\begin{equation} \label{fixed_point} 
a_{\mathrm{s}}
= \frac{\sigma^{2}}{R(-d_{\mathrm{s}}/\sigma^{2})}, 
\quad 
d_{\mathrm{s}}=\mathbb{E}\left[
 \{f_{\mathrm{opt}}(x_{1}+z_{\mathrm{s}}; a_{\mathrm{s}}) - x_{1}\}^{2}
\right],
\end{equation}
with $z_{\mathrm{s}}\sim\mathcal{N}(0,a_{\mathrm{s}})$, 
where $R(x)$ denotes the R-transform of the asymptotic eigenvalue distribution 
of $\boldsymbol{A}^{\mathrm{T}}\boldsymbol{A}$. 
\end{theorem}
\begin{IEEEproof}
See Appendix~\ref{proof_theorem_fixed_point}. 
\end{IEEEproof}

The fixed-point equations given in (\ref{fixed_point}) coincide with those for 
describing the asymptotic performance of the posterior mean estimator 
of the signal vector $\boldsymbol{x}$~\cite{Takeda06,Tulino13,Barbier18}.  
This coincidence implies that the CAMP with Bayes-optimal denoisers 
is Bayes-optimal if the SE equation~(\ref{SE_equation}) converges 
toward a fixed-point and if the fixed-point is unique. 
Thus, we refer to CAMP with Bayes-optimal denoisers as Bayes-optimal CAMP. 

\subsection{Implementation} \label{sec3E}
We summarize the implementation of the Bayes-optimal CAMP. We need to 
specify the sequence of denoisers $\{f_{t}\}$ and 
the tap coefficients $\{g_{\tau}, \theta_{\tau}\}$ in (\ref{xt}) and 
(\ref{zt}). For simplicity, assume $\theta_{\tau}=0$ for all $\tau>2$. 
To impose the condition $\Theta(a_{\mathrm{s}}/d_{\mathrm{s}})=1$ in 
Theorem~\ref{theorem_fixed_point}, we use $\theta_{0}=1$, 
$\theta_{1}=-\theta d_{\mathrm{s}}/a_{\mathrm{s}}$, and 
$\theta_{2}=\theta\in\mathbb{R}$, in which 
$a_{\mathrm{s}}$ and $d_{\mathrm{s}}$ are a solution to the fixed-point 
equations~(\ref{fixed_point}). In particular, the CAMP reduces to the 
original one in \cite{Takeuchi202} for $\theta=0$.  

For a given parameter $\theta$, the tap coefficients $\{g_{\tau}\}$ are 
determined via Theorem~\ref{theorem_solution}. More precisely, we use 
the coefficients $\{p_{\tau}, q_{\tau}\}$ in the rational generating function 
$G(z)=P(z)/Q(z)$. See Corollaries~\ref{corollary1}--\ref{corollary3} for 
examples of the coefficients. 

For given parameters $\{\theta, p_{\tau}, q_{\tau}\}$, we can solve the 
SE equation~(\ref{SE_equation}) numerically. The obtained parameter 
$a_{t,t}$ is used to determine the Bayes-optimal denoiser 
$f_{t}(\cdot)=f_{\mathrm{opt}}(\cdot; a_{t,t})$. 

Damping~\cite{Rangan191} is a well-known technique to improve the convergence 
property in finite-sized systems. In damped CAMP, the update 
rule~(\ref{xt}) is replaced by 
\begin{equation}
\boldsymbol{x}_{t+1} 
= \zeta f_{t}(\boldsymbol{x}_{t}+\boldsymbol{A}^{\mathrm{T}}\boldsymbol{z}_{t})
+ (1-\zeta)\boldsymbol{x}_{t},
\end{equation}
with damping factor $\zeta\in[0, 1]$. In solving the SE 
equation~(\ref{SE_equation}), the associated parameters $d_{t'+1,t+1}$ and 
$\bar{\xi}_{t}$ in (\ref{correlation}) and (\ref{xi_bar}) are damped as 
follows: 
\begin{IEEEeqnarray}{rl}
d_{t'+1,t+1} 
=& \zeta\mathbb{E}\left[
 \{ f_{t'}(x_{1}+z_{t'}) - x_{1}\}\{f_{t}(x_{1}+z_{t}) - x_{1}\}
\right] \nonumber \\
&+ (1-\zeta)d_{t',t},  
\end{IEEEeqnarray} 
\begin{equation}
\bar{\xi}_{t} 
= \zeta\mathbb{E}\left[
 f_{t}'(x_{1} + z_{t})
\right] + (1-\zeta)\bar{\xi}_{t-1}. 
\end{equation}
In particular, no damping is applied for $\zeta=1$. 

\begin{table}[t]
\begin{center}
\caption{
Complexity in $M\leq N$ and the number of iterations~$t$. 
}
\label{table1}
\begin{tabular}{|c|c|c|}
\hline
& Time complexity & Space complexity \\
\hline
CAMP & ${\cal O}(tMN + t^{2}M + t^{4})$ & ${\cal O}(MN + tM + t^{2})$ \\
\hline
AMP & ${\cal O}(tMN)$ & ${\cal O}(MN)$ \\
\hline
OAMP/VAMP &  ${\cal O}(M^{2}N+tMN)$ & ${\cal O}(N^{2}+MN)$ \\
\hline
\end{tabular}
\end{center}
\end{table}

Table~\ref{table1} lists time and space complexity of the CAMP, AMP, and 
OAMP/VAMP. Let $t$ denote the number of iterations. 
We assume that the scalar parameters in the CAMP can be computed in 
${\cal O}(t^{4})$ time. In particular, computation of $\{a_{t,t}\}$ via the 
SE equation~(\ref{SE_equation}) is dominant.  

To compute the update rule~(\ref{zt}) in the CAMP efficiently,  
the vectors $\boldsymbol{z}_{t}\in\mathbb{R}^{M}$ and 
$\boldsymbol{A}\boldsymbol{A}^{\mathrm{T}}\boldsymbol{z}_{t}\in\mathbb{R}^{M}$ 
are computed and stored in iteration~$t$. We need ${\cal O}(MN)$ space 
complexity to store the sensing matrix $\boldsymbol{A}$, which is dominant 
for the case $t\ll N$. Furthermore, the time complexity is dominated by 
matrix-vector multiplications. 

In the OAMP/VAMP, the SVD of $\boldsymbol{A}$ requires dominant complexity 
unless the sensing matrix has a special structure that enables efficient SVD 
computation. As a result, the OAMP/VAMP has higher complexity than the AMP 
and CAMP while the CAMP has comparable complexity to the AMP for $t\ll N$. 

\section{Numerical Results} \label{sec4}
\subsection{Simulation Conditions} 
The Bayes-optimal CAMP---called CAMP simply---is compared to the AMP and 
OAMP/VAMP. In all numerical results, $10^{5}$ independent trials were 
simulated. We assumed the AWGN noise $\boldsymbol{w}\sim\mathcal{N}(
\boldsymbol{0},\sigma^{2}\boldsymbol{I}_{M})$ and 
 i.i.d.\ Bernoulli-Gaussian signals with signal 
density $\rho\in[0, 1]$ in the measurement model~(\ref{model}). 
The probability density function (pdf) of $x_{n}$ is given by
\begin{equation} \label{BG}
p(x_{n}) = (1-\rho)\delta(x_{n}) + \frac{\rho}{\sqrt{2\pi/\rho}}
e^{-\frac{x_{n}^{2}}{2/\rho}}. 
\end{equation} 
Since $x_{n}$ has zero mean and unit variance, the signal-to-noise ratio (SNR) 
is equal to $1/\sigma^{2}$. See Appendix~\ref{proof_BG_signal} for evaluation 
of the correlation~$d_{t'+1,t+1}$ given in (\ref{correlation}).  

Corollary~\ref{corollary3} was used to simulate ill-conditioned sensing 
matrices $\boldsymbol{A}$. The non-zero singular values 
$\{\sigma_{m}\}$ of $\boldsymbol{A}$ are uniquely determined via the 
condition number~$\kappa$. To reduce the complexity of the OAMP/VAMP, 
we assumed the SVD structure 
$\boldsymbol{A}=\mathrm{diag}\{\sigma_{0},\ldots,\sigma_{M-1},
\boldsymbol{0}\}\boldsymbol{V}^{\mathrm{T}}$. Note that the CAMP does not 
require this SVD structure. The CAMP only needs the right-orthogonal 
invariance of $\boldsymbol{A}$. For a further reduction in the complexity, 
we used the Hadamard matrix~$\boldsymbol{V}^{\mathrm{T}}\in\mathcal{O}_{N}$ with 
the rows permuted uniformly and randomly. This matrix $\boldsymbol{A}$ 
is a practical alternative of right-orthogonally invariant matrices.  

\begin{figure}[t]
\begin{center}
\includegraphics[width=\hsize]{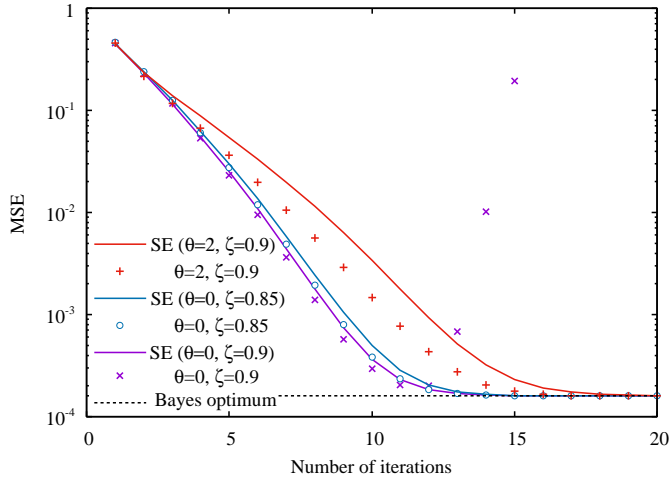}
\caption{
MSE versus the number of iterations~$t$ for the CAMP. $M=2^{12}$, $N=2^{13}$, 
$\rho=0.1$, $\kappa=5$, and $1/\sigma^{2}=30$~dB.  
}
\label{fig1} 
\end{center}
\end{figure}

We simulated damped AMP~\cite{Rangan191} with the same Bayes-optimal 
denoiser $f_{t}(\cdot)=f_{\mathrm{opt}}(\cdot; v_{t})$ as in the 
CAMP. The variance parameter $v_{t}$ was computed via the SE equation 
\begin{equation} \label{SE_equation_AMP}
v_{t} = \sigma^{2} + \frac{1}{\delta}\mathrm{MMSE}(v_{t-1}), 
\quad \mathrm{MMSE}(v_{-1})=1, 
\end{equation} 
with 
\begin{equation}
\mathrm{MMSE}(v) 
= \mathbb{E}\left[
 \{f_{\mathrm{opt}}(x_{1}+\sqrt{v}z; v) - x_{1}\}^{2}
\right], 
\end{equation}
where $z\sim\mathcal{N}(0,1)$ denotes the standard Gaussian random variable 
independent of $x_{1}$. The SE equation~(\ref{SE_equation_AMP}) was derived in 
\cite{Bayati11} 
under the assumption of zero-mean i.i.d.\ Gaussian sensing matrix with 
compression rate $\delta=M/N$. Furthermore, $\xi_{t}$ in (\ref{zt_AMP}) 
was replaced by the asymptotic value 
$\bar{\xi}_{t}=\mathrm{MMSE}(v_{t})/v_{t}$~\cite[Lemma~2]{Takeuchi201}. 
To improve the convergence 
property of the AMP, we replaced the update rule~(\ref{xt_AMP}) with 
the damped rule  
\begin{equation}
\boldsymbol{x}_{t+1} 
= \zeta f_{t}(\boldsymbol{x}_{t} + \boldsymbol{A}^{\mathrm{T}}\boldsymbol{z}_{t})
+ (1-\zeta)\boldsymbol{x}_{t}. 
\end{equation}
Note that SE cannot describe the exact 
dynamics of AMP when damping is employed.

For the OAMP/VAMP~\cite{Ma17,Rangan192}, we used the Bayes-optimal 
denoiser $f_{t}(\cdot)=f_{\mathrm{opt}}(\cdot;
\bar{v}_{\mathrm{A}\to\mathrm{B},t})$ computed via the SE 
equations~\cite{Takeuchi201}
\begin{equation}
\bar{v}_{\mathrm{A}\to \mathrm{B},t} 
= \bar{\gamma}_{t} - \bar{v}_{\mathrm{B}\to \mathrm{A},t}, \quad 
\bar{v}_{\mathrm{B}\to \mathrm{A},0}=1, 
\end{equation} 
\begin{equation}
\frac{1}{\bar{v}_{\mathrm{B}\to \mathrm{A},t+1}} 
= \frac{1}{\mathrm{MMSE}(\bar{v}_{\mathrm{A}\to \mathrm{B},t})} 
- \frac{1}{\bar{v}_{\mathrm{A}\to \mathrm{B},t}}, 
\end{equation}
with 
\begin{equation}
\bar{\gamma}_{t}^{-1} 
= \lim_{M=\delta N\to\infty}\frac{1}{N}
\sum_{m=0}^{M-1}\frac{\sigma_{m}^{2}}
{\sigma^{2} + \bar{v}_{\mathrm{B}\to \mathrm{A},t}\sigma_{m}^{2}}. 
\end{equation}
To improve the convergence property, we applied the damping technique: 
The messages $\boldsymbol{x}_{\mathrm{B}\to \mathrm{A},t+1}$ and 
$v_{\mathrm{B}\to \mathrm{A},t+1}$ in (\ref{module_B_mean}) 
were replaced by the damped messages  
$\zeta\boldsymbol{x}_{\mathrm{B}\to \mathrm{A},t+1}
+(1-\zeta)\boldsymbol{x}_{\mathrm{B}\to \mathrm{A},t}$ and 
$\zeta\bar{v}_{\mathrm{B}\to \mathrm{A},t+1}
+(1-\zeta)\bar{v}_{\mathrm{B}\to \mathrm{A},t}$, respectively.  
Note that damped EP cannot be described via SE.

\subsection{Ill-Conditioned Sensing Matrices}
We first consider the parameter $\theta$ in the CAMP defined in 
Section~\ref{sec3E}. From Theorem~\ref{theorem_fixed_point}, we know that 
the CAMP is Bayes-optimal for any $\theta$ if it converges. Thus, the 
parameter $\theta$ only affects the convergence property of the CAMP.  

Figure~\ref{fig1} shows the MSEs of the CAMP for a sensing matrix 
with condition number $\kappa=5$ defined in Corollary~\ref{corollary3}. 
As a baseline, we plotted the asymptotic MSE of the Bayes-optimal signal 
recovery~\cite{Takeda06,Tulino13,Barbier18}. 
The CAMP with $\theta=2$ and $\zeta=0.9$ converges to the Bayes-optimal 
performance more slowly than that with $\theta=0$ and $\zeta=0.85$. This 
observation does not necessarily imply that $\theta=0$ is the best option. 
When the damping factor $\zeta=0.9$ is used, the CAMP converges for $\theta=2$ 
in the finite-sized system while it diverges for $\theta=0$. Thus, we 
conclude that using non-zero $\theta\neq0$ improves the stability of the 
CAMP in finite-sized systems. 

\begin{figure}[t]
\begin{center}
\includegraphics[width=\hsize]{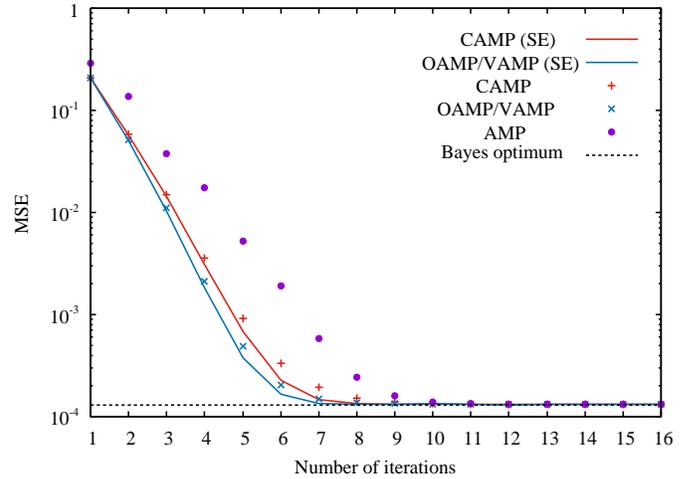}
\end{center}
\caption{
MSE versus the number of iterations~$t$ for the CAMP with $\theta=0$. 
$M=2^{11}$, $N=2^{12}$, $\rho=0.1$, $\kappa=1$, $1/\sigma^{2}=30$~dB, and 
$\zeta=1$. 
}
\label{fig2} 
\end{figure}

The CAMP is compared to the AMP and OAMP/VAMP for sensing matrices with 
unit condition number, i.e.\ orthogonal rows. As noted in Remark~\ref{remark1}, 
without loss of generality, we can use $\theta=0$ for this case. 
In this case, the 
OAMP/VAMP has comparable complexity to the AMP since the SVD of the sensing 
matrix is not required. Figure~\ref{fig2} shows that the OAMP/VAMP is the best 
in terms of the convergence speed among the three MP algorithms.

\begin{figure}[t]
\begin{center}
\includegraphics[width=\hsize]{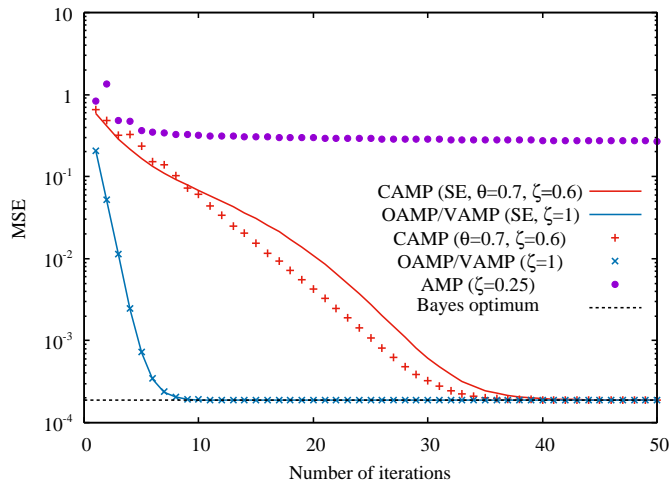}
\caption{
MSE versus the number of iterations~$t$ for the CAMP. 
$M=2^{13}$, $N=2^{14}$, $\rho=0.1$, $\kappa=10$, and $1/\sigma^{2}=30$~dB. 
}
\label{fig3} 
\end{center}
\end{figure}

We next consider a sensing matrix with condition number $\kappa=10$. 
As shown in Fig.~\ref{fig3}, the AMP cannot approach the Bayes-optimal 
performance. The CAMP converges to the Bayes-optimal performance more slowly 
than the OAMP/VAMP while the CAMP does not require high-complexity SVD of 
the sensing matrix. Especially in large systems, thus, the CAMP should need 
lower complexity to achieve the Bayes-optimal performance than the OAMP/VAMP. 

\begin{table}[t]
\begin{center}
\caption{
Parameters used in Fig.~\ref{fig4}. 
}
\label{table2}
\begin{tabular}{|c|c|c|}
\hline 
CAMP  & OAMP/VAMP  & AMP \\ 
\hline 
($\kappa$, $\theta$, $\zeta$) & ($\kappa$, $\zeta$) & ($\kappa$, $\zeta$) \\ 
\hline 
(1, 0, 0.8) & (1, 0.9) & (1, 1) \\
\hline 
(5, 1.65, 0.75) & (5, 0.75) & (2, 0.8) \\
\hline
(7.5, 1.1, 0.6) & (10, 0.7) & (2.5, 0.6) \\ 
\hline
(10, 0.75, 0.5) & (15, 0.7) & (3, 0.55) \\
\hline 
(12.5, 0.75, 0.45) & (20, 0.7) & (4, 0.45) \\
\hline 
(13.75, 0.35, 0.25) & (25, 0.7) & (5, 0.35) \\
\hline 
(14.375--14.6875, 0.35, 0.2) & (30, 0.7) & (6, 0.35) \\
\hline 
(15, 0.3, 0.2) &   & (7, 0.3) \\
\hline 
(17.5, 0.2, 0.1) &  & (8, 0.3) \\ 
\hline 
(20, 0.1, 0.05) & & \\
\hline
\end{tabular}
\end{center}
\end{table}

\begin{figure}[t]
\begin{center}
\includegraphics[width=\hsize]{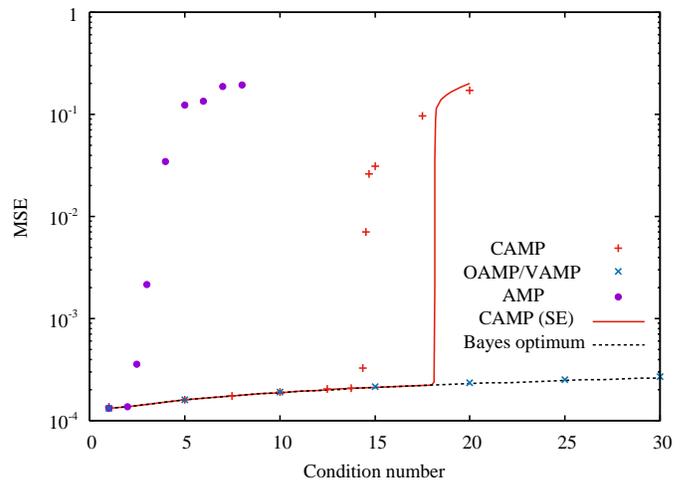}
\caption{
MSE versus the condition number~$\kappa$ for the CAMP. 
$M=512$, $N=1024$, $\rho=0.1$, $1/\sigma^{2}=30$~dB, and $150$~iterations. 
}
\label{fig4} 
\end{center}
\end{figure}

We investigate the influence of the condition number~$\kappa$ 
shown in Fig.~\ref{fig4}. In evaluating the SE of the CAMP as a baseline, 
the parameter $\theta$ was optimized for each condition number while no 
damping was employed. 
In particular, the parameter $\theta$ was set to $-0.7$ for 
$\kappa\geq17$. Otherwise, $\theta=0$ was used. 
See Table~\ref{table2} for the parameters used in the three algorithms, 
which were numerically optimized for each condition number. More precisely, 
the parameters were selected so as to achieve the fastest convergence among 
all possible parameters that approach the best MSE in the last iteration. 

The AMP has poor performance with the exception of small condition numbers. 
The CAMP achieves the Bayes-optimal performance for low-to-moderate condition 
numbers. However, it is inferior to the high-complexity OAMP/VAMP for large 
condition numbers. These observations are consistent with the SE results of 
the CAMP. The SE prediction of the MSE changes rapidly from the Bayes-optimal 
performance to a large value around a condition number $\kappa\approx18$ while 
the OAMP/VAMP still achieves the Bayes-optimal performance for $\kappa>18$.   
This is because the CAMP fails to converge for $\kappa>18$. As a result, 
we cannot use Theorem~\ref{theorem_fixed_point} to 
claim the Bayes-optimality of the CAMP. Thus, we conclude that the CAMP is 
Bayes-optimal in a strictly smaller class of sensing matrices than the 
OAMP/VAMP.  

\begin{figure}[t]
\begin{center}
\includegraphics[width=\hsize]{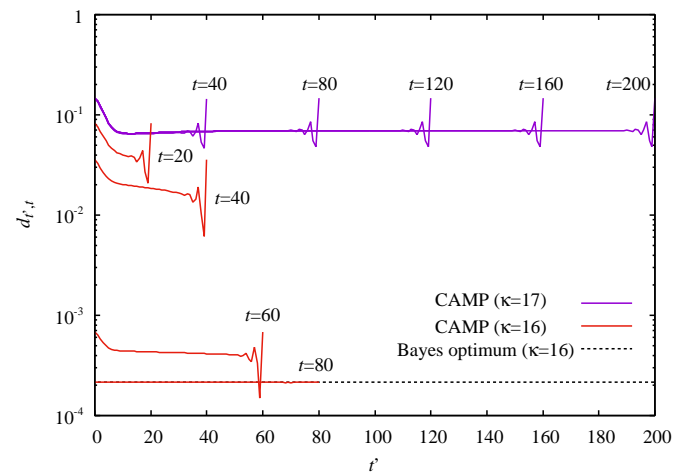}
\caption{
Correlation $d_{t',t}$ versus $t'=0,\ldots,t$ for the CAMP. 
$\delta=0.5$, $\rho=0.1$, $1/\sigma^{2}=30$~dB, $\theta=0$, and $\zeta=1$. 
}
\label{fig5} 
\end{center}
\end{figure}

Finally, we investigate the convergence properties of the CAMP for high 
condition numbers. Figure~\ref{fig5} shows the correlation $d_{t',t}$ in the 
CAMP for $t'=0,\ldots,t$. For the condition number $\kappa=16$, 
the correlation $d_{t',t}$ converges toward the Bayes-optimal MSE for all $t'$ 
as $t$ increases. This provides numerical evidence for the assumption in 
Theorem~\ref{theorem_fixed_point}: the convergence of the CAMP toward 
a fixed-point. 

The results for $\kappa=17$ imply that the CAMP fails to converge. 
A soliton-like quasi-steady wave propagates as $t$ 
grows, while the CAMP does not diverge. As implied from Fig.~\ref{fig4}, 
using non-zero $\theta\neq0$ allows us to avoid the occurrence of such a 
wave for $\kappa=17$. However, such waves occur for any $\theta$ when the 
condition number is larger than $\kappa\approx18$.  

Intuitively, the occurrence of soliton-like waves can be understood as 
follows: The SE equation~(\ref{SE_equation}) in time domain becomes unstable 
for high condition numbers, so that $a_{t',t}$ increases as $t$ grows.  
However, larger $a_{t',t}$ results in a geometrically smaller forgetting factor 
$\bar{\xi}_{t-\tau}^{(t-1)}$ in (\ref{SE_equation}), which suppresses the 
divergence of $a_{t',t}$. As a result, a soliton-like quasi-steady wave 
occurs for high condition numbers. 

\section{Conclusions} \label{sec5} 
The Bayes-optimal CAMP solves the disadvantages of AMP and OAMP/VAMP,  
and realizes their advantages for orthogonally invariant sensing matrices 
with low-to-moderate condition numbers: The Bayes-optimal CAMP is an efficient 
MP algorithm that has comparable complexity to AMP. Furthermore, the CAMP has 
been proved to be Bayes-optimal for all orthogonally invariant sensing 
matrices if it converges. High-complexity OAMP/VAMP is Bayes-optimal 
for this class of sensing matrices while AMP is not. The CAMP converges 
for sensing matrices with low-to-moderate condition numbers while it fails 
to converge for high condition numbers.

A disadvantage of CAMP is to need all moments of the asymptotic 
singular-value distribution of the sensing matrix. In general, computation 
of the moments requires high complexity unless their closed-form is available. 
To circumvent this issue, deep unfolding~\cite{Gregor10,Borgerding17} might  
be utilized to learn the tap coefficients in the Onsager correction without 
using the asymptotic singular-value distribution.  

The CAMP has a room for improvement especially in finite-sized and 
ill-conditioned sensing matrices. One option is a replacement of scalar  
parameters determined via the SE equation with empirical estimators that 
depend on the measurements, as considered in AMP and OAMP/VAMP. 

Another option is a damping technique that keeps the asymptotic Gaussianity 
of estimation errors. This paper used a heuristic damping technique to improve 
the convergence property of the CAMP. However, the heuristic damping breaks 
the asymptotic Gaussianity. Damped CAMP should be designed via 
Theorem~\ref{theorem_SE} to guarantee the asymptotic Gaussianity. 
A recent paper~\cite{Liu20} proposed long-memory damping in the MF-based 
interference cancellation to improve the convergence property of long-memory 
MP. A possible future work is to design CAMP with long-memory damping.

\appendices 

\section{Proof of Theorem~\ref{theorem_SE}}
\label{proof_theorem_SE}. 
\subsection{Formulation} 
We use Bolthausen's conditioning technique~\cite{Bolthausen14} to prove 
Theorem~\ref{theorem_SE}. In the technique, the random variables are 
classified into three groups: $\boldsymbol{V}$, 
$\mathfrak{F}=\{\boldsymbol{\lambda},\tilde{\boldsymbol{w}},\boldsymbol{x}\}$, 
and $\mathfrak{E}_{t,t'}=\{\boldsymbol{B}_{t'}, \tilde{\boldsymbol{M}}_{t'}, 
\boldsymbol{H}_{t}, \tilde{\boldsymbol{Q}}_{t+1}\}$ with 
$\tilde{\boldsymbol{Q}}_{t+1}=(\tilde{\boldsymbol{q}}_{0},\ldots,
\tilde{\boldsymbol{q}}_{t})$ and $\tilde{\boldsymbol{M}}_{t}
=(\tilde{\boldsymbol{m}}_{0},\ldots,\tilde{\boldsymbol{m}}_{t-1})$.  
The random variables in $\mathfrak{F}$ are fixed throughout the proof of 
Theorem~\ref{theorem_SE} while $\boldsymbol{V}$ is averaged out.  

The set $\mathfrak{E}_{t,t}$ contains all messages just before updating 
$\boldsymbol{b}_{t}=\boldsymbol{V}^{\mathrm{T}}\tilde{\boldsymbol{q}}_{t}$  
while $\mathfrak{E}_{t,t+1}$ includes all messages just before updating  
$\boldsymbol{h}_{t}=\boldsymbol{V}\tilde{\boldsymbol{m}}_{t}$. The main 
part in the conditioning technique is evaluation of  
the conditional distribution of $\boldsymbol{b}_{t}$ given 
$\mathfrak{E}_{t,t}$ and $\mathfrak{F}$ via that of $\boldsymbol{V}$. 

Theorem~\ref{theorem_SE} is proved by induction. More precisely, we prove 
a theorem obtained by adding several technical results to 
Theorem~\ref{theorem_SE}. Before presenting the theorem, 
we first define several notations. 

The notation $\boldsymbol{o}(1)$ denotes a finite-dimensional vector 
with vanishing norm. 
For a tall matrix $\boldsymbol{M}\in\mathbb{R}^{N\times t}$ with rank $r\leq t$, 
the SVD of $\boldsymbol{M}$ is denoted by 
$\boldsymbol{M}=\boldsymbol{\Phi}_{\boldsymbol{M}}
\boldsymbol{\Sigma}_{\boldsymbol{M}}\boldsymbol{\Psi}_{\boldsymbol{M}}^{\mathrm{T}}$, 
with $\boldsymbol{\Phi}_{\boldsymbol{M}}=
(\boldsymbol{\Phi}_{\boldsymbol{M}}^{\parallel}, 
\boldsymbol{\Phi}_{\boldsymbol{M}}^{\perp})$. The matrix  
$\boldsymbol{\Phi}_{\boldsymbol{M}}^{\parallel}\in\mathcal{O}_{N\times r}$ consists 
of all left-singular vectors corresponding to $r$ non-zero singular values 
while $\boldsymbol{\Phi}_{\boldsymbol{M}}^{\perp}\in\mathcal{O}_{N\times(N-r)}$ 
is composed of left-singular vectors corresponding to $N-r$ zero singular 
values. The matrix $\boldsymbol{P}_{\boldsymbol{M}}^{\parallel}
=\boldsymbol{M}(\boldsymbol{M}^{\mathrm{T}}\boldsymbol{M})^{-1}
\boldsymbol{M}^{\mathrm{T}}$ is the projection to the space spanned by the 
columns of $\boldsymbol{M}$ while $\boldsymbol{P}_{\boldsymbol{M}}^{\perp}
=\boldsymbol{I}-\boldsymbol{P}_{\boldsymbol{M}}^{\parallel}$ is the projection 
to the orthogonal complement. Note that $\boldsymbol{P}_{\boldsymbol{M}}^{\parallel}
=\boldsymbol{\Phi}_{\boldsymbol{M}}^{\parallel}
(\boldsymbol{\Phi}_{\boldsymbol{M}}^{\parallel})^{\mathrm{T}}$ and 
$\boldsymbol{P}_{\boldsymbol{M}}^{\perp}
=\boldsymbol{\Phi}_{\boldsymbol{M}}^{\perp}
(\boldsymbol{\Phi}_{\boldsymbol{M}}^{\perp})^{\mathrm{T}}$ hold. 

In the following theorem, we call the system with respect to 
$\{\boldsymbol{B}_{t}, \tilde{\boldsymbol{M}}_{t}\}$ module~A while 
we refer to that for $\{\boldsymbol{H}_{t}, \tilde{\boldsymbol{Q}}_{t+1}\}$ 
as module~B. 
\begin{theorem} \label{theorem_SE_tech} 
Suppose that Assumptions~\ref{assumption_x}--\ref{assumption_thresholding} 
hold. Then, the following properties in module A hold  
for all $\tau=0,1,\ldots$ in the large system limit.  
\begin{enumerate}[label=(A\arabic*)]
\item \label{property_A1}
Let $\boldsymbol{\beta}_{\tau}=(\tilde{\boldsymbol{Q}}_{\tau}^{\mathrm{T}}
\tilde{\boldsymbol{Q}}_{\tau})^{-1}\tilde{\boldsymbol{Q}}_{\tau}^{\mathrm{T}}
\tilde{\boldsymbol{q}}_{\tau}$, $\tilde{\boldsymbol{q}}_{\tau}^{\perp}
=\boldsymbol{P}_{\tilde{\boldsymbol{Q}}_{\tau}}^{\perp}
\tilde{\boldsymbol{q}}_{\tau}$, and 
\begin{equation} 
\tilde{\boldsymbol{\omega}}_{\tau} 
= \tilde{\boldsymbol{V}}^{\mathrm{T}}
(\boldsymbol{\Phi}_{(\tilde{\boldsymbol{Q}}_{\tau}, \boldsymbol{H}_{\tau})}^{\perp})^{\mathrm{T}}
\tilde{\boldsymbol{q}}_{\tau}, 
\end{equation}
where $\tilde{\boldsymbol{V}}\in\mathcal{O}_{N-2\tau}$ is a Haar orthogonal 
matrix and independent of $\mathfrak{F}$ and $\mathfrak{E}_{\tau,\tau}$. Then, 
for $\tau>0$ 
\begin{equation} \label{b_tau}
\boldsymbol{b}_{\tau}
\sim \boldsymbol{B}_{\tau}\boldsymbol{\beta}_{\tau} 
+ \tilde{\boldsymbol{M}}_{\tau}\boldsymbol{o}(1) 
+ \boldsymbol{B}_{\tau}\boldsymbol{o}(1) 
+ \boldsymbol{\Phi}_{(\boldsymbol{B}_{\tau}, \tilde{\boldsymbol{M}}_{\tau})}^{\perp}
\tilde{\boldsymbol{\omega}}_{\tau} 
\end{equation}
conditioned on $\mathfrak{F}$ and $\mathfrak{E}_{\tau,\tau}$  
in the large system limit, with 
\begin{equation} \label{omega_tilde_variance} 
\lim_{M=\delta N\to\infty}\frac{1}{N}\left\{
 \|\tilde{\boldsymbol{\omega}}_{\tau}\|^{2} 
 - \|\tilde{\boldsymbol{q}}_{\tau}^{\perp}\|^{2} 
\right\}
\aeq 0.
\end{equation}

\item \label{property_A2}
Suppose that $\tilde{\boldsymbol{\phi}}_{\tau}(\boldsymbol{B}_{\tau+1}, 
\tilde{\boldsymbol{w}}, \boldsymbol{\lambda}):
\mathbb{R}^{N\times(\tau+3)}\to\mathbb{R}^{N}$ 
is separable, pseudo-Lipschitz of order~$k$ with respect to 
the first $\tau+2$ variables, and proper. 
If $N^{-1}\tilde{\boldsymbol{q}}_{t}^{\mathrm{T}}\tilde{\boldsymbol{q}}_{t'}$ 
converges almost surely to some constant $\kappa_{t,t'}\in\mathbb{R}$ 
in the large system limit for all $t, t'=0,\ldots,\tau$, then   
\begin{equation} \label{property_A2_former}
\langle\tilde{\boldsymbol{\phi}}_{\tau}(\boldsymbol{B}_{\tau+1}, 
\tilde{\boldsymbol{w}};\boldsymbol{\lambda})\rangle
- \mathbb{E}\left[
 \langle\tilde{\boldsymbol{\phi}}_{\tau}(\tilde{\boldsymbol{Z}}_{\tau+1}, 
 \tilde{\boldsymbol{w}},\boldsymbol{\lambda})\rangle 
\right]\ato 0. 
\end{equation}
In (\ref{property_A2_former}), 
$\tilde{\boldsymbol{Z}}_{\tau+1}=(\tilde{\boldsymbol{z}}_{0},\ldots,
\tilde{\boldsymbol{z}}_{\tau})\in\mathbb{R}^{N\times(\tau+1)}$ denotes a zero-mean 
Gaussian random matrix with covariance $\mathbb{E}[\tilde{\boldsymbol{z}}_{t}
\tilde{\boldsymbol{z}}_{t'}^{\mathrm{T}}]
=\kappa_{t,t'}\boldsymbol{I}_{N}$ for all $t, t'=0,\ldots,\tau$. In evaluating 
the expectation in (\ref{property_A2_former}), 
$\boldsymbol{U}^{\mathrm{T}}\boldsymbol{w}$ in (\ref{w_tilde}) follows 
the zero-mean Gaussian distribution with covariance 
$\sigma^{2}\boldsymbol{I}_{M}$.  
In particular, for $k=1$ we have 
\begin{equation} \label{property_A2_latter}
\langle\partial_{\tau'}\tilde{\boldsymbol{\phi}}_{\tau}(\boldsymbol{B}_{\tau+1}, 
\tilde{\boldsymbol{w}};\boldsymbol{\lambda})\rangle
- \mathbb{E}\left[
 \langle\partial_{\tau'}\tilde{\boldsymbol{\phi}}_{\tau}
 (\tilde{\boldsymbol{Z}}_{\tau+1}, 
 \tilde{\boldsymbol{w}};\boldsymbol{\lambda})\rangle 
\right]\ato 0
\end{equation}
for all $\tau'=0,\ldots,\tau$. 

\item \label{property_A3} 
Suppose that $\tilde{\boldsymbol{\phi}}_{\tau}(\boldsymbol{B}_{\tau+1},
\tilde{\boldsymbol{w}};\boldsymbol{\lambda}):
\mathbb{R}^{N\times(\tau+3)}\to\mathbb{R}^{N}$ is separable, 
Lipschitz-continuous with respect to the first $\tau+2$ variables, 
and proper. Then, 
\begin{equation} \label{property_A3_eq}
\frac{1}{N}\boldsymbol{b}_{\tau'}^{\mathrm{T}}\left(
 \tilde{\boldsymbol{\phi}}_{\tau}
 - \sum_{t'=0}^{\tau}\left\langle
  \partial_{t'}\tilde{\boldsymbol{\phi}}_{\tau}
 \right\rangle\boldsymbol{b}_{t'}
\right)
\ato0 
\end{equation}
for all $\tau'=0,\ldots,\tau$. 

\item \label{property_A4}
The inner product 
$N^{-1}\tilde{\boldsymbol{m}}_{\tau'}^{\mathrm{T}}\tilde{\boldsymbol{m}}_{\tau}$ 
converges almost surely to some constant $\pi_{\tau',\tau}\in\mathbb{R}$ for 
all $\tau'=0,\ldots,\tau$. 

\item \label{property_A5}
For some $\epsilon>0$ and $C>0$, 
\begin{equation} \label{4th_moment_m}
\lim_{M=\delta N\to\infty}\mathbb{E}\left[
 |\tilde{m}_{\tau,n}|^{2k-2+\epsilon}
\right]<\infty, 
\end{equation} 
\begin{equation} \label{norm_M}
\liminf_{M=\delta N\to\infty}\lambda_{\mathrm{min}}\left(
 \frac{1}{N}\tilde{\boldsymbol{M}}_{\tau+1}^{\mathrm{T}}
 \tilde{\boldsymbol{M}}_{\tau+1}
\right)
\ag C.  
\end{equation}
\end{enumerate}

The following properties in module~B hold for all $\tau=0,1,\ldots$ 
in the large system limit. 
\begin{enumerate}[label=(B\arabic*)]
\item \label{property_B1} 
Let $\boldsymbol{\alpha}_{\tau}=(\tilde{\boldsymbol{M}}_{\tau}^{\mathrm{T}}
\tilde{\boldsymbol{M}}_{\tau})^{-1}\tilde{\boldsymbol{M}}_{\tau}^{\mathrm{T}}
\tilde{\boldsymbol{m}}_{\tau}$, $\tilde{\boldsymbol{m}}_{0}^{\perp}
=\tilde{\boldsymbol{m}}_{0}$, $\tilde{\boldsymbol{m}}_{\tau}^{\perp}
=\boldsymbol{P}_{\tilde{\boldsymbol{M}}_{\tau}}^{\perp}
\tilde{\boldsymbol{m}}_{\tau}$, and  
\begin{equation} 
\boldsymbol{\omega}_{\tau} 
= \left\{
\begin{array}{cl}
\tilde{\boldsymbol{V}}(\boldsymbol{\Phi}_{\boldsymbol{b}_{0}}^{\perp})^{\mathrm{T}}
\tilde{\boldsymbol{m}}_{0} & \hbox{for $\tau=0$,} \\
\tilde{\boldsymbol{V}}
(\boldsymbol{\Phi}_{(\tilde{\boldsymbol{M}}_{\tau}, 
\boldsymbol{B}_{\tau+1})}^{\perp})^{\mathrm{T}}
\tilde{\boldsymbol{m}}_{\tau} & \hbox{for $\tau>0$,} \\
\end{array}
\right. 
\end{equation}
where $\tilde{\boldsymbol{V}}\in\mathcal{O}_{N-(2\tau+1)}$ is a Haar orthogonal 
matrix and independent of $\mathfrak{F}$ and $\mathfrak{E}_{\tau,\tau+1}$. Then, 
we have 
\begin{equation} \label{h0}
\boldsymbol{h}_{0}
\sim o(1)\tilde{\boldsymbol{q}}_{0} 
+ \boldsymbol{\Phi}_{\tilde{\boldsymbol{q}}_{0}}^{\perp}
\boldsymbol{\omega}_{\tau}, 
\end{equation}
conditioned on $\mathfrak{F}$ and 
$\mathfrak{E}_{0,1}=\{\boldsymbol{b}_{0}, \tilde{\boldsymbol{m}}_{0}, 
\tilde{\boldsymbol{q}}_{0}\}$ in the large system limit. For $\tau>0$ 
\begin{equation} 
\boldsymbol{h}_{\tau}
\sim \boldsymbol{H}_{\tau}\boldsymbol{\alpha}_{\tau} 
+ \tilde{\boldsymbol{Q}}_{\tau+1}\boldsymbol{o}(1) 
+ \boldsymbol{H}_{\tau}\boldsymbol{o}(1)  
+ \boldsymbol{\Phi}_{(\boldsymbol{H}_{\tau}, \tilde{\boldsymbol{Q}}_{\tau+1})}^{\perp}
\boldsymbol{\omega}_{\tau}, 
\end{equation}
conditioned on $\mathfrak{F}$ and $\mathfrak{E}_{\tau,\tau+1}$ 
in the large system limit, with 
\begin{equation} \label{omega_variance}
\lim_{M=\delta N\to\infty}\frac{1}{N}\left\{
 \|\boldsymbol{\omega}_{\tau}\|^{2}  
 - \|\tilde{\boldsymbol{m}}_{\tau}^{\perp}\|^{2}
\right\} \aeq 0.  
\end{equation}

\item \label{property_B2}
Suppose that $\tilde{\boldsymbol{\psi}}_{\tau}(\boldsymbol{H}_{\tau+1}, 
\boldsymbol{x}):\mathbb{R}^{N\times(\tau+2)}\to\mathbb{R}^{N}$ 
is a separable and proper pseudo-Lipschitz function of order~$k$.  
If $N^{-1}\tilde{\boldsymbol{m}}_{t}^{\mathrm{T}}\tilde{\boldsymbol{m}}_{t'}$ 
converges almost surely to some constant $\pi_{t,t'}\in\mathbb{R}$ 
in the large system limit for all $t, t'=0,\ldots,\tau$, then   
\begin{equation} \label{property_B2_former}
\langle\tilde{\boldsymbol{\psi}}_{\tau}(\boldsymbol{H}_{\tau+1}, 
\boldsymbol{x})\rangle
- \mathbb{E}\left[
 \langle\tilde{\boldsymbol{\psi}}_{\tau}(\boldsymbol{Z}_{\tau+1}, 
 \boldsymbol{x})\rangle 
\right]\ato 0, 
\end{equation}
where $\boldsymbol{Z}_{\tau+1}=(\boldsymbol{z}_{0},\ldots,\boldsymbol{z}_{\tau})
\in\mathbb{R}^{N\times(\tau+1)}$ denotes a zero-mean Gaussian random matrix with 
covariance $\mathbb{E}[\boldsymbol{z}_{t}\boldsymbol{z}_{t'}^{\mathrm{T}}]
=\pi_{t,t'}\boldsymbol{I}_{N}$ for all $t, t'=0,\ldots,\tau$. 
In particular, for $k=1$ we have 
\begin{equation} \label{property_B2_latter}
\langle\partial_{\tau'}\tilde{\boldsymbol{\psi}}_{\tau}(\boldsymbol{H}_{\tau+1}, 
\boldsymbol{x})\rangle
- \mathbb{E}\left[
 \langle\partial_{\tau'}\tilde{\boldsymbol{\psi}}_{\tau}(\boldsymbol{Z}_{\tau+1}, 
 \boldsymbol{x})\rangle 
\right]\ato 0
\end{equation}
for all $\tau'=0,\ldots,\tau$. 

\item \label{property_B3}
Suppose that $\tilde{\boldsymbol{\psi}}_{\tau}(\boldsymbol{H}_{\tau+1},
\boldsymbol{x}):\mathbb{R}^{N\times(\tau+2)}\to\mathbb{R}^{N}$ is a 
separable and proper Lipschitz-continuous function. Then, 
\begin{equation} \label{property_B3_eq} 
\frac{1}{N}\boldsymbol{h}_{\tau'}^{\mathrm{T}}\left(
 \tilde{\boldsymbol{\psi}}_{\tau}
 - \sum_{t'=0}^{\tau}\left\langle
  \partial_{t'}\tilde{\boldsymbol{\psi}}_{\tau}
 \right\rangle\boldsymbol{h}_{t'}
\right)
\ato0 
\end{equation}
for all $\tau'=0,\ldots,\tau$. 

\item \label{property_B4} 
The inner product 
$N^{-1}\tilde{\boldsymbol{q}}_{\tau'}^{\mathrm{T}}\tilde{\boldsymbol{q}}_{\tau+1}$ 
converges almost surely to some constant $\pi_{\tau',\tau+1}\in\mathbb{R}$ for 
all $\tau'=0,\ldots,\tau+1$. 

\item \label{property_B5}
For some $\epsilon>0$ and $C>0$, 
\begin{equation} \label{4th_moment_q}
\lim_{M=\delta N\to\infty}\mathbb{E}\left[
 |\tilde{q}_{\tau+1,n}|^{2+\epsilon}
\right]<\infty,  
\end{equation} 
\begin{equation} \label{norm_Q}
\liminf_{M=\delta N\to\infty}\lambda_{\mathrm{min}}\left(
 \frac{1}{N}\tilde{\boldsymbol{Q}}_{\tau+2}^{\mathrm{T}}
 \tilde{\boldsymbol{Q}}_{\tau+2}
\right)
\ag C.
\end{equation}
\end{enumerate}
\end{theorem}

We summarize useful lemmas used in the proof of Theorem~\ref{theorem_SE_tech} 
by induction. 
\begin{lemma}[\cite{Rangan192,Takeuchi201}] 
\label{lemma_conditioning} 
Suppose that $\boldsymbol{X}\in\mathbb{R}^{N\times t}$ has full rank for 
$0<t<N$, and consider the noiseless and compressed observation 
$\boldsymbol{Y}\in\mathbb{R}^{N\times t}$ of $\boldsymbol{V}$ given by 
\begin{equation} \label{constraint}
\boldsymbol{Y} = \boldsymbol{V}\boldsymbol{X}. 
\end{equation}
Then, the conditional distribution of the Haar orthogonal matrix 
$\boldsymbol{V}$ given $\boldsymbol{X}$ and $\boldsymbol{Y}$ satisfies 
\begin{equation} \label{V_conditional_distribution} 
\boldsymbol{V}|_{\boldsymbol{X}, \boldsymbol{Y}} 
\sim \boldsymbol{Y}(\boldsymbol{Y}^{\mathrm{T}}\boldsymbol{Y})^{-1}
\boldsymbol{X}^{\mathrm{T}} + \boldsymbol{\Phi}_{\boldsymbol{Y}}^{\perp}
\tilde{\boldsymbol{V}}(\boldsymbol{\Phi}_{\boldsymbol{X}}^{\perp})^{\mathrm{T}}, 
\end{equation}
where $\tilde{\boldsymbol{V}}\in\mathcal{O}_{N-t}$ is a Haar orthogonal matrix 
independent of $\boldsymbol{X}$ and $\boldsymbol{Y}$. 
\end{lemma}

The following lemma is a generalization of Stein's lemma.  
The lemma is proved under a different assumption from in \cite{Campese15}. 

\begin{lemma} \label{lemma_Stein}
Let $\boldsymbol{z}=(z_{1},\ldots,z_{t})^{\mathrm{T}}
\sim\mathcal{N}(\boldsymbol{0},\boldsymbol{\Sigma})$ for any positive 
definite covariance matrix $\boldsymbol{\Sigma}$. 
If $f:\mathbb{R}^{t}\to\mathbb{R}$ is Lipschitz-continuous, 
then we have 
\begin{equation} \label{Stein}
\mathbb{E}[z_{1}f(\boldsymbol{z})] 
= \sum_{t'=1}^{t}\mathbb{E}[z_{1}z_{t'}]\mathbb{E}[\partial_{t'}f(\boldsymbol{z})]. 
\end{equation}
\end{lemma}
\begin{IEEEproof}
We first confirm that both sides of (\ref{Stein}) are bounded. 
For the left-hand side (LHS), we find 
$f(\boldsymbol{z})={\cal O}(\|\boldsymbol{z}\|)$ as 
$\|\boldsymbol{z}\|\to\infty$ since $f$ is Lipschitz-continuous. 
Thus, $\mathbb{E}[z_{1}f(\boldsymbol{z})]$ is 
bounded for $\boldsymbol{z}\sim
\mathcal{N}(\boldsymbol{0},\boldsymbol{\Sigma})$. 

For the RHS, we use the Lipschitz-continuity of $f$ to find that 
there is some Lipschitz-constant $L>0$ such that 
\begin{equation}
\left|
 \frac{f(\boldsymbol{z}+\Delta\boldsymbol{e}_{t'}) - f(\boldsymbol{z})}{\Delta}
\right|
\leq L
\end{equation}
holds for any $\Delta\neq0$, 
where $\boldsymbol{e}_{t'}\in\mathbb{R}^{t}$ 
is the $t'$th column of $\boldsymbol{I}_{t}$. This implies that each 
partial derivative $\partial_{t'}f$ is bounded almost everywhere since the 
partial derivatives of any Lipschitz-continuous function exist almost 
everywhere. Thus, $\mathbb{E}[\partial_{t'}f(\boldsymbol{z})]$ is 
bounded. These observations indicate the boundedness of both sides 
in (\ref{Stein}). 

For the eigen-decomposition 
$\boldsymbol{\Sigma}=\boldsymbol{\Phi}\boldsymbol{\Lambda}
\boldsymbol{\Phi}^{\mathrm{T}}$, we use the change of variables 
$\tilde{\boldsymbol{z}}=\boldsymbol{\Phi}^{\mathrm{T}}\boldsymbol{z}$ to obtain 
\begin{equation}
\mathbb{E}[z_{1}f(\boldsymbol{z})] 
= \sum_{\tau=1}^{t}[\boldsymbol{\Phi}]_{1,\tau}
\mathbb{E}[\tilde{z}_{\tau}f(\boldsymbol{\Phi}\tilde{\boldsymbol{z}})] 
= \sum_{\tau=1}^{t}[\boldsymbol{\Phi}]_{1,\tau}
\mathbb{E}[\tilde{z}_{\tau}g(\tilde{z}_{\tau})], 
\end{equation}
with $g(\tilde{z}_{\tau})
=\mathbb{E}[f(\boldsymbol{\Phi}\tilde{\boldsymbol{z}})|\tilde{z}_{\tau}]$. 

We prove that $g$ is Lipschitz-continuous. 
Let $\tilde{\boldsymbol{z}}_{x}$ denote the vector obtained by replacing 
$\tilde{z}_{\tau}$ in $\tilde{\boldsymbol{z}}$ with $x$. 
Since $\tilde{\boldsymbol{z}}\sim\mathcal{N}(\boldsymbol{0},
\boldsymbol{\Lambda})$ has independent elements, we have 
\begin{IEEEeqnarray}{rl}
|g(x) - g(y)| 
\leq& \mathbb{E}\left[
 \left|
  f(\boldsymbol{\Phi}\tilde{\boldsymbol{z}}_{x}) 
  - f(\boldsymbol{\Phi}\tilde{\boldsymbol{z}}_{y}) 
 \right|
\right] \nonumber \\
\leq& L\mathbb{E}\left[
 \|\boldsymbol{\Phi}(\tilde{\boldsymbol{z}}_{x} - \tilde{\boldsymbol{z}}_{y})\|
\right] \nonumber \\
=& L\mathbb{E}\left[
 \|\tilde{\boldsymbol{z}}_{x} - \tilde{\boldsymbol{z}}_{y}\|
\right]
= L|x-y|, 
\end{IEEEeqnarray}
where the second inequality follows from the Lipschitz-continuity of $f$ with 
a Lipschitz-constant $L>0$. 
Thus, $g(\tilde{z}_{\tau})$ is Lipschitz-continuous, so that 
$g(\tilde{z}_{\tau})$ is differentiable almost everywhere. 

Since $\tilde{\boldsymbol{z}}\sim\mathcal{N}(\boldsymbol{0},
\boldsymbol{\Lambda})$ holds, Stein's lemma~\cite{Stein72} yields 
\begin{IEEEeqnarray}{rl}
\mathbb{E}[z_{1}f(\boldsymbol{z})] 
&= \sum_{\tau=1}^{t}[\boldsymbol{\Phi}]_{1,\tau}
\mathbb{E}[\tilde{z}_{\tau}^{2}]\mathbb{E}\left[
 g'(\tilde{z}_{\tau})
\right] \nonumber \\
=&\sum_{\tau=1}^{t}[\boldsymbol{\Phi}]_{1,\tau}
[\boldsymbol{\Lambda}]_{\tau,\tau}\mathbb{E}\left[
 \sum_{t'=1}^{t}[\boldsymbol{\Phi}]_{t',\tau}\partial_{t'}
 f(\boldsymbol{z})
\right]. 
\end{IEEEeqnarray}
Using the identity 
\begin{equation}
\sum_{\tau=1}^{t}[\boldsymbol{\Phi}]_{1,\tau}
[\boldsymbol{\Lambda}]_{\tau,\tau}[\boldsymbol{\Phi}]_{t',\tau}
= [\boldsymbol{\Phi}\boldsymbol{\Lambda}\boldsymbol{\Phi}^{\mathrm{T}}
]_{1,t'}
= \mathbb{E}[z_{1}z_{t'}],  
\end{equation}
we arrive at Lemma~\ref{lemma_Stein}. 
\end{IEEEproof}

\begin{lemma}[\cite{Takeuchi201}] \label{lemma_SLLN}
For $t\in\mathbb{N}$, 
suppose that $\boldsymbol{f}:\mathbb{R}^{N\times(t+1)}\to\mathbb{R}^{N}$ is 
separable and pseudo-Lipschitz of order $k$. Let $L_{n}>0$ denote 
a Lipschitz constant of the $n$th element $[\boldsymbol{f}]_{n}$. The 
sequence of Lipschitz constants is assumed to satisfy 
\begin{equation} \label{Lipschitz_constant}
\limsup_{N\to\infty}\frac{1}{N}\sum_{n=1}^{N}L_{n}^{2}<\infty. 
\end{equation}
Let $\boldsymbol{\epsilon}=(\epsilon_{1},\ldots,\epsilon_{N})^{\mathrm{T}}
\in\mathbb{R}^{N}$ denote a vector that satisfies  
\begin{equation} \label{assumption_a1}
\lim_{N\to\infty}\frac{1}{N}\sum_{n=1}^{N}L_{n}\epsilon_{n}^{2}\aeq0, 
\end{equation}
\begin{equation} \label{assumption_a2}
\limsup_{N\to\infty}\frac{1}{N}\sum_{n=1}^{N}L_{n}\epsilon_{n}^{2k-2}\al \infty. 
\end{equation} 
Suppose that $\boldsymbol{A}_{t+1}=(\boldsymbol{a}_{0},\ldots,\boldsymbol{a}_{t})
\in\mathbb{R}^{N\times(t+1)}$  satisfies 
\begin{equation} \label{assumption_a3}
\limsup_{N\to\infty}\frac{1}{N}\sum_{n=1}^{N}L_{n}^{i}a_{t',n}^{2k-2}\al \infty
\quad \hbox{for $i=1, 2$.}  
\end{equation} 
For $t'>0$, let $\boldsymbol{E}=(\boldsymbol{e}_{1}^{\mathrm{T}},\ldots, 
\boldsymbol{e}_{N}^{\mathrm{T}})^{\mathrm{T}}\in\mathbb{R}^{N\times t'}$ denote 
a matrix that satisfies   
\begin{equation} \label{assumption_E}
\limsup_{N\to\infty}
\frac{1}{N}\sum_{n=1}^{N}L_{n}\|\boldsymbol{e}_{n}\|^{\max\{2, 2k-2\}}\al\infty, 
\end{equation}
\begin{equation} \label{norm_E}
\liminf_{N\to\infty}\lambda_{\mathrm{min}}\left(
 \frac{1}{N}\boldsymbol{E}^{\mathrm{H}}\boldsymbol{E}
\right) \ag C 
\end{equation}
for some constant $C>0$. 
Suppose that $\boldsymbol{\omega}\in\mathbb{R}^{N-t'}$ is an orthogonally 
invariant random vector conditioned on 
$\boldsymbol{\epsilon}$, $\boldsymbol{A}_{t+1}$, and $\boldsymbol{E}$. 
For some $v>0$, postulate the following: 
\begin{equation} \label{convergence_X}
\lim_{N\to\infty}\frac{1}{N}\|\boldsymbol{\omega}\|^{2}\aeq v>0. 
\end{equation}
Let $\boldsymbol{z}\sim\mathcal{N}(\boldsymbol{0},v\boldsymbol{I}_{N})$ 
denote a standard Gaussian random vector independent of the other 
random variables. Then,  
\begin{equation}
\lim_{N\to\infty}\left\langle
 \boldsymbol{f}(\boldsymbol{A}_{t},
 \boldsymbol{a}_{t}+\boldsymbol{\epsilon}
 +\boldsymbol{\Phi}_{\boldsymbol{E}}^{\perp}\boldsymbol{\omega})
 - \mathbb{E}_{\boldsymbol{z}}[\boldsymbol{f}(
 \boldsymbol{A}_{t}, 
 \boldsymbol{a}_{t}+\boldsymbol{z})]
\right\rangle \aeq 0.
\end{equation}  
\end{lemma}

\subsection{Module~A for $\tau=0$}
\begin{IEEEproof}[Proof of Property~\ref{property_A2} for $\tau=0$]
The latter property~(\ref{property_A2_latter}) follows from the 
former property~(\ref{property_A2_former}) and a technical result proved 
in \cite[Lemma 5]{Bayati11}. Thus, we only prove the former  
property for $\tau=0$. 

Property~(\ref{property_A2_former}) follows from Lemma~\ref{lemma_SLLN} 
for $\boldsymbol{f}(\tilde{\boldsymbol{w}},\tilde{\boldsymbol{b}}_{0})
=\tilde{\boldsymbol{\phi}}_{0}(\tilde{\boldsymbol{b}}_{0},
\tilde{\boldsymbol{w}};\boldsymbol{\lambda})$ with 
$\boldsymbol{a}_{0}=\tilde{\boldsymbol{w}}$, 
$\boldsymbol{a}_{1}+\boldsymbol{\epsilon}=\boldsymbol{0}$, 
$\boldsymbol{\Phi}_{\boldsymbol{E}}^{\perp}=\boldsymbol{I}_{N}$, and  
$\boldsymbol{\omega}=\tilde{\boldsymbol{b}}_{0}$. 
We confirm all conditions in Lemma~\ref{lemma_SLLN}. 
Applying H\"older's inequality for any $\epsilon>0$, we have    
\begin{equation}
\frac{1}{N}\sum_{n=1}^{N}L_{n}^{i}\tilde{w}_{n}^{2k-2}
\leq \left(
 \frac{1}{N}\sum_{n=1}^{N}L_{n}^{ip}
\right)^{1/p}\left(
 \frac{1}{N}\sum_{n=1}^{N}\tilde{w}_{n}^{2k-2+\epsilon}
\right)^{1/q}
\end{equation}
for $i=1, 2$, with $q=1+\epsilon/(2k-2)$ and $p^{-1}=1-q^{-1}$, which is bounded 
because of Assumption~\ref{assumption_w}. Furthermore, the definition 
$\boldsymbol{b}_{0}=-\boldsymbol{V}^{\mathrm{T}}\boldsymbol{x}$ implies the 
orthogonal invariance and $N^{-1}\|\boldsymbol{b}_{0}\|^{2} \ato 1$. 
Thus, all conditions in Lemma~\ref{lemma_SLLN} hold. 
Using Lemma~\ref{lemma_SLLN}, we obtain 
\begin{equation}
\langle\tilde{\boldsymbol{\phi}}_{0}(\tilde{\boldsymbol{b}}_{0},
\tilde{\boldsymbol{w}}; \boldsymbol{\lambda})\rangle 
- \mathbb{E}_{\tilde{\boldsymbol{z}}_{0}}\left[
 \langle\tilde{\boldsymbol{\phi}}_{0}(\tilde{\boldsymbol{z}}_{0},
 \tilde{\boldsymbol{w}};\boldsymbol{\lambda})\rangle 
\right]\ato 0, 
\end{equation}
with $\tilde{\boldsymbol{z}}_{0}\sim\mathcal{N}(\boldsymbol{0},
\boldsymbol{I}_{N})$. 

We repeat the use of Lemma~\ref{lemma_SLLN} for 
$\boldsymbol{f}(\tilde{\boldsymbol{z}}_{0},\tilde{\boldsymbol{w}})
=\tilde{\boldsymbol{\phi}}_{0}(\tilde{\boldsymbol{z}}_{0},
\tilde{\boldsymbol{w}}; \boldsymbol{\lambda})$ with 
$\boldsymbol{a}_{0}=\tilde{\boldsymbol{z}}_{0}$ and 
$\boldsymbol{\omega}=\tilde{\boldsymbol{w}}$. Using Lemma~\ref{lemma_SLLN} 
from Assumption~\ref{assumption_w} and applying Assumption~\ref{assumption_A}, 
we obtain 
\begin{equation}
\langle\tilde{\boldsymbol{\phi}}_{0}(\tilde{\boldsymbol{z}}_{0},
\tilde{\boldsymbol{w}};\boldsymbol{\lambda})\rangle 
- \mathbb{E}\left[
 \langle\tilde{\boldsymbol{\phi}}_{0}(\tilde{\boldsymbol{z}}_{0},
 \tilde{\boldsymbol{w}};\boldsymbol{\lambda})\rangle 
\right]\ato 0. 
\end{equation}
In evaluating the expectation over $\tilde{\boldsymbol{w}}$, 
the first $M$ elements $\boldsymbol{U}^{\mathrm{T}}\boldsymbol{w}$ 
in (\ref{w_tilde}) follow 
$\mathcal{N}(\boldsymbol{0},\sigma^{2}\boldsymbol{I}_{M})$. 
Combining these results, we arrive at (\ref{property_A2_former}) for $\tau=0$. 
\end{IEEEproof}

\begin{IEEEproof}[Proof of \ref{property_A3} for $\tau=0$] 
The LHS of (\ref{property_A3_eq}) is a separable and 
proper pseudo-Lipschitz function of order~$2$. We can use 
(\ref{property_A2_former}) for $\tau=0$ to find that the LHS of 
(\ref{property_A3_eq}) converges almost surely to its expectation in which 
$\boldsymbol{b}_{0}$ and $\langle\partial_{0}\tilde{\boldsymbol{\phi}}_{0}
\rangle$ are replaced by 
$\tilde{\boldsymbol{z}}_{0}\sim\mathcal{N}(\boldsymbol{0},
\boldsymbol{I}_{N})$ and the expected one, 
respectively. Thus, it is sufficient to evaluate the expectation.  

The function $\boldsymbol{f}(\tilde{\boldsymbol{z}}_{0}; 
\tilde{\boldsymbol{w}}, \boldsymbol{\lambda})
=\tilde{\boldsymbol{\phi}}_{0}(\tilde{\boldsymbol{z}}_{0},
\tilde{\boldsymbol{w}};\boldsymbol{\lambda}) 
- \mathbb{E}[ \langle\partial_{0}\tilde{\boldsymbol{\phi}}_{0}\rangle]
\tilde{\boldsymbol{z}}_{0}$ is a separable Lipschitz-continuous function of 
$\tilde{\boldsymbol{z}}_{0}$. 
Thus, we can use Lemma~\ref{lemma_Stein} to obtain 
\begin{IEEEeqnarray}{rl}
&\frac{1}{N}\mathbb{E}\left[
 \tilde{\boldsymbol{z}}_{0}^{\mathrm{T}}\left(
  \tilde{\boldsymbol{\phi}}_{0}
  - \mathbb{E}\left[
   \left\langle
    \partial_{0}\tilde{\boldsymbol{\phi}}_{0}
   \right\rangle
  \right]\tilde{\boldsymbol{z}}_{0}
 \right)
\right] \nonumber \\
=& \frac{1}{N}\sum_{n=1}^{N}\mathbb{E}\left[
 \tilde{z}_{0,n}^{2}
\right]\mathbb{E}\left[
 \partial_{0}\tilde{\phi}_{0,n}
\right]
- \mathbb{E}\left[
 \left\langle
  \partial_{0}\tilde{\boldsymbol{\phi}}_{0}
 \right\rangle
\right]=0. 
\end{IEEEeqnarray}
Thus, (\ref{property_A3_eq}) holds for $\tau=0$. 
\end{IEEEproof}

\begin{IEEEproof}[Proof of \ref{property_A4} for $\tau=0$] 
From the definition~(\ref{h}) of $\tilde{\boldsymbol{m}}_{0}$ and 
(\ref{property_A3_eq}), we find the orthogonality 
$N^{-1}\boldsymbol{b}_{0}^{\mathrm{T}}\tilde{\boldsymbol{m}}_{0}\ato0$. 
Using this orthogonality and (\ref{property_A2_latter}) for $\tau=0$ yields 
\begin{IEEEeqnarray}{rl}
&\frac{1}{N}\|\tilde{\boldsymbol{m}}_{0}\|^{2}
\aeq \frac{1}{N}\boldsymbol{m}_{0}^{\mathrm{T}}\tilde{\boldsymbol{m}}_{0} + o(1) 
\nonumber \\
=& \frac{1}{N}\boldsymbol{m}_{0}^{\mathrm{T}}\boldsymbol{m}_{0}
- \mathbb{E}\left[
 \langle\partial_{0}\boldsymbol{\phi}_{0}\rangle
\right]\frac{\boldsymbol{m}_{0}^{\mathrm{T}}\boldsymbol{b}_{0}}{N}
+ o(1). 
\end{IEEEeqnarray}

The first and second terms are separable and proper pseudo-Lipschitz functions 
of order~$2$. From (\ref{property_A2_former}) for $\tau=0$, they converge  
almost surely to their expected terms. Thus, 
$N^{-1}\|\tilde{\boldsymbol{m}}_{0}\|^{2}$ converges almost surely to a 
constant.  
\end{IEEEproof}

\begin{IEEEproof}[Proof of Property~\ref{property_A5} for $\tau=0$]
The latter property~(\ref{norm_M}) for $\tau=0$ follows from the nonlinearity 
of $\boldsymbol{\phi}_{0}$ in Assumption~\ref{assumption_thresholding}. 
Thus, we only prove the former property~(\ref{4th_moment_m}) for $\tau=0$. 

The proper Lipschitz-continuity in Assumption~\ref{assumption_thresholding} 
implies the upper bound $|\tilde{m}_{0,n}|\leq C_{n}(1+|b_{0,n}|
+|\tilde{w}_{0,n}|)$ for some $\lambda_{n}$-dependent constant $C_{n}$. 
From Assumptions~\ref{assumption_x} and \ref{assumption_w}, we find that 
$\boldsymbol{b}_{0}$ and $\tilde{\boldsymbol{w}}$ have 
bounded $(2k-2+\epsilon)$th moments for some $\epsilon>0$. Thus, 
we obtain the former property~(\ref{4th_moment_m}) for $\tau=0$.  
\end{IEEEproof}

\subsection{Module~B for $\tau=0$} 
\begin{IEEEproof}[Proof of Property~\ref{property_B1} for $\tau=0$]
Lemma~\ref{lemma_conditioning} for the constraint 
$\boldsymbol{V}\boldsymbol{b}_{0}=\tilde{\boldsymbol{q}}_{0}$ implies  
\begin{equation}
\boldsymbol{V}\sim
\frac{\tilde{\boldsymbol{q}}_{0}\boldsymbol{b}_{0}^{\mathrm{T}}}
{\|\tilde{\boldsymbol{q}}_{0}\|^{2}} 
+ \boldsymbol{\Phi}_{\tilde{\boldsymbol{q}}_{0}}^{\perp}\tilde{\boldsymbol{V}}
(\boldsymbol{\Phi}_{\boldsymbol{b}_{0}}^{\perp})^{\mathrm{T}}  
\end{equation} 
conditioned on $\mathfrak{F}$ and $\mathfrak{E}_{0,0}$, where 
$\tilde{\boldsymbol{V}}\in\mathcal{O}_{N-1}$ is Haar orthogonal and independent 
of $\boldsymbol{b}_{0}$ and $\tilde{\boldsymbol{q}}_{0}$. 
Using the definition~(\ref{h}) of $\boldsymbol{h}_{0}$ and 
the orthogonality $N^{-1}\boldsymbol{b}_{0}^{\mathrm{T}}
\tilde{\boldsymbol{m}}_{0}\ato0$ obtained from Property~\ref{property_A3} 
for $\tau=0$, we obtain (\ref{h0}).  

To complete the proof of Property~\ref{property_B1} for $\tau=0$, we prove 
(\ref{omega_variance}) for $\tau=0$. By definition, 
\begin{equation}
\frac{1}{N}\|\tilde{\boldsymbol{\omega}}_{0}\|^{2} 
= \frac{1}{N}
\tilde{\boldsymbol{m}}_{0}^{\mathrm{T}}\boldsymbol{P}_{\boldsymbol{b}_{0}}^{\perp}
\tilde{\boldsymbol{m}}_{0}
\aeq\frac{1}{N}\|\tilde{\boldsymbol{m}}_{0}\|^{2},  
\end{equation}
where the last equality follows from the orthogonality 
$N^{-1}\boldsymbol{b}_{0}^{\mathrm{T}}\tilde{\boldsymbol{m}}_{0}\ato0$. 
Thus, (\ref{omega_variance}) holds for $\tau=0$, because of 
the notational convention $\tilde{\boldsymbol{m}}_{0}^{\perp}
=\tilde{\boldsymbol{m}}_{0}$. 
\end{IEEEproof}

\begin{IEEEproof}[Proof of Property~\ref{property_B2} for $\tau=0$] 
Since the latter property~(\ref{property_B2_latter}) follows from the 
former property~(\ref{property_B2_former}), we only prove the former  
property for $\tau=0$. Using Property~\ref{property_B1} for $\tau=0$ and 
Lemma~\ref{lemma_SLLN} for $\boldsymbol{f}(\boldsymbol{x},\boldsymbol{h}_{0})
=\tilde{\boldsymbol{\psi}}_{0}(\boldsymbol{h}_{0},\boldsymbol{x})$ with 
$\boldsymbol{a}_{0}=\boldsymbol{x}$, $\boldsymbol{a}_{1}=\boldsymbol{0}$, 
$\boldsymbol{\epsilon}=o(1)\tilde{\boldsymbol{q}}_{0}$, 
$\boldsymbol{E}=\tilde{\boldsymbol{q}}_{0}$, and 
$\boldsymbol{\omega}=\boldsymbol{\omega}_{0}$, we obtain 
\begin{equation}
\langle\tilde{\boldsymbol{\psi}}_{0}(\boldsymbol{h}_{0}, 
\boldsymbol{x})\rangle
- \mathbb{E}_{\boldsymbol{z}_{0}}\left[
 \langle\tilde{\boldsymbol{\psi}}_{0}(\boldsymbol{z}_{0}, 
 \boldsymbol{x})\rangle 
\right]\ato 0, 
\end{equation} 
with $\boldsymbol{z}_{0}\sim\mathcal{N}(\boldsymbol{0},
\pi_{0,0}\boldsymbol{I}_{N})$. Applying Assumption~\ref{assumption_x} to 
the second term, we arrive at (\ref{property_B2_former}) for $\tau=0$. 
\end{IEEEproof}

\begin{IEEEproof}[Proof of Properties~\ref{property_B3} and 
\ref{property_B4} for $\tau=0$]
Repeat the proofs of Properties~\ref{property_A3} and 
\ref{property_A4} for $\tau=0$. 
\end{IEEEproof}

\begin{IEEEproof}[Proof of Property~\ref{property_B5} for $\tau=0$]
The former property~(\ref{4th_moment_q}) for $\tau=0$ is obtained by repeating 
the proof of (\ref{4th_moment_m}) for $\tau=0$. See \cite[p.~377]{Takeuchi201} 
for the proof of the latter property~(\ref{norm_Q}) for $\tau=0$. 
\end{IEEEproof}

\subsection{Proof by Induction} 
Suppose that Theorem~\ref{theorem_SE_tech} is correct for all 
$\tau<t$. In a proof by induction we need to prove all properties in 
modules~A and B for $\tau=t$. Since the properties for module~B can be proved 
by repeating the proofs for module~A, we only prove the properties for 
module~A. 

\begin{IEEEproof}[Proof of Property~\ref{property_A1} for $\tau=t$]
The matrix $(\boldsymbol{B}_{t}, \tilde{\boldsymbol{M}}_{t})$ has full rank 
from the induction hypotheses~(\ref{norm_M}) and (\ref{norm_Q}) for 
$\tau=t-1$, as well as the orthogonality $N^{-1}\boldsymbol{b}_{\tau}^{\mathrm{T}}
\tilde{\boldsymbol{m}}_{\tau'}\ato0$ for all $\tau, \tau'<t$.  
Using Lemma~\ref{lemma_conditioning} for the constraint 
$(\tilde{\boldsymbol{Q}}_{t}, \boldsymbol{H}_{t}) 
=\boldsymbol{V}(\boldsymbol{B}_{t}, \tilde{\boldsymbol{M}}_{t})$, we obtain  
\begin{IEEEeqnarray}{rl}
\boldsymbol{V} 
=& (\tilde{\boldsymbol{Q}}_{t},\boldsymbol{H}_{t})
\begin{bmatrix}
\tilde{\boldsymbol{Q}}_{t}^{\mathrm{T}}\tilde{\boldsymbol{Q}}_{t} 
 &\tilde{\boldsymbol{Q}}_{t}^{\mathrm{T}}\boldsymbol{H}_{t} \\
\boldsymbol{H}_{t}^{\mathrm{T}}\tilde{\boldsymbol{Q}}_{t} &
\boldsymbol{H}_{t}^{\mathrm{T}}\boldsymbol{H}_{t}
\end{bmatrix}^{-1}
\begin{bmatrix}
\boldsymbol{B}_{t}^{\mathrm{T}} \\ 
\tilde{\boldsymbol{M}}_{t}^{\mathrm{T}}
\end{bmatrix} \nonumber \\
&+ \boldsymbol{\Phi}_{(\tilde{\boldsymbol{Q}}_{t}, \boldsymbol{H}_{t})}^{\perp}
\tilde{\boldsymbol{V}}(\boldsymbol{\Phi}_{(\boldsymbol{B}_{t}, 
\tilde{\boldsymbol{M}}_{t})}^{\perp})^{\mathrm{T}} 
\end{IEEEeqnarray}
conditioned on $\mathfrak{F}$ and $\mathfrak{E}_{t,t}$. 
Applying the orthogonality $N^{-1}\boldsymbol{b}_{\tau}^{\mathrm{T}}
\tilde{\boldsymbol{m}}_{\tau'}\ato0$ and $N^{-1}\boldsymbol{h}_{\tau}^{\mathrm{T}}
\tilde{\boldsymbol{q}}_{\tau'}\ato0$ obtained from the induction 
hypotheses~\ref{property_A3} and \ref{property_B3} for $\tau<t$, 
as well as the definition~(\ref{b}) of $\boldsymbol{b}_{t}$, we have 
\begin{IEEEeqnarray}{rl}
\boldsymbol{b}_{t}
\sim& \boldsymbol{B}_{t}(\tilde{\boldsymbol{Q}}_{t}^{\mathrm{T}}
\tilde{\boldsymbol{Q}}_{t})^{-1}\tilde{\boldsymbol{Q}}_{t}^{\mathrm{T}}
\tilde{\boldsymbol{q}}_{t}  
+ \boldsymbol{B}_{t}\boldsymbol{o}(1) 
+ \tilde{\boldsymbol{M}}_{t}\boldsymbol{o}(1) 
\nonumber \\
&+ \boldsymbol{\Phi}_{(\boldsymbol{B}_{t}, 
\tilde{\boldsymbol{M}}_{t})}^{\perp}\tilde{\boldsymbol{V}}^{\mathrm{T}}
(\boldsymbol{\Phi}_{(\tilde{\boldsymbol{Q}}_{t}, \boldsymbol{H}_{t})}^{\perp})^{\mathrm{T}}
\tilde{\boldsymbol{q}}_{t} 
\end{IEEEeqnarray}
conditioned on $\mathfrak{F}$ and $\mathfrak{E}_{t,t}$, which is 
equivalent to (\ref{b_tau}) for $\tau=t$. 

To complete the proof of Property~\ref{property_A1} for $\tau=t$, 
we shall prove (\ref{omega_tilde_variance}). By definition, 
\begin{equation}
\frac{\|\tilde{\boldsymbol{\omega}}_{t}\|^{2}}{N} 
= \frac{\tilde{\boldsymbol{q}}_{t}^{\mathrm{T}}
\boldsymbol{P}_{(\tilde{\boldsymbol{Q}}_{t},\boldsymbol{H}_{t})}^{\perp}
\tilde{\boldsymbol{q}}_{t}}{N}
\aeq \frac{\tilde{\boldsymbol{q}}_{t}^{\mathrm{T}}
\boldsymbol{P}_{\tilde{\boldsymbol{Q}}_{t}}^{\perp}
\tilde{\boldsymbol{q}}_{t}}{N}+ o(1),  
\end{equation}
where the last equality follows from the orthogonality 
$N^{-1}\boldsymbol{h}_{\tau}^{\mathrm{T}}\tilde{\boldsymbol{q}}_{\tau'}\ato0$. 
Thus, (\ref{omega_tilde_variance}) holds for $\tau=t$.  
\end{IEEEproof}

\begin{IEEEproof}[Proof of Property~\ref{property_A2} for $\tau=t$]
Since the latter property~(\ref{property_A2_latter}) follows from the 
former property~(\ref{property_A2_former}), we only prove the former  
property for $\tau=t$. 

We use Property~\ref{property_A1} for $\tau=t$ and 
Lemma~\ref{lemma_SLLN} for the function $\boldsymbol{f}(\tilde{\boldsymbol{w}}, 
\boldsymbol{B}_{t}, \boldsymbol{b}_{t}) 
=\tilde{\boldsymbol{\phi}}_{t}(\boldsymbol{B}_{t+1},\tilde{\boldsymbol{w}};
\boldsymbol{\lambda})$ with $\boldsymbol{A}_{t+1}=(\tilde{\boldsymbol{w}}, 
\boldsymbol{B}_{t})$, $\boldsymbol{a}_{t+1}
=\boldsymbol{B}_{t}\boldsymbol{\beta}_{t}$, 
$\boldsymbol{\epsilon}=\tilde{\boldsymbol{M}}_{t}\boldsymbol{o}(1) 
+ \boldsymbol{B}_{t}\boldsymbol{o}(1)$, 
$\boldsymbol{E}=(\boldsymbol{B}_{t},\tilde{\boldsymbol{M}}_{t})$, and 
$\boldsymbol{\omega}=\tilde{\boldsymbol{\omega}}$. Then, 
\begin{equation}
\langle\tilde{\boldsymbol{\phi}}_{t}(\boldsymbol{B}_{t+1}, 
\tilde{\boldsymbol{w}};\boldsymbol{\lambda})\rangle
- \mathbb{E}_{\tilde{\boldsymbol{z}}_{t}}\left[
 \langle\tilde{\boldsymbol{\phi}}_{t}(\boldsymbol{B}_{t}, 
 \boldsymbol{B}_{t}\boldsymbol{\beta}_{t}+\tilde{\boldsymbol{z}}_{t}, 
 \tilde{\boldsymbol{w}},\boldsymbol{\lambda})\rangle 
\right]\ato 0,
\end{equation}
where $\tilde{\boldsymbol{z}}_{t}$ has independent zero-mean Gaussian 
elements with variance $\mu_{t}\aeq N^{-1}
\|\tilde{\boldsymbol{q}}_{t}^{\perp}\|^{2}$. Repeating this argument yields 
\begin{equation}
\langle\tilde{\boldsymbol{\phi}}_{t}(\boldsymbol{B}_{t+1}, 
\tilde{\boldsymbol{w}};\boldsymbol{\lambda})\rangle
- \mathbb{E}\left[
 \langle\tilde{\boldsymbol{\phi}}_{t}(\tilde{\boldsymbol{Z}}_{t+1}, 
 \tilde{\boldsymbol{w}},\boldsymbol{\lambda})\rangle 
\right]\ato 0,
\end{equation}
where $\tilde{\boldsymbol{Z}}_{t+1}$ is a zero-mean Gaussian random matrix 
having independent elements. In evaluating the expectation over 
$\tilde{\boldsymbol{w}}$,  
$\boldsymbol{U}^{\mathrm{T}}\boldsymbol{w}$ in (\ref{w_tilde}) follows 
the zero-mean Gaussian distribution with covariance 
$\sigma^{2}\boldsymbol{I}_{M}$. 

To complete the proof of (\ref{property_A2_former}) for $\tau=t$, we evaluate 
the covariance of $\boldsymbol{Z}_{t+1}$. By construction, we have 
$N^{-1}\mathbb{E}[\boldsymbol{z}_{\tau}^{\mathrm{T}}\boldsymbol{z}_{\tau'}]
= N^{-1}\boldsymbol{b}_{\tau}^{\mathrm{T}}\boldsymbol{b}_{\tau'}
\aeq \kappa_{\tau,\tau'}+o(1)$. Thus, the former 
property~(\ref{property_A2_former}) is correct for $\tau=t$. 
\end{IEEEproof}

\begin{IEEEproof}[Proof of Property~\ref{property_A3} for $\tau=t$]
The LHS of (\ref{property_A3_eq}) is a separable and proper 
pseudo-Lipschitz function of order~$2$. We can use 
(\ref{property_A2_former}) for $\tau=t$ to find that the LHS of 
(\ref{property_A3_eq}) converges almost surely to its expectation in which 
$\boldsymbol{B}_{t+1}$ and $\langle\partial_{t'}\tilde{\boldsymbol{\phi}}_{t}
\rangle$ are replaced by $\tilde{\boldsymbol{Z}}_{t+1}$ and the expected one, 
respectively. Thus, it is sufficient to evaluate the expectation.  

Since the function $\boldsymbol{f}(\tilde{\boldsymbol{Z}}_{t+1}; 
\tilde{\boldsymbol{w}}, \boldsymbol{\lambda})
=\tilde{\boldsymbol{\phi}}_{t}(\tilde{\boldsymbol{Z}}_{t+1},
\tilde{\boldsymbol{w}};\boldsymbol{\lambda}) - \sum_{t'=0}^{t}
\mathbb{E}[ \langle\partial_{t'}\tilde{\boldsymbol{\phi}}_{t}\rangle]
\tilde{\boldsymbol{z}}_{t'}$ is separable and Lipschitz-continuous with 
respect to $\tilde{\boldsymbol{Z}}_{t+1}$, 
we can use Lemma~\ref{lemma_Stein} to obtain 
\begin{IEEEeqnarray}{rl}
&\frac{1}{N}\mathbb{E}\left[
 \tilde{\boldsymbol{z}}_{\tau'}^{\mathrm{T}}\left(
  \tilde{\boldsymbol{\phi}}_{t}
  - \sum_{t'=0}^{t}\mathbb{E}\left[
   \left\langle
    \partial_{t'}\tilde{\boldsymbol{\phi}}_{t}
   \right\rangle
  \right]\tilde{\boldsymbol{z}}_{t'}
 \right)
\right] \nonumber \\
=& \frac{1}{N}\sum_{n=1}^{N}\sum_{t'=0}^{t}
\mathbb{E}[\tilde{z}_{\tau',n}\tilde{z}_{t,n}]
\mathbb{E}\left[
 \partial_{t'}\tilde{\phi}_{t,n}
\right] \nonumber \\
&- \sum_{t'=0}^{t}\mathbb{E}\left[
 \left\langle
  \partial_{t'}\tilde{\boldsymbol{\phi}}_{t}
 \right\rangle
\right]\frac{\mathbb{E}[\tilde{\boldsymbol{z}}_{\tau'}^{\mathrm{T}}
\tilde{\boldsymbol{z}}_{t'}]}{N}=0. 
\end{IEEEeqnarray}
Thus, (\ref{property_A3_eq}) holds for $\tau=t$. 
\end{IEEEproof}

\begin{IEEEproof}[Proof of Properties~\ref{property_A4} and 
\ref{property_A5} for $\tau=t$]
Repeat the proofs of Properties~\ref{property_A4} and 
\ref{property_A5} for $\tau=0$. In particular, see \cite[p.~378]{Takeuchi201} 
for the proof of (\ref{norm_M}) for $\tau=t$. 
\end{IEEEproof}

\section{Proof of Theorem~\ref{theorem_CAMP}}
\label{proof_theorem_CAMP} 
In evaluating the derivative in $g_{t',t}^{(j)}$, the parameter $\xi_{t}$ 
requires a careful treatment since it depends on $\boldsymbol{B}_{t+1}$ via 
$\boldsymbol{h}_{t}$. If the general error model contained the error model of 
the CAMP, we could use (\ref{deriv_psi_tilde}) in Theorem~\ref{theorem_SE} to 
prove that $\xi_{t}$ converges almost surely to a 
$\boldsymbol{B}_{t+1}$-independent constant $\bar{\xi}_{t}$ in the large system 
limit. To use Theorem~\ref{theorem_SE}, however, we have to prove the 
inclusion of the CAMP error model into the general error model. 
To circumvent this dilemma, we prove $g_{t-\tau,t}^{(j)}\aeq 
\xi_{t-\tau}^{(t-1)}g_{\tau}^{(j)}+o(1)$ for all $t$ and $\tau=0,\ldots,t$ 
by induction. 

We consider the case $\tau=0$, in which the expression~(\ref{m_CAMP}) 
requires no special treatments in computing the derivative. 
Differentiating (\ref{m_CAMP}) with respect to the $t$th variable yields 
\begin{equation} \label{g_0j}
g_{t,t}^{(j)} = \mu_{j+1} - \mu_{j},  
\end{equation}
where $\mu_{j}$ denotes the $j$th moment~(\ref{moment}) of the asymptotic 
eigenvalue distribution of $\boldsymbol{A}^{\mathrm{T}}\boldsymbol{A}$. 
Comparing (\ref{g_0}) and (\ref{g_0j}), we have 
$g_{t,t}^{(j)}=g_{0}^{(j)}$ for all $t$. 

Suppose that there is some $t>0$ such that 
$g_{t'-\tau,t'}^{(j)}\aeq \xi_{t'-\tau}^{(t'-1)}g_{\tau}^{(j)}+o(1)$ is correct 
for all $t'<t$ and $\tau=0,\ldots,t'$. Then, (\ref{deriv_psi_tilde}) in 
Theorem~\ref{theorem_SE} implies that $\xi_{t'}$ converges almost 
surely to a constant $\bar{\xi}_{t'}$ for any $t'<t$. 
We need to prove $g_{t-\tau,t}^{(j)}\aeq \xi_{t-\tau}^{(t-1)}g_{\tau}^{(j)}+o(1)$ 
for all $\tau=0,\ldots,t$. 

We first consider the case $\tau=1$ since we have already proved the case 
$\tau=0$. Differentiating (\ref{m_CAMP}) with respect to the $(t-1)$th 
variable yields  
\begin{IEEEeqnarray}{rl}
g_{t-1,t}^{(j)}
=& \bar{\xi}_{t-1}(g_{t-1,t-1}^{(j)} - g_{t-1,t-1}^{(j+1)})
- \bar{\xi}_{t-1}g_{1}(g_{t-1,t-1}^{(j)} + \mu_{j})
\nonumber \\
&+ \bar{\xi}_{t-1}\theta_{1}(g_{t-1,t-1}^{(j+1)} + \mu_{j+1}).  
\end{IEEEeqnarray}
Using $g_{t,t}^{(j)}=g_{0}^{(j)}$ and (\ref{g_1}), we arrive at 
$g_{t-1,t}^{(j)}\aeq\xi_{t-1}g_{1}^{(j)}+o(1)$.  

We next consider the case $\tau>1$. 
Differentiating (\ref{m_CAMP}) with respect to the $(t-\tau)$th variable, 
we have 
\begin{IEEEeqnarray}{l}
g_{t-\tau,t}^{(j)} 
= \bar{\xi}_{t-1}(g_{t-\tau,t-1}^{(j)} - g_{t-\tau,t-1}^{(j+1)})
\nonumber \\
+ \sum_{\tau'=t-\tau}^{t-1}\bar{\xi}_{\tau'}^{(t-1)}(\theta_{t-\tau'}g_{t-\tau,\tau'}^{(j+1)}
- g_{t-\tau'}g_{t-\tau,\tau'}^{(j)})
\nonumber \\
- \sum_{\tau'=t-\tau+1}^{t-1}\bar{\xi}_{\tau'-1}^{(t-1)}
(\theta_{t-\tau'}g_{t-\tau,\tau'-1}^{(j+1)} - g_{t-\tau'}g_{t-\tau,\tau'-1}^{(j)}) 
\nonumber \\ 
+ \bar{\xi}_{t-\tau}^{(t-1)}(\theta_{\tau}\mu_{j+1} - g_{\tau}\mu_{j}). 
\end{IEEEeqnarray}
Using (\ref{g_tau}) and the induction hypothesis 
$g_{t'-\tau,t'}^{(j)}\aeq\xi_{t'-\tau}^{(t'-1)}
g_{\tau}^{(j)}+o(1)$ for all $t'<t$ and $\tau=0,\ldots,t'$, 
we find $g_{t-\tau,t}^{(j)}\aeq \xi_{t-\tau}^{(t-1)}g_{\tau}^{(j)}+o(1)$.  

\section{Proof of Theorem~\ref{theorem_solution}}
\label{proof_theorem_solution}
Let $G(x,z)$ denote the generating function of $\{g_{\tau}^{(j)}\}$ given by  
\begin{equation}
G(x,z) = \sum_{j=0}^{\infty}G_{j}(z)x^{j}, 
\end{equation}
with 
\begin{equation}
G_{j}(z) = \sum_{\tau=0}^{\infty}g_{\tau}^{(j)}z^{-\tau}.  
\end{equation}
It is possible to prove that $G(x,z)$ is given by 
\begin{equation} \label{G}
G(x,z) = \frac{\{\Theta(z) - xG(z)\}\eta(-x) - \Theta(z)}
{x\tilde{G}(z) + 1 - \tilde{\Theta}(z)},  
\end{equation}
with $\tilde{G}(z)=(1-z^{-1})G(z)$ and $\tilde{\Theta}(z)=(1-z^{-1})\Theta(z)$. 
Let $-x^{*}$ denote a pole of the generating function, i.e.\ 
$x^{*} = [1 - \tilde{\Theta}(z)]/\tilde{G}(z)$. Since the generating function 
is analytical, the numerator of (\ref{G}) at $x=-x^{*}$ must be zero. 
\begin{equation}
\{\Theta(z) + x^{*}G(z)\}\eta(x^{*}) - \Theta(z) = 0, 
\end{equation}
which is equivalent to (\ref{closed_form}). 
 
To complete the proof of Theorem~\ref{theorem_solution}, 
we prove (\ref{G}). The proof is a simple exercise of the Z-transform. 
We first compute $G_{j}(z)$ given by 
\begin{equation}
G_{j}(z) 
= g_{0}^{(j)} + g_{1}^{(j)}z^{-1}
+ \sum_{\tau=2}^{\infty}g_{\tau}^{(j)}z^{-\tau}. 
\end{equation}
To evaluate the last term with (\ref{g_tau}), we note 
\begin{equation}
\sum_{\tau=2}^{\infty}g_{\tau-1}^{(j)}z^{-\tau}
= z^{-1}\sum_{\tau=1}^{\infty}g_{\tau}^{(j)}z^{-\tau}
= z^{-1}\left\{
 G_{j}(z) - g_{0}^{(j)} 
\right\}, 
\end{equation}
\begin{IEEEeqnarray}{rl}
&\sum_{\tau=2}^{\infty}\sum_{\tau'=0}^{\tau-1}g_{\tau-\tau'}g_{\tau'}^{(j)}z^{-\tau} 
\nonumber \\
=& g_{0}^{(j)}\sum_{\tau=2}^{\infty}g_{\tau}z^{-\tau} 
+ \sum_{\tau'=1}^{\infty}\sum_{\tau=\tau'+1}^{\infty}
g_{\tau-\tau'}g_{\tau'}^{(j)}z^{-\tau} \nonumber \\
=& \left[
 G(z) - 1
\right]G_{j}(z) - g_{1}g_{0}^{(j)}z^{-1}, 
\end{IEEEeqnarray}
\begin{IEEEeqnarray}{rl}
&\sum_{\tau=2}^{\infty}\sum_{\tau'=1}^{\tau-1}g_{\tau-\tau'}g_{\tau'-1}^{(j)}z^{-\tau} 
\nonumber \\
=& \sum_{\tau'=1}^{\infty}\sum_{\tau=\tau'+1}^{\infty}
g_{\tau-\tau'}g_{\tau'-1}^{(j)}z^{-\tau} \nonumber \\
=& \left[
 G(z) - 1
\right]z^{-1}G_{j}(z). 
\end{IEEEeqnarray}
Combining (\ref{g_0}), (\ref{g_1}), (\ref{g_tau}), and these results, 
we arrive at  
\begin{IEEEeqnarray}{rl}
G_{j}(z) 
=& [1-\tilde{G}(z)]G_{j}(z) - [1-\tilde{\Theta}(z)]G_{j+1}(z) \nonumber \\
&- \mu_{j}G(z) + \mu_{j+1}\Theta(z). \label{G_j}
\end{IEEEeqnarray}

We next evaluate $G(x,z)$. Substituting (\ref{G_j}) into the definition 
of $G(x,z)$ yields 
\begin{IEEEeqnarray}{rl}
G(x,z)   
=& [1-\tilde{G}(z)]G(x,z) - [1 - \tilde{\Theta}(z)]\frac{G(x,z)}{x} \nonumber \\
&- \eta(-x)G(z) + \frac{\eta(-x)-1}{x}\Theta(z),   
\end{IEEEeqnarray}
where we have used the definition~(\ref{eta_transform}) and 
the identity $G_{0}(z)=0$ obtained from 
Theorem~\ref{theorem_CAMP}. Solving this equation with respect to 
$G(x,z)$, we obtain (\ref{G}).

\section{Proof of Theorem~\ref{theorem_CAMP_SE}}
\label{proof_theorem_CAMP_SE} 
\subsection{SE Equations}
The proof of Theorem~\ref{theorem_CAMP_SE} consists of four steps: 
A first step is a derivation of the SE equations, which is a dynamical 
system that describes the dynamics of five variables with three indices.  
A second step is evaluation of the generating functions for the five 
variables. The step is a simple exercise of the Z-transform. 
In a third step, we evaluate the obtained generating functions at poles 
to prove the SE equation~(\ref{SE_equation_generating_tmp}) in terms of 
the generating functions. The last step is a derivation of the SE 
equation~(\ref{SE_equation}) in time domain via the inverse Z-transform. 

Let $a_{t',t}^{(j)}=N^{-1}\boldsymbol{m}_{t'}^{\mathrm{T}}
\boldsymbol{\Lambda}^{j}\boldsymbol{m}_{t}$, 
$b_{t',t}^{(j)}=N^{-1}\boldsymbol{b}_{t'}^{\mathrm{T}}
\boldsymbol{\Lambda}^{j}\boldsymbol{m}_{t}$, 
$c_{t',t}=N^{-1}\tilde{\boldsymbol{q}}_{t'}^{\mathrm{T}}
\tilde{\boldsymbol{q}}_{t}$, $d_{t',t}=N^{-1}\boldsymbol{q}_{t'}^{\mathrm{T}}
\boldsymbol{q}_{t}$, and $e_{t}^{(j)}=N^{-1}\boldsymbol{w}^{\mathrm{T}}\boldsymbol{U}
\boldsymbol{\Sigma}\boldsymbol{\Lambda}^{j}\boldsymbol{m}_{t}$. 
Theorem~\ref{theorem_CAMP} implies the asymptotic orthogonality between 
$\boldsymbol{b}_{t'}$ and $\boldsymbol{m}_{t}$.  
We use the definition~(\ref{m_CAMP}) to obtain  
\begin{IEEEeqnarray}{l}
a_{t',t}^{(j)} 
\aeq b_{t,t'}^{(j)} - b_{t,t'}^{(j+1)} 
+ \bar{\xi}_{t-1}(a_{t',t-1}^{(j)} - a_{t',t-1}^{(j+1)})
+ e_{t'}^{(j)} 
\nonumber \\
+ \sum_{\tau=0}^{t-1}\bar{\xi}_{\tau}^{(t-1)}\theta_{t-\tau}
(a_{t',\tau}^{(j+1)} - b_{\tau,t'}^{(j+1)} - \bar{\xi}_{\tau-1}a_{t',\tau-1}^{(j+1)})
\nonumber \\ 
- \sum_{\tau=0}^{t-1}\bar{\xi}_{\tau}^{(t-1)}g_{t-\tau}
(a_{t',\tau}^{(j)} - b_{\tau,t'}^{(j)} - \bar{\xi}_{\tau-1}a_{t',\tau-1}^{(j)})
+ o(1), 
\label{a}
\end{IEEEeqnarray}
where we have replaced $\xi_{t}$ with the asymptotic value $\bar{\xi}_{t}$. 
Applying (\ref{b_orthogonality}) in Theorem~\ref{theorem_SE} and 
(\ref{b}) yields 
\begin{IEEEeqnarray}{rl}
b_{t',t}^{(j)} 
\aeq& (\mu_{j} - \mu_{j+1})c_{t',t}  
+ \bar{\xi}_{t-1}(b_{t',t-1}^{(j)} - b_{t',t-1}^{(j+1)}) + o(1)
\nonumber \\
+& \sum_{\tau=0}^{t-1}\bar{\xi}_{\tau}^{(t-1)}\theta_{t-\tau}
(b_{t',\tau}^{(j+1)} - \mu_{j+1}c_{t',\tau} - \bar{\xi}_{\tau-1}b_{t',\tau-1}^{(j+1)})
\nonumber \\
-& \sum_{\tau=0}^{t-1}\bar{\xi}_{\tau}^{(t-1)}g_{t-\tau}
(b_{t',\tau}^{(j)} - \mu_{j}c_{t',\tau} - \bar{\xi}_{\tau-1}b_{t',\tau-1}^{(j)}).  
\end{IEEEeqnarray}
Using (\ref{h_orthogonality}) in Theorem~\ref{theorem_SE}, 
(\ref{q_tilde_CAMP}), and (\ref{h}), we have  
\begin{equation}
c_{t'+1,t+1} 
\aeq \frac{\boldsymbol{q}_{t'+1}^{\mathrm{T}}\tilde{\boldsymbol{q}}_{t+1}}{N}
+ o(1) 
\aeq d_{t'+1,t+1} - \bar{\xi}_{t}\bar{\xi}_{t'}a_{t',t}^{(0)}
+ o(1).  
\end{equation}
Applying (\ref{psi_tilde}) in Theorem~\ref{theorem_SE} yields 
\begin{equation} \label{d}
d_{t'+1,t+1} 
\ato \mathbb{E}\left[
 \{f_{t'}(x_{1}+z_{t'}) - x_{1}\}
 \{f_{t}(x_{1}+z_{t}) - x_{1}\}
\right], 
\end{equation}
where $\{z_{t}\}$ are zero-mean Gaussian random variables with 
covariance $\mathbb{E}[z_{t'}z_{t}]=a_{t',t}^{(0)}$. 
Finally, we use (\ref{b_orthogonality}) in Theorem~\ref{theorem_SE} to 
obtain  
\begin{IEEEeqnarray}{rl}
e_{t}^{(j)} 
\aeq& \bar{\xi}_{t-1}(e_{t-1}^{(j)} - e_{t-1}^{(j+1)})
+ \sigma^{2}\mu_{j+1} + o(1)
\nonumber \\
&+ \sum_{\tau=0}^{t-1}\bar{\xi}_{\tau}^{(t-1)}\theta_{t-\tau}
(e_{\tau}^{(j+1)} - \bar{\xi}_{\tau-1}e_{\tau-1}^{(j+1)})
\nonumber \\
&- \sum_{\tau=0}^{t-1}\bar{\xi}_{\tau}^{(t-1)}g_{t-\tau}
(e_{\tau}^{(j)} - \bar{\xi}_{\tau-1}e_{\tau-1}^{(j)}). \label{e} 
\end{IEEEeqnarray}

To transform the summations in these equations to convolution, 
we use the change of variables 
$a_{t',t}^{(j)}=\bar{\xi}_{0}^{(t'-1)}\bar{\xi}_{0}^{(t-1)}\tilde{a}_{t',t}^{(j)}$. 
Similarly, we define $\tilde{b}_{t',t}^{(j)}$, $\tilde{c}_{t',t}$, and 
$\tilde{d}_{t',t}$ while we use 
$e_{t'}^{(j)}=\bar{\xi}_{0}^{(t'-1)}\bar{\xi}_{0}^{(t-1)}\tilde{e}_{t',t}^{(j)}$. 
Then, the SE equations~(\ref{a})--(\ref{e}) reduce to 
\begin{IEEEeqnarray}{rl}
\tilde{a}_{t',t}^{(j)} 
\aeq& \tilde{b}_{t,t'}^{(j)} - \tilde{b}_{t,t'}^{(j+1)} 
+ \tilde{a}_{t',t-1}^{(j)} - \tilde{a}_{t',t-1}^{(j+1)}
+ \tilde{e}_{t',t}^{(j)} 
\nonumber \\
&+ \sum_{\tau=0}^{t-1}\theta_{t-\tau}
(\tilde{a}_{t',\tau}^{(j+1)} - \tilde{b}_{\tau,t'}^{(j+1)} 
- \tilde{a}_{t',\tau-1}^{(j+1)}) \nonumber \\ 
&- \sum_{\tau=0}^{t-1}g_{t-\tau}
(\tilde{a}_{t',\tau}^{(j)} - \tilde{b}_{\tau,t'}^{(j)} 
- \tilde{a}_{t',\tau-1}^{(j)}) + o(1),  \label{a_tilde}
\end{IEEEeqnarray}
\begin{IEEEeqnarray}{rl}
\tilde{b}_{t',t}^{(j)} 
\aeq& (\mu_{j} - \mu_{j+1})\tilde{c}_{t',t}  
+ \tilde{b}_{t',t-1}^{(j)} - \tilde{b}_{t',t-1}^{(j+1)} + o(1) 
\nonumber \\
&+ \sum_{\tau=0}^{t-1}\theta_{t-\tau}
(\tilde{b}_{t',\tau}^{(j+1)} - \mu_{j+1}\tilde{c}_{t',\tau} 
- \tilde{b}_{t',\tau-1}^{(j+1)})
\nonumber \\
&- \sum_{\tau=0}^{t-1}g_{t-\tau}
(\tilde{b}_{t',\tau}^{(j)} - \mu_{j}\tilde{c}_{t',\tau} 
- \tilde{b}_{t',\tau-1}^{(j)}), 
\end{IEEEeqnarray}
\begin{equation}
\tilde{c}_{t'+1,t+1} 
\aeq \tilde{d}_{t'+1,t+1} - \tilde{a}_{t',t}^{(0)} + o(1), 
\end{equation}
\begin{IEEEeqnarray}{rl}
\tilde{e}_{t',t}^{(j)} 
\aeq& \tilde{e}_{t'-1,t}^{(j)} - \tilde{e}_{t'-1,t}^{(j+1)}
+ \mu_{j+1}\sigma_{t',t}^{2} + o(1)
\nonumber \\
&+ \sum_{\tau=0}^{t'-1}\theta_{t'-\tau}
(\tilde{e}_{\tau,t}^{(j+1)} - \tilde{e}_{\tau-1,t}^{(j+1)})
\nonumber \\
&- \sum_{\tau=0}^{t'-1}g_{t'-\tau}
(\tilde{e}_{\tau,t}^{(j)} - \tilde{e}_{\tau-1,t}^{(j)}), 
\label{e_tilde}
\end{IEEEeqnarray}
with 
\begin{equation}
\sigma_{t',t}^{2} = \frac{\sigma^{2}}{\bar{\xi}_{0}^{(t'-1)}\bar{\xi}_{0}^{(t-1)}}.  
\end{equation}

In principle, it is possible to solve the coupled dynamical 
system~(\ref{d}), (\ref{a_tilde})--(\ref{e_tilde}) numerically. 
However, numerical evaluation is a challenging task due to instability 
against numerical errors.  

\subsection{Generating Functions}
We solve the coupled dynamical system via the Z-transform. 
Define the generating function of $\tilde{a}_{t',t}^{(j)}$ as  
\begin{equation} \label{generating_A}
A(x,y,z) = \sum_{j=0}^{\infty}x^{j}A_{j}(y,z),  
\end{equation}
with 
\begin{equation}
A_{j}(y,z) = \sum_{t',t=0}^{\infty}\tilde{a}_{t',t}^{(j)}y^{-t'}z^{-t}.  
\end{equation}
Similarly, we write the generating functions of $\{\tilde{b}_{t',t}^{(j)}\}$, 
$\{\tilde{c}_{t',t}\}$, $\{\tilde{d}_{t',t}\}$, $\{\tilde{e}_{t',t}^{(j)}\}$, and 
$\{\sigma_{t',t}^{2}\}$ as $B(x,y,z)$, $C(y,z)$, $D(y,z)$, $E(x,y,z)$, and 
$\Sigma(y,z)$, respectively. 

To evaluate the generating function $A_{j}(y,z)$, we utilize  
\begin{IEEEeqnarray}{rl}
&\sum_{t'=0}^{\infty}y^{-t'}\sum_{t=1}^{\infty}z^{-t}\sum_{\tau=0}^{t-1}g_{t-\tau}
\tilde{a}_{t',\tau-k}^{(j)} \nonumber \\
=& \sum_{t'=0}^{\infty}y^{-t'}\sum_{\tau=0}^{\infty}\sum_{t=\tau+1}^{\infty}
z^{-t}g_{t-\tau}\tilde{a}_{t',\tau-k}^{(j)} \nonumber \\ 
=&z^{-k}\left[
 G(z) - 1
\right]A_{j}(y,z)   
\end{IEEEeqnarray}
for any integer $k$, where we have used the definition~(\ref{generating_g}) 
of $G(z)$. From (\ref{a_tilde}), we have 
\begin{IEEEeqnarray}{l}
A_{j}(y,z) 
\aeq B_{j}(z,y) - B_{j+1}(z,y) + \frac{A_{j}(y,z)}{z} - \frac{A_{j+1}(y,z)}{z}  
\nonumber \\
+ \left[
 \Theta(z) - 1
\right]\left\{
 A_{j+1}(y,z) - B_{j+1}(z,y) - \frac{A_{j+1}(y,z)}{z} 
\right\} \nonumber \\
- \left[
 G(z) - 1
\right]\left\{
 A_{j}(y,z) - B_{j}(z,y) - \frac{A_{j}(y,z)}{z}
\right\}
\nonumber \\ 
+ E_{j}(y,z).
\label{A_j}
\end{IEEEeqnarray}
Similarly, we can derive 
\begin{IEEEeqnarray}{l}
B_{j}(y,z) 
\aeq (\mu_{j} - \mu_{j+1})C(y,z)  
+ \frac{B_{j}(y,z)}{z} - \frac{B_{j+1}(y,z)}{z} 
\nonumber \\
+ \left[
 \Theta(z) - 1
\right]\left\{
 B_{j+1}(y,z) - \mu_{j+1}C(y,z) - \frac{B_{j+1}(y,z)}{z} 
\right\}
\nonumber \\
- \left[
 G(z) - 1
\right]\left\{
 B_{j}(y,z) - \mu_{j}C(y,z) - \frac{B_{j}(y,z)}{z} 
\right\} + o(1), \nonumber \\ 
\label{B_j}
\end{IEEEeqnarray}
\begin{equation} \label{C}
C(y,z) \aeq D(y,z) - (yz)^{-1}A_{0}(y,z) + o(1), 
\end{equation}
\begin{IEEEeqnarray}{rl}
E_{j}(y,z)
\aeq& \frac{E_{j}(y,z)}{y} - \frac{E_{j+1}(y,z)}{y} 
+ \mu_{j+1}\Sigma(y,z) + o(1)
\nonumber \\
&+ (1-y^{-1})[\Theta(y) - 1]E_{j+1}(y,z) \nonumber \\ 
&- (1-y^{-1})[G(y) - 1]E_{j}(y,z). 
\end{IEEEeqnarray}

We next substitute (\ref{A_j}) into (\ref{generating_A}) to obtain 
\begin{IEEEeqnarray}{l}
\left\{
 x\tilde{G}(z) + 1 - \tilde{\Theta}(z) 
\right\}A(x,y,z) 
\aeq  [1 - \tilde{\Theta}(z)]A_{0}(y,z) \nonumber \\
+ \left\{
 x\tilde{G}(z) - \tilde{\Theta}(z) 
\right\}\frac{B(x,z,y)}{1-z^{-1}}   
+ xE(x,y,z) + o(1), \label{A}
\end{IEEEeqnarray}
with $\tilde{G}(z)=(1-z^{-1})G(z)$ and $\tilde{\Theta}(z)=(1-z^{-1})\Theta(z)$, 
where we have used the identity $B_{0}(y,z)\aeq o(1)$ obtained from the 
asymptotic orthogonality between $\boldsymbol{b}_{t'}$ and 
$\boldsymbol{m}_{t}$. Similarly, we use (\ref{eta_transform}) and 
(\ref{B_j}) to obtain 
\begin{equation} \label{B}
B(x,y,z) 
\aeq \frac{[x\tilde{G}(z) - \tilde{\Theta}(z)]\eta(-x) + \tilde{\Theta}(z)}
{x\tilde{G}(z) + 1 - \tilde{\Theta}(z)}\frac{C(y,z)}{1-z^{-1}}
+ o(1). 
\end{equation}
Furthermore, we have 
\begin{IEEEeqnarray}{rl}
E(x,y,z)
\aeq& \frac{1-\tilde{\Theta}(y)}
{x\tilde{G}(y) + 1 - \tilde{\Theta}(y)}E_{0}(y,z) 
\nonumber \\ 
&+ \frac{\eta(-x)-1}{x\tilde{G}(y) + 1 - \tilde{\Theta}(y)}\Sigma(y,z) 
+ o(1). \label{E}
\end{IEEEeqnarray}

\subsection{Evaluation at Poles}
The equations (\ref{C}), (\ref{A}), (\ref{B}), and (\ref{E}) 
provide all information about the generating functions. However, 
we are interested only in those at $x=0$. To extract this information,   
we focus on the poles of $A(x,y,z)$ and $E(x,y,z)$. Let $-x^{*}$ denote 
the pole of $A(x,y,z)$ given by 
\begin{equation} \label{pole}
x^{*} = \frac{1-\tilde{\Theta}(z)}{\tilde{G}(z)}. 
\end{equation}
Since $A(x,y,z)$ is analytical, the RHS of (\ref{A}) has to be 
zero at $x=-x^{*}$. 
\begin{equation} \label{A*}
\frac{B(-x^{*},z,y)}{1-z^{-1}} 
\aeq [1-\tilde{\Theta}(z)]A_{0}(y,z) - x^{*}E(-x^{*},y,z)
+ o(1). 
\end{equation}

Similarly, we use (\ref{E}) and Theorem~\ref{theorem_solution} 
to obtain  
\begin{equation}
E_{0}(y,z) \aeq \Sigma(y,z) + o(1). 
\end{equation}
Thus, (\ref{E}) reduces to  
\begin{IEEEeqnarray}{rl}
&E(-x^{*},y,z) \nonumber \\
\aeq& \frac{[\tilde{\Theta}(z)  - \tilde{\Theta}(y)]\tilde{G}(z)\Sigma(y,z)}
{\tilde{G}(y)\tilde{\Theta}(z) - \tilde{\Theta}(y)\tilde{G}(z)
+ \tilde{G}(z) - \tilde{G}(y) } + o(1).  
\label{E*}
\end{IEEEeqnarray} 

Evaluating $B(x,z,y)$ given via (\ref{B}) at $x=-x^{*}$ yields 
\begin{IEEEeqnarray}{rl}
\frac{B(-x^{*},z,y)}{1-z^{-1}} 
\aeq& \frac{\Theta(y)G(z) - G(y)\Theta(z)}
{\tilde{G}(y)\tilde{\Theta}(z) - \tilde{\Theta}(y)\tilde{G}(z) 
+ \tilde{G}(z) - \tilde{G}(y)} 
\nonumber \\
&\cdot[1-\tilde{\Theta}(z)]C(y,z) + o(1),  \label{B*}
\end{IEEEeqnarray}
where we have used $\tilde{\Theta}(z)=(1-z^{-1})\Theta(z)$, 
$\tilde{G}(z)=(1-z^{-1})G(z)$, and the symmetry $C(z,y)=C(y,z)$.  
Substituting (\ref{C}), (\ref{E*}), and (\ref{B*}) into (\ref{A*}),  
we obtain 
\begin{IEEEeqnarray}{l} 
F_{G,\Theta}(y,z)A_{0}(y,z) 
\aeq \frac{\Theta(y)G(z) - G(y)\Theta(z)}{y^{-1} - z^{-1}}D(y,z)  \nonumber \\
+ \frac{(1-z^{-1})\Theta(z) - (1-y^{-1})\Theta(y)}{y^{-1} - z^{-1}}\Sigma(y,z) 
+ o(1),  \label{SE_equation_generating_tmp0}
\end{IEEEeqnarray}
with 
\begin{IEEEeqnarray}{rl}
F_{G,\Theta}(y,z) 
=& \frac{(y^{-1}+z^{-1}-1)[\Theta(y)G(z) - G(y)\Theta(z)]}{y^{-1} - z^{-1}} 
\nonumber \\
&+ \frac{(1-z^{-1})G(z) - (1-y^{-1})G(y)}{y^{-1} - z^{-1}}. 
\end{IEEEeqnarray}

We transform the SE equation~(\ref{SE_equation_generating_tmp0}) into another 
generating-function representation that is suited for deriving time-domain 
representation. 
Let $S$ denote the generating function of some sequence $\{s_{t}\}$. 
We use the notations $S_{1}(z)=z^{-1}S(z)$, $\Delta_{S}$, and $\Delta_{S_{1}}$, 
given by 
\begin{equation} \label{Delta_S}
\Delta_{S} 
= \frac{S(y) - S(z)}{y^{-1}-z^{-1}}, 
\end{equation}
which is a function of $y$ and $z$. The inverse Z-transform of these 
generating functions can be evaluated straightforwardly, as shown shortly. 
We use these notations to re-write the SE 
equation~(\ref{SE_equation_generating_tmp0}) as  
\begin{IEEEeqnarray}{rl}
F_{G,\Theta}(y,z)A_{0}(y,z) 
\aeq& \left\{
 G(z)\Delta_{\Theta} - \Theta(z)\Delta_{G} 
\right\}D(y,z) 
\nonumber \\
+& \left(
 \Delta_{\Theta_{1}}  - \Delta_{\Theta}   
\right)\Sigma(y,z) + o(1), 
\label{SE_equation_generating}
\end{IEEEeqnarray}
with  
\begin{IEEEeqnarray}{rl}
F_{G,\Theta}(y,z)
=& (y^{-1}+z^{-1}-1)[G(z)\Delta_{\Theta} - \Theta(z)\Delta_{G}]
\nonumber \\ 
&+ \Delta_{G_{1}} - \Delta_{G},  
\end{IEEEeqnarray}
where $G_{1}(z)=z^{-1}G(z)$ and $\Theta_{1}(z)=z^{-1}\Theta(z)$ 
are defined in the same manner as in $S_{1}(z)$. The SE 
equation~(\ref{SE_equation_generating}) is equivalent to the former 
statement in Theorem~\ref{theorem_CAMP_SE}. 

\subsection{Time-Domain Representation} 
We transform the SE equation~(\ref{SE_equation_generating}) into 
a time-domain representation that is suitable for numerical evaluation. 
Suppose that $G(z)$ is represented as $G(z)=P(z)/Q(z)$. 
Let $R(z)$ denote the generating function of $\{r_{t}\}$, i.e.\ 
$R(z)=Q(z)\Theta(z)$. 
We multiply both sides of the SE equation~(\ref{SE_equation_generating}) 
by $Q(y)Q(z)$ to obtain  
\begin{IEEEeqnarray}{rl}
&F_{P,Q,\Theta}(y,z)A_{0}(y,z) \nonumber \\ 
\aeq& \left\{
 P(z)\Delta_{R} - R(z)\Delta_{P} 
\right\}D(y,z) 
\nonumber \\
+& Q(y)Q(z)\left(
 \Delta_{\Theta_{1}}  - \Delta_{\Theta}   
\right)\Sigma(y,z) + o(1), 
\label{SE_equation_generating_PQ}
\end{IEEEeqnarray}
with  
\begin{IEEEeqnarray}{rl}
F_{P,Q,\Theta}(y,z)
=& [\Delta_{P_{1}} - \Delta_{P}]Q(z) + (1 - z^{-1})P(z)\Delta_{Q} 
\nonumber \\ 
&+ (z^{-1}-1)[P(z)\Delta_{R} - R(z)\Delta_{P}] \nonumber \\
&+ y^{-1}[P(z)\Delta_{R} - R(z)\Delta_{P}].  
\end{IEEEeqnarray}

It is possible to evaluate the inverse Z-transform of $S_{1}(z)$, 
$\Delta_{S}$, $\Delta_{S_{1}}$, and $z^{-1}\Delta_{S}$ 
for any generating function $S(z)$. 
By definition, we have 
\begin{equation}
S_{1}(z) = \sum_{t=0}^{\infty}s_{t}z^{-(t+1)} 
= \sum_{t=0}^{\infty}s_{t-1}z^{-t}, 
\end{equation}
where the convention $s_{-1}=0$ has been used. Thus, $S_{1}(z)$ is the 
generating function of the sequence $\{s_{t-1}\}$. 

For $\Delta_{S}$, we obtain  
\begin{IEEEeqnarray}{rl}
&\Delta_{S} 
= \sum_{\tau=1}^{\infty}s_{\tau}\frac{y^{-\tau} - z^{-\tau}}{y^{-1} - z^{-1}}
= \sum_{\tau=1}^{\infty}\sum_{\tau'=0}^{\tau-1}s_{\tau}y^{-\tau'}z^{-(\tau-\tau'-1)} 
\nonumber \\
=& \sum_{\tau'=0}^{\infty}\sum_{\tau=\tau'+1}^{\infty}s_{\tau}y^{-\tau'}z^{-(\tau-\tau'-1)}
= \sum_{\tau'=0}^{\infty}\sum_{\tau=0}^{\infty}s_{\tau'+\tau+1}y^{-\tau'}z^{-\tau}, 
\nonumber \\
\end{IEEEeqnarray} 
which implies that $\Delta_{S}$ is the generating function of 
the two-dimensional array $s_{t',t}=s_{t'+t+1}$. 

We combine these results to evaluate the inverse Z-transform of the remaining  
generating functions. For $S_{1}(z)$, $\Delta_{S_{1}}$ is 
the generating function of $\{s_{t'+t}\}$. 
Since $y^{-1}$ is the generating function of 
$\delta_{t',1}\delta_{t,0}$ and since $\Delta_{S}$ is the generating function 
of $s_{t',t}=s_{t'+t+1}$, $y^{-1}\Delta_{S}$ is the generating function of 
the two-dimensional convolution:
\begin{equation}
(\delta_{t',1}\delta_{t,0})*s_{t',t} 
= s_{t'-1,t}
= s_{t'+t} - \delta_{t',0}s_{t},
\end{equation}  
where the last expression is due to the convention $s_{-1,t}=0$. 
See Table~\ref{table_Z_transform} for a summary of these results. 

\begin{table}[t]
\begin{center}
\caption{
Z-transform of 2-dimensional arrays. 
}
\label{table_Z_transform}
\begin{tabular}{|c|c|}
\hline
 Array~$s_{t',t}$ & Z-transform \\
\hline
$\delta_{t',0}s_{t-1}$ & $S_{1}(z)$ \\
\hline
$s_{t'+t+1}$ & $\Delta_{S}$ \\
\hline
$s_{t'+t}$ & $\Delta_{S_{1}}$ \\
\hline
 $s_{t'+t} - \delta_{t',0}s_{t}$ & $y^{-1}\Delta_{S}$ \\
\hline
\end{tabular}
\end{center}
\end{table}

We evaluate the inverse Z-transform of (\ref{SE_equation_generating_PQ}). 
It is a simple exercise to confirm that 
(\ref{SE_equation_generating_PQ}) is equal to the Z-transform of 
the following difference equation:  
\begin{IEEEeqnarray}{rl}
\mathfrak{D}_{t',t}*\tilde{a}_{t',t}^{(0)} 
&\aeq (p_{t}*r_{t'+t+1} - r_{t}*p_{t'+t+1})*\tilde{d}_{t',t}
 \nonumber \\
+& (q_{t'}q_{t})*(\theta_{t'+t} - \theta_{t'+t+1})*\sigma_{t',t}^{2}
+ o(1),  
\label{SE_equation_tmp}   
\end{IEEEeqnarray}
with 
\begin{IEEEeqnarray}{rl}
\mathfrak{D}_{t',t}
=& (p_{t'+t} - p_{t'+t+1})*q_{t} + (p_{t} - p_{t-1})*q_{t'+t+1} 
\nonumber \\
+& (p_{t-1}-p_{t})*r_{t'+t+1} + (r_{t} - r_{t-1})*p_{t'+t+1}
 \nonumber \\
+& p_{t}*(r_{t'+t} - \delta_{t',0}r_{t}) 
- r_{t}*(p_{t'+t} - \delta_{t',0}p_{t}), 
\end{IEEEeqnarray}
where all variables with negative indices are set to zero. 
Multiplying (\ref{SE_equation_tmp}) by 
$\bar{\xi}_{0}^{(t'-1)}\bar{\xi}_{0}^{(t-1)}$ and 
using the definitions $\tilde{a}_{\tau',\tau}^{(0)}=a_{\tau',\tau}^{(0)}/
(\bar{\xi}_{0}^{(\tau'-1)}\bar{\xi}_{0}^{(\tau-1)})$, 
$\tilde{d}_{\tau',\tau}^{(0)}=d_{\tau',\tau}/
(\bar{\xi}_{0}^{(\tau'-1)}\bar{\xi}_{0}^{(\tau-1)})$, and 
$\sigma_{\tau',\tau}^{2}=\sigma^{2}/(\bar{\xi}_{0}^{(\tau'-1)}\bar{\xi}_{0}^{(\tau-1)})$, 
we arrive at the SE equation~(\ref{SE_equation}) in time domain, with 
the superscript in $a_{\tau',\tau}^{(0)}$ omitted. 

Finally, we use the notational convention $f_{-1}(\cdot)=0$ to obtain initial 
and boundary conditions. From the definition~(\ref{correlation}) of 
$d_{t'+1,t+1}$, we have the initial condition 
$d_{0,0}=\mathbb{E}[x_{1}^{2}]=1$. Similarly, we use 
(\ref{correlation}) to obtain the boundary condition  
$d_{0,\tau+1}= -\mathbb{E}[x_{1}\{f_{\tau}(x_{1}+z_{\tau}) - x_{1}\}]$. 
The boundary condition $d_{\tau+1,0}=d_{0,\tau+1}$ follows from the symmetry.

\section{Proof of Theorem~\ref{theorem_fixed_point}} 
\label{proof_theorem_fixed_point} 
Without the loss of generality, we assume $p_{t}=g_{t}$ and $q_{t}=\delta_{t,0}$. 
Then, the SE equation~(\ref{SE_equation}) in time domain reduces to 
\begin{IEEEeqnarray}{rl}
\sum_{\tau'=0}^{t'}\sum_{\tau=0}^{t}\bar{\xi}_{t'-\tau'}^{(t'-1)}\bar{\xi}_{t-\tau}^{(t-1)}
\Big\{ \mathfrak{D}_{\tau',\tau}a_{t'-\tau',t-\tau}&  
\nonumber \\
- (g_{\tau}*\theta_{\tau'+\tau+1} - \theta_{\tau}*g_{\tau'+\tau+1})d_{t'-\tau',t-\tau}&
\nonumber \\
- \sigma^{2}\left(
 \theta_{\tau'+\tau} - \theta_{\tau'+\tau+1}
\right)\Big\}& 
=0, \label{reduced_SE_equation}  
\end{IEEEeqnarray}
with
\begin{IEEEeqnarray}{rl}
\mathfrak{D}_{\tau',\tau}
=& g_{\tau'+\tau} - g_{\tau'+\tau+1} +(g_{\tau-1}-g_{\tau})*\theta_{\tau'+\tau+1}  
\nonumber \\
&+ (\theta_{\tau} - \theta_{\tau-1})*g_{\tau'+\tau+1}
+ g_{\tau}*(\theta_{\tau'+\tau} - \delta_{\tau',0}\theta_{\tau}) 
 \nonumber \\
&- \theta_{\tau}*(g_{\tau'+\tau} - \delta_{\tau',0}g_{\tau}). 
\end{IEEEeqnarray}
 
We evaluate a fixed-point of the reduced SE 
equation~(\ref{reduced_SE_equation}) for the Bayes-optimal denoiser 
$f_{\mathrm{opt}}$. 
Suppose that $\lim_{t',t\to\infty}a_{t',t}=a_{\mathrm{s}}$, 
$\lim_{t',t\to\infty}d_{t',t}=d_{\mathrm{s}}$, and 
$\lim_{t\to\infty}\bar{\xi}_{t}=\xi_{\mathrm{s}}$ hold. 
The main feature of the Bayes-optimal denoiser is the identity 
$\xi_{\mathrm{s}}=d_{\mathrm{s}}/a_{\mathrm{s}}$~\cite[Lemma~2]{Takeuchi201}. 
We use this identity and the assumptions in Theorem~\ref{theorem_fixed_point} 
to prove the fixed-point~(\ref{fixed_point}). 

Taking the limits $t',t\to\infty$ in (\ref{reduced_SE_equation}) yields 
\begin{IEEEeqnarray}{rl}
&a_{\mathrm{s}}\sum_{\tau',\tau=0}^{\infty}
\mathfrak{D}_{\tau',\tau}(\xi_{\mathrm{s}}^{-1})^{-\tau'-\tau}  \nonumber \\
=& d_{\mathrm{s}}\sum_{\tau',\tau=0}^{\infty}
(g_{\tau}*\theta_{\tau'+\tau+1} - \theta_{\tau}*g_{\tau'+\tau+1})
(\xi_{\mathrm{s}}^{-1})^{-\tau'-\tau} \nonumber \\
&+ \sigma^{2}\sum_{\tau',\tau=0}^{\infty}
(\theta_{\tau'+\tau} - \theta_{\tau'+\tau+1})
(\xi_{\mathrm{s}}^{-1})^{-\tau'-\tau}. 
\end{IEEEeqnarray}
We use the properties of the Z-transform in Table~\ref{table_Z_transform} 
and the identity $\xi_{\mathrm{s}}=d_{\mathrm{s}}/a_{\mathrm{s}}$ 
to find 
\begin{equation} 
F_{G,\Theta}(y,z)\frac{d_{\mathrm{s}}}{\xi_{\mathrm{s}}}
=\{G(z)\Delta_{\Theta} - \Theta(z)\Delta_{G}\}d_{\mathrm{s}}  
+ (\Delta_{\Theta_{1}} - \Delta_{\Theta})\sigma^{2}  
\label{fixed_point_tmp}
\end{equation}
in the limit $y, z\to\xi_{\mathrm{s}}^{-1}$, where $F_{G,\Theta}$ is given by 
(\ref{Operator}). 

Series-expanding $\Delta_{S}$ with respect to $z^{-1}$ at $z=y$ up to the 
first order yields  
\begin{equation}
\lim_{y,z\to\xi_{\mathrm{s}}^{-1}}\Delta_{S}
= \frac{dS}{dz^{-1}}(\xi_{\mathrm{s}}^{-1}).  
\end{equation}
Similarly, we have 
\begin{equation}
\lim_{y,z\to\xi_{\mathrm{s}}^{-1}}\Delta_{S_{1}}
= S(\xi_{\mathrm{s}}^{-1}) + \xi_{\mathrm{s}}\frac{dS}{dz^{-1}}(\xi_{\mathrm{s}}^{-1}), 
\end{equation}
Applying these results to (\ref{fixed_point_tmp}) with (\ref{Operator}) yields 
\begin{equation}
\left\{
 1 + (\xi_{\mathrm{s}}-1)
 \frac{d\Theta}{dz^{-1}}(\xi_{\mathrm{s}}^{-1}) 
\right\}\left\{
 \frac{G(\xi_{\mathrm{s}}^{-1})d_{\mathrm{s}}}{\xi_{\mathrm{s}}}
 - \sigma^{2}
\right\} 
= 0,
\end{equation}
where we have used the assumption $\Theta(\xi_{\mathrm{s}}^{-1})=1$. 
Since $1+(\xi_{\mathrm{s}}-1)d\Theta(\xi_{\mathrm{s}}^{-1})/(dz^{-1})\neq0$ 
has been assumed, we arrive at 
\begin{equation} \label{G_idenity}
\frac{G(\xi_{\mathrm{s}}^{-1})}{\xi_{\mathrm{s}}} 
= \frac{\sigma^{2}}{d_{\mathrm{s}}}.  
\end{equation}

To prove the fixed-point~(\ref{fixed_point}), we use the 
relationship~(\ref{relationship}) between the $\eta$-transform and 
the R-transform. Evaluating (\ref{relationship}) at 
$x=x^{*}$ given in (\ref{pole}) and using Theorem~\ref{theorem_solution}, 
we obtain 
\begin{equation}
G(z) 
= \Theta(z) R\left(
 -\frac{1 - (1-z^{-1})\Theta(z)}{G(z)}\Theta(z)
\right).
\end{equation}
Letting $z=\xi_{\mathrm{s}}^{-1}$ and applying the assumption 
$\Theta(\xi_{\mathrm{s}}^{-1})=1$ yields 
\begin{equation}
G(\xi_{\mathrm{s}}^{-1}) 
= R\left(
 - \frac{\xi_{\mathrm{s}}}{G(\xi_{\mathrm{s}}^{-1})} 
\right). 
\end{equation}
Substituting (\ref{G_idenity}) into this identity and using 
$\xi_{\mathrm{s}}=d_{\mathrm{s}}/a_{\mathrm{s}}$, we arrive at 
\begin{equation} 
a_{\mathrm{s}}
= \frac{\sigma^{2}}{R(- d_{\mathrm{s}}/\sigma^{2})}. 
\end{equation}

\section{Evaluation of (\ref{correlation}) for Bernoulli-Gaussian signals} 
\label{proof_BG_signal}
\subsection{Summary}
We evaluate the correlation~(\ref{correlation}) for the Bernoulli-Gaussian 
signals. This appendix is organized as an independent section of the other 
parts. Thus, we use different notations from the other parts. 

Let $A\in\{0, 1\}$ denote a Bernoulli random variable taking $1$ with 
probability $\rho\in[0, 1]$. Suppose that $Z\sim\mathcal{N}(0,\rho^{-1})$ is 
independent of $A$ and a zero-mean Gaussian random variable with 
variance $\rho^{-1}$. We consider estimation of a Bernoulli-Gaussian signal 
$X=AZ$ on the basis of two dependent noisy observations,
\begin{equation} 
Y_{t'}=X+W_{t'}, \quad Y_{t}=X+W_{t},
\end{equation} 
with 
\begin{equation}
\begin{pmatrix}
W_{t'} \\
W_{t} 
\end{pmatrix}
\sim\mathcal{N}(\boldsymbol{0},\boldsymbol{\Sigma}), 
\quad \boldsymbol{\Sigma}
=\begin{pmatrix}
a_{t',t'} & a_{t',t} \\
a_{t',t} & a_{t,t} 
\end{pmatrix},
\end{equation}
where $\boldsymbol{\Sigma}$ is positive definite. 
The goal of this appendix is to evaluate the correlation $d_{t'+1,t+1}$ of 
the estimation errors for the Bayes-optimal denoiser 
$f_{\mathrm{opt}}(Y_{t};a_{t,t})=\mathbb{E}[X|Y_{t}]$,   
\begin{equation} \label{correlation_dif}
d_{t'+1,t+1}=\mathbb{E}[\{f_{\mathrm{opt}}(Y_{t'};a_{t',t'}) - X\}
\{f_{\mathrm{opt}}(Y_{t};a_{t,t}) - X\}].   
\end{equation}

Before presenting the derived expression of the 
correlation~(\ref{correlation_dif}), we first introduce 
several definitions. We write the pdf of a zero-mean Gaussian random variable 
$Y$ with variance $\sigma^{2}$ as $p_{\mathrm{G}}(y;\sigma^{2})$, with 
\begin{equation}
p_{\mathrm{G}}(y; \sigma^{2}) 
= \frac{1}{\sqrt{2\pi\sigma^{2}}}\exp\left(
 -\frac{y^{2}}{2\sigma^{2}}
\right). 
\end{equation}
The pdf of a Gaussian mixture is defined as 
\begin{equation} \label{p_Y} 
p_{\mathrm{GM}}(y;a_{t,t}) 
= \rho p_{\mathrm{G}}(y;\rho^{-1}+a_{t,t})
+ (1-\rho)p_{\mathrm{G}}(y;a_{t,t}), 
\end{equation}
which is used to represent the marginal pdf of $Y_{t}$. As proved in 
Appendix~\ref{appen_denoiser}, the probability of 
$A=1$ given $Y_{t}$ is given by 
$\mathrm{Pr}(A=1|Y_{t}=y)=\pi(y;a_{t,t})$, with 
\begin{equation} \label{conditional_A}
\pi(y,a_{t,t}) = \frac{\rho p_{\mathrm{G}}(y;\rho^{-1}+a_{t,t})}
{p_{\mathrm{GM}}(y;a_{t,t})}. 
\end{equation}
The Bayes-optimal denoiser $f_{\mathrm{opt}}(Y_{t};a_{t,t})$ 
is derived in the same appendix: 
\begin{equation} \label{Bayes_opt_denoiser}
f_{\mathrm{opt}}(y;a_{t,t}) 
= \frac{y}{1+\rho a_{t,t}}\pi(y,a_{t,t}),  
\end{equation}
where the conditional probability $\pi(y,a_{t,t})$ is 
given by (\ref{conditional_A}). 

We write the MSE function $\mathrm{MSE}(a_{t,t})$ as 
\begin{IEEEeqnarray}{rl}
&\mathrm{MSE}(a_{t,t}) \nonumber \\
=& \frac{a_{t,t}}{1+\rho a_{t,t}} 
+ \mathbb{E}[\{1-\pi(Y_{t},a_{t,t})\}\{f_{\mathrm{opt}}(Y_{t};a_{t,t})\}^{2}]  
\nonumber \\
+& \mathbb{E}\left[
 \pi(Y_{t},a_{t,t})
 \left\{
  \frac{Y_{t}}{1+\rho a_{t,t}} 
  - f_{\mathrm{opt}}(Y_{t};a_{t,t})
 \right\}^{2}
\right], \label{MSE} 
\end{IEEEeqnarray}
where the Bayes-optimal denoiser $f_{\mathrm{opt}}$ is given in 
(\ref{Bayes_opt_denoiser}). 
In (\ref{MSE}), the expectation  is over 
$Y_{t}\sim p_{\mathrm{GM}}(y;a_{t,t})$ given in (\ref{p_Y}).

The joint pdf of $\{Y_{t'}, Y_{t}\}$ is represented as  
\begin{equation} \label{joint_pdf}
p(Y_{t'}, Y_{t})=\rho p(Y_{t'}, Y_{t}|A=1) + (1-\rho)p(Y_{t'}, Y_{t}|A=0). 
\end{equation}
As proved in Appendix~\ref{appen_conditional}, 
the conditional pdf $p(Y_{t'}, Y_{t}|A)$ is given by
\begin{IEEEeqnarray}{rl}
&p(Y_{t'}, Y_{t}|A=a) 
= p_{\mathrm{G}}(Y_{t'};\rho^{-1}a+a_{t',t'}) \nonumber \\
&\cdot p_{\mathrm{G}}\left(
 Y_{t} - \frac{a+\rho a_{t',t}}{a+\rho a_{t',t'}}Y_{t'}; 
 \frac{a+\rho a_{t,t}}{\rho} - \frac{(a+\rho a_{t',t})^{2}}{\rho(a+\rho a_{t',t'})}
\right) \nonumber \\
\label{conditional_y}
\end{IEEEeqnarray}
for $a=0, 1$. 

\begin{proposition} \label{proposition1}
\begin{itemize}
\item Let $\mathrm{MSE}(a_{t,t})$ denote the MSE function~(\ref{MSE}). Then,  
\begin{equation} \label{d_tt} 
d_{t+1,t+1} = \mathrm{MSE}(a_{t,t}). 
\end{equation}

\item
For $t'\neq t$, let 
\begin{equation} \label{variance}
v_{t',t} = \frac{a_{t',t'}a_{t,t}-a_{t',t}^{2}}
{a_{t',t'} + a_{t,t} - 2a_{t',t}}. 
\end{equation}
Then, the correlation $d_{t'+1,t+1}$ for $t'\neq t$ is given by 
\begin{IEEEeqnarray}{l}
d_{t'+1,t+1}
= \mathbb{E}[f_{\mathrm{opt}}(Y_{t'};a_{t',t'})f_{\mathrm{opt}}(Y_{t};a_{t,t})] 
\nonumber \\
+ \mathbb{E}\left[
 \pi(Y_{t',t};v_{t',t})\left\{
  \left(
   \frac{Y_{t',t}}{1 + \rho v_{t',t}}
  \right)^{2}
  + \frac{\rho^{-1}v_{t',t}}{\rho^{-1}+v_{t',t}} 
 \right.
\right. \nonumber \\
\left.
 \left.
  - \frac{Y_{t',t}[f_{\mathrm{opt}}(Y_{t'};a_{t',t'})+f_{\mathrm{opt}}(Y_{t};a_{t,t})]}
  {1 + \rho v_{t',t}} 
 \right\}
\right],  
\label{d_t't}
\end{IEEEeqnarray}
with 
\begin{equation} \label{Y}
Y_{t',t} = \frac{(a_{t,t} - a_{t',t})Y_{t'} + (a_{t',t'} - a_{t',t})Y_{t}}
{a_{t',t'} + a_{t,t} - 2a_{t',t}}, 
\end{equation}
where the expectation in (\ref{d_t't}) over $\{Y_{t'},Y_{t}\}$ is evaluated 
via the joint pdf~(\ref{joint_pdf}).  
\end{itemize}
\end{proposition}
\begin{IEEEproof}
See from Appendix~\ref{appen_denoiser} to Appendix~\ref{appen_conditional}. 
\end{IEEEproof}

Proposition~\ref{proposition1} implies that $d_{t'+1,t+1}$ for $t'\neq t$ 
requires numerical computation of the double integrals.  

\subsection{Bayes-Optimal Denoiser} \label{appen_denoiser} 
We compute the Bayes-optimal denoiser 
$f_{\mathrm{opt}}(Y_{t};a_{t,t})=\mathbb{E}[X|Y_{t}]$, given by 
\begin{IEEEeqnarray}{rl}
f_{\mathrm{opt}}(Y_{t};a_{t,t}) 
=& \mathbb{E}\left[
 \left.
  \mathbb{E}[AZ|Y_{t}, A]
 \right| Y_{t} 
\right] \nonumber \\
=& \mathbb{E}[Z|Y_{t}, A=1]\mathrm{Pr}(A=1 | Y_{t}). 
\end{IEEEeqnarray}
Note that $f_{\mathrm{opt}}$ is different from the true posterior mean estimator 
(PME) $\mathbb{E}[X|Y_{t'}, Y_{t}]$. 

We first evaluate the former factor $\mathbb{E}[Z|Y_{t}, A=1]$. 
Since $Y_{t}=Z+W_{t}$ given $A=1$ is the AWGN observation of 
$Z\sim\mathcal{N}(0,\rho^{-1})$, we obtain the well-known LMMSE estimator 
\begin{equation} \label{LMMSE}
\mathbb{E}[Z|Y_{t}, A=1] 
= \frac{\rho^{-1}Y_{t}}{\rho^{-1}+a_{t,t}},
\end{equation}
which implies the Bayes-optimal denoiser~(\ref{Bayes_opt_denoiser}). 

We next prove that the latter factor $\mathrm{Pr}(A=1 | Y_{t})$ is equal to 
$\pi(Y_{t};a_{t,t})$ given in (\ref{conditional_A}). By definition,  
\begin{equation}
\mathrm{Pr}(A=1 | Y_{t})
= \frac{\rho p(Y_{t}|A=1)}{p(Y_{t})}. 
\end{equation}
For the numerator, we have  
\begin{IEEEeqnarray}{rl}
&p(Y_{t}|A=1) 
= \mathbb{E}_{Z}[p(Y_{t}|A=1, Z)]\nonumber \\
=& \mathbb{E}_{Z}[p_{\mathrm{G}}(Y_{t}-Z;a_{t,t})]  
= p_{\mathrm{G}}(Y_{t};\rho^{-1}+a_{t,t}), 
\end{IEEEeqnarray}
where the last equality follows from the fact that 
$Z+W_{t}$ is a zero-mean Gaussian random variable with 
variance $\rho^{-1}+a_{t,t}$. 

The denominator $p(Y_{t})$ is computed in the same manner,   
\begin{IEEEeqnarray}{rl}
&p(Y_{t}) = \rho p(Y_{t} | A=1) + (1-\rho)p(Y_{t} | A=0) \nonumber \\
=& \rho p_{\mathrm{G}}(Y_{t};\rho^{-1}+a_{t,t})
+ (1-\rho)p_{\mathrm{G}}(Y_{t};a_{t,t}), 
\end{IEEEeqnarray}
which is equal to $p_{\mathrm{GM}}(Y_{t};a_{t,t})$ given in (\ref{p_Y}). 
Combining these results, we arrive at $\mathrm{Pr}(A=1|Y_{t})
=\pi(Y_{t};a_{t,t})$ given in (\ref{conditional_A}). 

\subsection{MSE} 
To evaluate the MSE 
$d_{t+1,t+1}=\mathbb{E}[\{X-f_{\mathrm{opt}}(Y_{t};a_{t,t})\}^{2}]$, we focus on 
the posterior variance  
$\mathbb{E}[\{X-f_{\mathrm{opt}}(Y_{t};a_{t,t})\}^{2}| Y_{t}]$. 
By definition, 
\begin{IEEEeqnarray}{rl}
&\mathbb{E}[\{X-f_{\mathrm{opt}}(Y_{t};a_{t,t})\}^{2}| Y_{t}] \nonumber \\
=& \mathrm{Pr}(A=1|Y_{t})
\mathbb{E}[\{Z-f_{\mathrm{opt}}(Y_{t};a_{t,t})\}^{2}| Y_{t}, A=1] \nonumber \\
&+ \{1 - \mathrm{Pr}(A=1|Y_{t})\}\{f_{\mathrm{opt}}(Y_{t};a_{t,t})\}^{2}, 
\end{IEEEeqnarray}
with $\mathrm{Pr}(A=1|Y_{t})=\pi(Y_{t},a_{t,t})$ given in (\ref{conditional_A}). 

Let $\mathbb{E}[Z|Y_{t}, A=1]$ denote the PME of $Z$ conditioned on 
$Y_{t}$ and $A=1$, given in (\ref{LMMSE}). 
The conditional expectation in the first term can be evaluated as follows: 
\begin{IEEEeqnarray}{rl}
&\mathbb{E}[\{Z-f_{\mathrm{opt}}(Y_{t};a_{t,t})\}^{2}| Y_{t}, A=1] 
= \mathbb{E}[\{Z-\mathbb{E}[Z| Y_{t}, A=1] \nonumber \\
&\left.
  \left.
   \left.
   + \mathbb{E}[Z|Y_{t}, A=1 ] -f_{\mathrm{opt}}(Y_{t};a_{t,t})
   \right\}^{2}
 \right| Y_{t}, A=1 
\right] \nonumber \\
&= \frac{\rho^{-1}a_{t,t}}{\rho^{-1}+a_{t,t}} 
+ \left\{
 \frac{Y_{t}}{1+\rho a_{t,t}} - f_{\mathrm{opt}}(Y_{t};a_{t,t})
\right\}^{2}. 
\end{IEEEeqnarray}
Combining these results and taking the expectation over 
$Y_{t}\sim p(Y_{t})=p_{\mathrm{GM}}(Y_{t};a_{t,t})$ given in (\ref{p_Y}), 
we arrive at the MSE~(\ref{d_tt}).  

\subsection{Sufficient Statistic}
As a preliminary step for computing the correlation~(\ref{correlation_dif}) 
for $t'\neq t$, 
we derive a sufficient statistic of $X$ based on 
the two correlated observations $\{Y_{t'}, Y_{t}\}$. 

Let $\boldsymbol{\Sigma}^{-1/2}$ denote a square root of 
$\boldsymbol{\Sigma}^{-1}$, 
i.e.\ $(\boldsymbol{\Sigma}^{-1/2})^{2}=\boldsymbol{\Sigma}^{-1}$. Applying 
the noise whitening filter $\boldsymbol{\Sigma}^{-1/2}$ to 
the observation vector $(Y_{t'}, Y_{t})^{\mathrm{T}}$ yields 
\begin{equation} \label{sufficient_statistic_tmp}
\boldsymbol{\Sigma}^{-1/2}
\begin{pmatrix}
Y_{t'} \\
Y_{t} 
\end{pmatrix} 
= \boldsymbol{\Sigma}^{-1/2}\boldsymbol{1}_{2}X
+ \boldsymbol{\Sigma}^{-1/2}
\begin{pmatrix}
W_{t'} \\
W_{t} 
\end{pmatrix},
\end{equation}
with $\boldsymbol{1}_{2}=(1, 1)^{\mathrm{T}}$. 
Note that the effective noise vector---the second term on the RHS---follows 
the standard Gaussian distribution. It is well-known that the MF output is 
a sufficient statistic of $X$ when the effective noise vector has zero-mean 
i.i.d.\ Gaussian elements. Applying the MF  
$(\boldsymbol{\Sigma}^{-1/2}\boldsymbol{1}_{2})^{\mathrm{T}}
/\boldsymbol{1}_{2}^{\mathrm{T}}\boldsymbol{\Sigma}^{-1}\boldsymbol{1}_{2}$ 
to (\ref{sufficient_statistic_tmp}), 
we arrive at a sufficient statistic $Y_{t',t}$, given by 
\begin{equation} \label{sufficient_statistic}
Y_{t',t} = \frac{\boldsymbol{1}_{2}^{\mathrm{T}}\boldsymbol{\Sigma}^{-1}}
{\boldsymbol{1}_{2}^{\mathrm{T}}\boldsymbol{\Sigma}^{-1}\boldsymbol{1}_{2}}
\begin{pmatrix}
Y_{t'} \\
Y_{t} 
\end{pmatrix} 
= X + W_{t',t}, 
\end{equation}
with 
\begin{equation}
W_{t',t} = \frac{\boldsymbol{1}_{2}^{\mathrm{T}}\boldsymbol{\Sigma}^{-1}}
{\boldsymbol{1}_{2}^{\mathrm{T}}\boldsymbol{\Sigma}^{-1}\boldsymbol{1}_{2}}
\begin{pmatrix}
W_{t'} \\
W_{t} 
\end{pmatrix}. 
\end{equation}
It is straightforward to confirm that 
the sufficient statistic~(\ref{sufficient_statistic}) reduces to 
(\ref{Y}). Furthermore, we find $W_{t',t}\sim\mathcal{N}(0,v_{t',t})$, 
with $v_{t',t}=(\boldsymbol{1}_{2}^{\mathrm{T}}
\boldsymbol{\Sigma}^{-1}\boldsymbol{1}_{2})^{-1}$, which reduces to 
(\ref{variance}). 

\subsection{Correlation} \label{appen_correlation} 
To evaluate the correlation~(\ref{correlation_dif}) for $t'\neq t$, 
we first derive a few quantities associated with the sufficient 
statistic~(\ref{sufficient_statistic}). 

The probability of $A=1$ given $Y_{t'}$ and $Y_{t}$ is equal to that of $A=1$ 
given the sufficient statistic~(\ref{sufficient_statistic}). Thus, 
repeating the derivation of $\mathrm{Pr}(A=1|Y_{t})=\pi(Y_{t};a_{t,t})$ given 
in (\ref{conditional_A}), we have 
\begin{equation} \label{conditional_A_Y}
\mathrm{Pr}(A=1|Y_{t'},Y_{t}) = \pi(Y_{t',t}; v_{t',t}), 
\end{equation} 
where $Y_{t',t}$ and $v_{t',t}$ are given by (\ref{Y}) and (\ref{variance}). 
Similarly, repeating the derivation of (\ref{LMMSE}) implies that 
the PME $\mathbb{E}[Z|Y_{t'}, Y_{t}, A=1]$ reduces to 
\begin{equation} \label{true_LMMSE} 
\mathbb{E}[Z|Y_{t'}, Y_{t}, A=1] 
= \frac{Y_{t',t}}{1 + \rho v_{t',t}}. 
\end{equation}
Furthermore, the true PME $\mathbb{E}[X|Y_{t'},Y_{y}]$ is given by 
\begin{equation} \label{true_PME}
\mathbb{E}[X|Y_{t'},Y_{t}]
= f_{\mathrm{opt}}(Y_{t',t}; v_{t',t}).
\end{equation}

We next evaluate the posterior covariance 
\begin{IEEEeqnarray}{rl}
& \mathbb{E}[\{f_{\mathrm{opt}}(Y_{t'};a_{t',t'}) - X\}
\{f_{\mathrm{opt}}(Y_{t};a_{t,t}) - X\} | Y_{t'}, Y_{t}] \nonumber \\
=& \mathrm{Pr}(A=0|Y_{t'}, Y_{t})f_{\mathrm{opt}}(Y_{t'};a_{t',t'})
f_{\mathrm{opt}}(Y_{t};a_{t,t}) \nonumber \\
&+ \mathrm{Pr}(A=1|Y_{t'}, Y_{t})\mathbb{E}[\{f_{\mathrm{opt}}(Y_{t'};a_{t',t'}) - Z\}
\nonumber \\
&\cdot\{f_{\mathrm{opt}}(Y_{t};a_{t,t}) - Z\} | Y_{t'}, Y_{t}, A=1 ].  
\label{posterior_covariance} 
\end{IEEEeqnarray}
Substituting (\ref{conditional_A_Y}) into (\ref{posterior_covariance}) and 
using $f_{\mathrm{opt}}(Y_{\tau};a_{\tau,\tau}) - Z
= \{f_{\mathrm{opt}}(Y_{\tau};a_{\tau,\tau}) - \mathbb{E}[Z|Y_{t'},Y_{t},A=1]\} 
+ \{\mathbb{E}[Z|Y_{t'},Y_{t},A=1] - Z\}$ with (\ref{true_LMMSE}) for 
$\tau=t', t$, we have 
\begin{IEEEeqnarray}{rl}
&\mathbb{E}[\{f_{\mathrm{opt}}(Y_{t'};a_{t',t'}) - X\}
\{f_{\mathrm{opt}}(Y_{t};a_{t,t}) - X\} | Y_{t'}, Y_{t}] \nonumber \\
=& \{1 - \pi(Y_{t',t};v_{t',t})\}f_{\mathrm{opt}}(Y_{t'};a_{t',t'})
f_{\mathrm{opt}}(Y_{t};a_{t,t}) \nonumber \\
&+ \pi(Y_{t',t};v_{t',t})\left\{
 \left[
  f_{\mathrm{opt}}(Y_{t'};a_{t',t'}) - \frac{Y_{t',t}}{1 + \rho v_{t',t}}
 \right]
\right. \nonumber \\
&\left.
 \cdot\left[
  f_{\mathrm{opt}}(Y_{t};a_{t,t}) - \frac{Y_{t',t}}{1 + \rho v_{t',t}}
 \right]
 + \frac{\rho^{-1}v_{t',t}}{\rho^{-1}+v_{t',t}}
\right\},  \label{posterior_covariance_tmp} 
\end{IEEEeqnarray}
where $Y_{t',t}$ is computed with $\{Y_{t'}, Y_{t}\}$, as given in (\ref{Y}). 

Finally, we derive the correlation~(\ref{correlation_dif}). 
Taking the expectation of the posterior 
covariance~(\ref{posterior_covariance_tmp}) over $Y_{t'}$ and $Y_{t}$,  
we arrive at (\ref{d_t't}).  

\subsection{Joint pdf} \label{appen_conditional}
To compute the expectation in (\ref{d_t't}), 
we need the conditional pdf $p(Y_{t'}, Y_{t}|A)$ 
in the joint pdf~(\ref{joint_pdf}) of $\{Y_{t'}, Y_{t}\}$.  

We first evaluate the conditional distribution of $W_{t}$ given $W_{t'}$. 
Let 
\begin{equation} \label{w2}
W_{t} = \alpha W_{t'} + \sqrt{\beta}\tilde{W}, 
\end{equation}
with some constants $\alpha\in\mathbb{R}$ and $\beta>0$,  
where $\tilde{W}$ is a standard Gaussian random variable independent of 
$W_{t'}$. Computing the correlation $\mathbb{E}[W_{t'}W_{t}]$ and 
variance $\mathbb{E}[W_{t}^{2}]$, we obtain 
\begin{equation}
\mathbb{E}[W_{t'}W_{t}]
= \alpha\mathbb{E}[W_{t'}^{2}], 
\end{equation}
\begin{equation}
\mathbb{E}[W_{t}^{2}] 
= \alpha^{2}\mathbb{E}[W_{t'}^{2}] + \beta. 
\end{equation}
We use the definitions $\mathbb{E}[W_{\tau}^{2}]=a_{\tau,\tau}$ for $\tau=t', t$ 
and $\mathbb{E}[W_{t'}W_{t}]=a_{t',t}$ to have  
$\alpha=a_{t',t}/a_{t',t'}$ and $\beta= a_{t,t} - a_{t',t}^{2}/a_{t',t'}$. 
Thus, (\ref{w2}) implies  
\begin{equation} \label{conditional_W}
W_{t}\;\hbox{conditioned on $W_{t'}$}\sim 
\mathcal{N}\left(
 \frac{a_{t',t}W_{t'}}{a_{t',t'}}, 
 a_{t,t} - \frac{a_{t',t}^{2}}{a_{t',t'}}
\right). 
\end{equation}

We next evaluate the conditional pdf $p(Y_{t'}, Y_{t}|A)$ for $A=0$.  
Since $Y_{\tau}=W_{\tau}$ holds for $A=0$, we have 
\begin{IEEEeqnarray}{rl}
&p(Y_{t'}, Y_{t}|A=0)
= p(W_{t'}=Y_{t'}, W_{t}=Y_{t})  \nonumber \\
=& p_{\mathrm{G}}\left(
 Y_{t} - \frac{a_{t',t}}{a_{t',t'}}Y_{t'};  
 a_{t,t} - \frac{a_{t',t}^{2}}{a_{t',t'}}
\right)p_{\mathrm{G}}(Y_{t'}; a_{t',t'}). 
\end{IEEEeqnarray}
For $A=1$, we use $Y_{\tau}=Z+W_{\tau}$ to find that 
$\{Y_{t'}, Y_{t}\}$ given $A=1$ are zero-mean Gaussian random variables 
with covariance, 
\begin{equation}
\mathbb{E}[Y_{\tau}^{2} | A=1] = \rho^{-1} + a_{\tau,\tau}
\quad \hbox{for $\tau=t', t$,} 
\end{equation}  
\begin{equation}
\mathbb{E}[Y_{t'}Y_{t} | A=1] = \rho^{-1} + a_{t',t}. 
\end{equation}
Repeating the derivation of (\ref{conditional_W}), we obtain 
\begin{IEEEeqnarray}{rl}
&p(Y_{t'}, Y_{t}|A=1) 
= p_{\mathrm{G}}(Y_{t'};\rho^{-1}+a_{t',t'}) \nonumber \\
&\cdot p_{\mathrm{G}}\left(
 Y_{t} - \frac{\rho^{-1}+a_{t',t}}{\rho^{-1}+a_{t',t'}}Y_{t'}; 
 \rho^{-1}+a_{t,t} - \frac{(\rho^{-1}+a_{t',t})^{2}}{\rho^{-1}+a_{t',t'}}
\right). \nonumber \\
\end{IEEEeqnarray}
Combining these results, we arrive at the conditional 
pdf~(\ref{conditional_y}).

\section*{Acknowledgment}
The author thanks the anonymous reviewers for their suggestions that have 
improved the quality of the manuscript greatly.

\ifCLASSOPTIONcaptionsoff
  \newpage
\fi

\balance



\bibliographystyle{IEEEtran}
\bibliography{IEEEabrv,kt-it2020}
\end{document}